\documentclass[%
reprint,
superscriptaddress,
prd,
 amsmath,
 amssymb,
 aps,
 longbibliography,
 nofootinbib,
]{revtex4-2}
\usepackage{graphicx}
\usepackage{lipsum}
\usepackage{lipsum, babel}
\usepackage{graphicx} % For including images
\usepackage{float}
\usepackage[titletoc]{appendix}
\usepackage{cancel}
\usepackage{graphicx}% Include figure files
\usepackage{dcolumn}% Align table columns on decimal point
\usepackage{bm}% bold math
\usepackage[usenames,dvipsnames]{color}
\usepackage{xcolor}
\usepackage{soul}
\usepackage{subcaption}
\usepackage{enumitem}
\usepackage{xspace}
\usepackage{booktabs}
\usepackage{lipsum}
\usepackage{siunitx} % Add support for units
\usepackage[ISO]{diffcoeff}
\usepackage{subcaption}
\usepackage{ragged2e}
\usepackage{mathrsfs}
\usepackage{hyperref}
\usepackage{url}
\usepackage[capitalise]{cleveref}
\usepackage[normalem]{ulem}
\usepackage{amsmath}
\usepackage{wrapfig}
\usepackage{esvect} %for correct vectors using \vv
\usepackage{titlesec}
\usepackage{titletoc}
\usepackage{chngcntr}
\usepackage{fontawesome5}
\usepackage{mleftright}
\usepackage[nodayofweek]{datetime}
\usepackage{comment}
\usepackage{adjustbox}

\DeclareSIUnit\year{yr} 
\def\arraystretch{2}\tabcolsep=10pt

\DeclareCaptionJustification{justified}{\justifying}
\definecolor{firebrick}{HTML}{B22222}

\hypersetup{
    colorlinks=true,       % false: boxed links; true: colored links
    linkcolor=firebrick,          % color of internal linkshttps://www.overleaf.com/project/65dd05895a901323fe62bed3
    citecolor=firebrick,        % color of links to bibliography
    filecolor=firebrick,      % color of file links
    urlcolor=firebrick           % color of external links
}

\definecolor{orcid-green}{RGB} {166, 206, 57}
\newcommand{\MYhref}[3][blue]{\href{#2}{\color{#1}{#3}}}%

\titleclass{\mysection}{straight}[\section]
\titleformat{\mysection}[runin]
  {\itshape}{\thesection}{}{}[.---]
\titlespacing{\mysection}{1em}{1em}{0em}

\newcommand\funop[1]{\mathop{{}#1}}
\newcommand{\dd}{\mathop{}\!\mathrm{d}}
\newcommand{\mrm}[1]{\mathrm{#1}}
\newcommand{\usim}{\mathord{\sim}}

\DeclareSIUnit\clight{c}

\begin{document}

\title{\boldmath  
 Dipole Radiation and Kinetic Mixing from Dark Photon Solitons}

\author{Enrico D.~Schiappacasse\,\MYhref[orcid-green]{
https://orcid.org/0000-0002-6136-1358}{\faOrcid}}
\email{enrico.schiappacasse@uss.cl}
\affiliation{Facultad de Ingenier\'ia, Universidad San Sebasti\'an, Bellavista 7, Santiago 8420524, Chile}

\author{Moira Venegas\,\MYhref[orcid-green]{https://orcid.org/0000-0002-8678-9051}{\faOrcid}}

\affiliation{Department of Physics and Astronomy, Rice University, MS-315,
Houston, TX, 77005, U.S.A.}

\date{\today}

\begin{abstract}
Wave-like dark matter composed of spin-1 particles, known as dark photons, is theorized to form clumps called ``vector solitons". These solitons are compact astrophysical objects that exhibit coherent oscillations and a high concentration relative to the local dark matter density.  A significant portion of dark matter in galactic halos today may consist of these solitons. This study explores how photons can be produced from these vector solitons by the influence of external electromagnetic fields or charge densities, via a dimension-6 dark photon-photon coupling and a kinetic mixing, respectively. We further explore the astrophysical implications of these phenomena, highlighting a novel avenue for dark matter discovery that our research provides.
\end{abstract}
\maketitle

\tableofcontents

%%%%%%%%%%%%%%%%%%%%%%%%%%%%%%%%%%%%%%%%%%%%%%%%%%%%%%%%%%%%%%%%%%%%%%%
\section{Introduction}

Dark matter (DM) is the predominant component of non-relativistic matter in our universe. However, our comprehension of its fundamental characteristics remains surprisingly limited. While it is known to interact through gravity, crucial properties such as its mass, spin, and potential additional interactions are still unknown. \,\cite{Bertone:2016nfn,Freese:2017idy,Cirelli:2024ssz}. Astrophysical studies indicate that its mass could range from $10^{-19}$ eV to a few solar masses, covering nearly 90 orders of magnitude\,\cite{Brandt:2016aco,Dalal:2022rmp,Amin:2022nlh}. The ultralight regime for DM, where its mass is $\leq$ 10 eV, has been gaining considerable attention in recent years\,\cite{Essig:2013lka, Antypas:2022asj}.

In the ultralight regime, DM particles must be bosonic to match the observed DM density, leading to wavelike properties. In this regime, we can describe the related field classically due to its high occupancy number and the overlap of de Broglie wavelengths. Well-known ultralight candidates consist of the QCD axion\,\cite{1978PhRvL..40..223W, Wilczek:1977pj,DiLuzio:2020wdo}, axionlike particles and other scalar particles\,\cite{Arvanitaki:2009fg,Arias:2012az, Ringwald:2014vqa, Ferreira:2020fam}, as well as higher-spin bosonic particles like vector\,\cite{Jaeckel:2012mjv, Fabbrichesi:2020wbt, Caputo:2021eaa} and tensor particles\,\cite{Gorji:2025tos, Blas:2024kps, Gorji:2023cmz}. In this study, the light dark matter candidate we focus on is a vector particle, or dark photon, which can arise from a symmetry similar to that of the Standard Model photon.

Light dark photon dark matter can be produced through several mechanisms in the early universe, including production from inflationary perturbations\,\cite{Graham:2015rva}, decay of topological defects like cosmic strings\,\cite{Long:2019lwl, Kitajima:2022lre}, and instabilities from coupling with fields in the dark sector\,\cite{Zhang:2025lwr}, among others.

The light dark photon is linked to intriguing and novel phenomena arising from its wavelike characteristics, including interference, coherence, and resonance. One of the most compelling forecasts concerning light bosonic dark matter involves the emergence of long-lasting, coherent structures that are spatially confined, referred to as solitons\,\cite{Jain:2021pnk, Zhang:2021xxa, Zhang:2024bjo}. Recently, these solitons have attracted considerable interest from the physics community due to their distinctive characteristics, suggesting that they might serve as prospective astrophysical laboratories in galaxies.

 Dark photon solitons may arise through various processes. For instance, inflationary isocurvature fluctuations of the dark photon field can undergo gravitational collapse around the time of matter-radiation equality, leading to the creation of solitons\,\cite{Gorghetto:2022sue}. Alternatively, dark photons residing within virialized dark matter halos can gravitationally condense via kinetic relaxation\,\cite{Amin:2022pzv, Jain:2023ojg},
 similar to how solitons composed of spin-0\,\cite{Levkov:2018kau, Hertzberg:2020hsz, Yin:2024xov} or spin-2\,\cite{Jain:2021pnk, Schiappacasse:2025mao} do.

 Solitons contain a much higher density of dark matter compared to the local dark matter density and show significant resilience against disruption by galactic tides. These features make them an ideal tool for developing indirect dark matter search strategies\,\cite{Jain:2021pnk, Schiappacasse:2025mao}. 

In the absence of external electromagnetic fields or charged particles, dark photon solitons have demonstrated the ability to undergo photon parametric resonance \,\cite{Amin:2023imi}, resulting in brief, narrow, and spectrally constrained high-energy surges. These surges can possess energies that considerably surpass those of the previously observed fast radio bursts\,\cite{Petroff:2016tcr,Katz:2018xiu, Petroff:2019tty}, offering a significant observational pathway for detecting spin-1 dark matter through transient radio signals, provided the mass of the dark photon lies within the range of $(10^{-6}-10^{-4})$ eV\,\cite{Amaral:2025fcd}.

In this study, within the framework of effective field theory (EFT), we primarily focus on investigating the radiative effects generated by dark photon solitons when they are influenced by external electromagnetic fields or charge distributions. In particular:  

(i) We analyze the dipole radiation effect caused by a dimension-6 operator that connects dark photons to ordinary photons as vector solitons traverse external electromagnetic fields. The soliton and the electromagnetic background fields induce effective charge and current densities that both source the radiation emitted.

(ii) When vector solitons interact with external charge densities, we examine the radiative effects associated with gauge kinetic mixing, which results from the interaction between the light dark photon field and charged matter. Charge particles, such as electrons, feel a Lorentz force inside the soliton, which comes from the dark electromagnetic field generated by the oscillating vector boson field. The generated oscillating current density is associated with a time- and space-dependent charge density, and both source the radiation emitted.

Using analytical calculations, we estimate the radiation power and spectral flux density associated with these phenomena. Our findings indicate that the predicted radiated signal in both cases is strongly dependent on the vector soliton's shape and on plasma effects. When plasma effects are neglected, the signal is exponentially suppressed in diluted soliton configurations. However, when we account for an effective photon mass, the signal is greatly enhanced in the resonance regime. Favorable scenarios include the presence of dark photon solitons within the magnetospheres of highly magnetized compact objects, such as neutron stars and white dwarfs.

The article is structured as follows. Section\,\ref{Sec:EFT} provides a brief introduction to the key characteristics of dark photon solitons and the dark photon-photon interactions that are our focus. In Sects.\,\ref{Sec:DR} and\,\ref{Sec:KM}, we utilize analytical methods to determine the spectrum of electromagnetic radiation produced by dark photon solitons when subjected to an external electromagnetic field or charge densities, respectively. In addition, we address the influence of plasma effects. In Sect.\,\ref{Sec:AS}, we examine potential observational signatures, highlighting the distinct features of the anticipated signal and the practicality of detection. Finally, in Sect.\,\ref{Sec.Soutlook}, we summarize the main conclusions of this work and discuss potential future research avenues.

%%%%%%%%%%%%%%%%%%%%%%%%%%%%%%%%%%%%%%%%%%%%%%%%%%%%%%%%%%%%%%%%%%%%%%%
\section{Effective non-relativistic theory}
\label{Sec:EFT}

The effective non-relativistic theory for a spin-1 field has been thoroughly addressed in the literature. In this section, we will highlight the key points of interest. For more detailed information, please refer to sources such as~\cite{Adshead:2021kvl, Jain:2021pnk, Amin:2022pzv}. We utilize natural units where $\hbar = c = 1$ and define the reduced Planck mass as $m_{\text{pl}}=1/\sqrt{8\pi G_{\text{N}}}$. Additionally, we adopt the metric signature (-+++).

%%%%%%%%%%%%%%%%%%%%%%%%%%%%%%%%%%%%%%%%%%%%%%%%%%%%%%%%%%%%%%%%%%%%%%%
\subsection{General Action}  

Consider a massive real vector field denoted as \( W_\mu(x) \), which is referred to as the dark photon field, that interacts with electromagnetism. The characteristics and interactions of this field are captured in the action\,\cite{Amin:2022pzv, Amin:2023imi}
\begin{align}
\label{eq:action}
    &S[W_\mu(x), A_\mu(x), \text{g}_{\mu\nu}(x)]=\int \! \text{d}^4 x \, \sqrt{-\text{g}} \biggl[ - \frac{1}{4} W_{\mu\nu} W^{\mu\nu} \biggr. \nonumber\\
    &\textcolor{white}{XXX}\biggl.- \frac{1}{2} m^2 W_\mu W^\mu - \frac{1}{4} F_{\mu\nu} F^{\mu\nu} + \frac{1}{2} m_{\text{pl}}^2 \text{R} + \mathcal{L}_\mathrm{int} \biggr]\,    .
\end{align}
The dark photon field strength tensor is defined as \( W_{\mu\nu} = \nabla_\mu W_\nu - \nabla_\nu W_\mu \), while the electromagnetic field strength tensor is given by \( F_{\mu\nu} = \nabla_\mu A_\nu - \nabla_\nu A_\mu \). Additionally, \( \text{R} \) represents the Ricci scalar, and the indices are raised and lowered using the metric \( \text{g}_{\mu\nu}(x) \). Building upon previous research regarding dark photons, we introduce interactions between $W_{\mu}(x)$ and the electromagnetic field $A_{\mu}(x)$ or the
fourth vector current of ordinary matter $J_{\mu}(x)$, represented by $\mathcal{L}_{\text{int}}$. 

We are interested in spin-1 massive particles as cold dark matter. This means that their typical wavenumbers are much smaller than the dark photon mass, or, equivalently, their de Broglie wavelength is much greater than the inverse of the dark photon mass, i.e., \( \lambda \gg 2\pi/m \). This inequality enables us to work in the zero spatial gradient approximation locally, implying that \( \nabla W_{\mu} = 0 \). If we temporarily disregard both the gravitational and non-gravitational interactions of the field in Eq.\,(\ref{eq:action}), we note that the temporal component of the dark photon field is not dynamical. Therefore, \( W_0(x) = (\nabla^2 - m^2)^{-1}(\bm\nabla \cdot {\bf W} )\) can be safely set to zero in leading order of our gradient expansion.

%%%%%%%%%%%%%%%%%%%%%%%%%%%%%%%%%%%%%%%%%%%%%%%%%%%%%%%%%%%%%%%%%%%%%%%
\subsection{Interactions with electromagnetism} 

In the framework of EFT, our goal is to explore radiative effects caused by vector solitons under the influence of external electromagnetic fields (dipole radiation phenomenon\,\cite{Amin:2021tnq}) or charge distributions (gauge kinetic mixing phenomenon\,\cite{Holdom:1985ag, Foot:1991kb}) through interactions between the dark photon field $W_{\mu}$ and the electromagnetic field $A_{\mu}$ or the fourth vector current of ordinary matter $J_{\mu}$, respectively.  

The dipole radiation phenomenon within oscillons and axion stars embedded in a background electromagnetic field was studied in Ref.~\cite{Amin:2021tnq}. The associated coherent photon emission in such a case is supported by an interacting Lagrangian density
$\mathcal{L}_{\text{int}} = -(1/4) g_{a\gamma} a F_{\mu\nu}\tilde{F}^{\mu\nu}$, where $a(\bm x,t)$ is the axion field and $g_{a\gamma}$ is its coupling with photons.
We are interested in extending this phenomenon to vector-boson solitons. We will see that such compact objects behave as charge densities and currents, emitting electromagnetic radiation when immersed in external electromagnetic fields. Following Ref.\,\cite{Jain:2021pnk}, we work with an electromagnetic gauge invariant dimension-6 operator $\mathcal{O}$, such that the interacting Lagrangian density in Eq.\,(\ref{eq:action}) reads as
\begin{equation}
\mathcal{L}^{(6)}_{\text{int}} =  g^2_{W\gamma} \mathcal{O} = -\frac{1}{4} g^2_{W\gamma} W_{\mu}W^{\mu} F_{\alpha \beta} \tilde{F}^{\alpha \beta}\,.
\label{Eq:Odipoleradiation}
\end{equation}
Here $g^2_{W\gamma}$ is the coupling and $g^2_{W\gamma}\bar{W}^2 \ll 1$ is required for the validity of the EFT\,\cite{Amin:2023imi}, where $\bar{W}$ is understood as the typical amplitude of the dark photon field ${\bm{W}}(t, \bm x)$. The Lorentz indices are contracted through various combinations of the diagonal inverse Minkowski metric $\eta^{\mu\nu}=(-,+,+,+)$ and the completely antisymmetric Levi-Civita symbol $\epsilon^{0123}=1$. Note that the dark photon field contracts itself in Eq.\,(\ref{Eq:Odipoleradiation}), behaving as the square of a spin-0 field in this sense and \textit{emulating} from this perspective the operator used in Ref.~\cite{Amin:2021tnq} for the case of axionic fields. This similarity will help us to compare our results with the previous one available in the literature\,\footnote{For a complete list of dimension-6 operators coupling spin-1 particles and electromagnetism in the framework of EFT and obeying electromagnetic gauge invariance, see Sec. 2.2 in Ref.\,\cite{Amin:2023imi}. 
\textcolor{black}{In principle, operators depending only on the scalar combination $W^\mu W_\mu$ are expected to yield qualitatively similar behavior to the one studied here, including the absence of }\textcolor{black}{radiation for exactly circularly polarized configurations (see Sec.\,\ref{sec:PEDR}). Other operators, for instance those involving derivatives, may instead be sensitive to the phase and orientation of the soliton field and could lead to radiation even in circularly polarized cases, as well as different angular emission patterns. A detailed analysis of these possibilities is left for future work.}
}.

The gauge kinetic mixing arises from a unique electromagnetic gauge invariant dimension-4 operator of the form $\mathcal{L}^{(4)}_{\text{int}}  \supset F_{\mu\nu}W_{\alpha\beta}$,
where combinations of $\eta^{\mu\nu}$ and the $\epsilon^{0123}$ are used to contract Lorentz indices as previously. By applying a field redefinition, the gauge kinetic mixing can be transformed into a coupling with charged matter as\,\cite{Fabbrichesi:2020wbt}
\begin{equation}
 \mathcal{L}^{(4)}_{\text{int}} = -e\epsilon J_{\mu}W^{\mu}\,,   \label{eq:LKM}
\end{equation}
where $\epsilon$ is the arbitrary mixing parameter and $J^{\mu}$ is the fourth vector current of ordinary matter.

%%%%%%%%%%%%%%%%%%%%%%%%%%%%%%%%%%%%%%%%%%%%%%%%%%%%%%%%%%%%%%%%%%%%%%%
\subsection{Effective non-relativistic action} 
 
Turning off the non-gravitational interactions, 
in the effective non-relativistic theory, consider ${\bf{W}}(t,\bm{x})$ as the physical degrees of freedom of a spin-1 field and express it in terms of a slowly varying field ${\bf{\Psi}}$ as\,\cite{Jain:2021pnk} 
\begin{equation}
{\bf{W}} = \frac{1}{\sqrt{2m}} \left[ e^{-imt}{\bf{\Psi}}(t,{\bm{x}}) + h.c. \right]\,,\label{eq:W}    
\end{equation}
where $h.c.$ indicates hermitian conjugate, $m$ is the vector boson mass, and ${\bf{\Psi}}$ has dimensions of $[\text{length}]^{-3/2}$. 

The effective non-relativistic action and the associated multi-component Schr$\ddot{\text{o}}$dinger-Poisson (SP) equations in flat space-time read as follows\,\cite{Jain:2021pnk} 
\begin{equation}
\mathcal{S}^{\text{eff}}_{\text{nr}} = \int d^4 x \left[ \frac{i}{2}\left( \text{Tr}\left[{\bf{\Psi}^{\dagger}{\bf{\dot{\Psi}}}}\right] -  \text{Tr}\left[{\bf{\Psi}{\bf{\dot{\Psi}}}^{\dagger}}\right] \right) + m_{\text{pl}}^2\Phi_{\text{N}} \nabla^2 \Phi_{\text{N}}\,\right.\nonumber
\label{Eq:Snr}
\end{equation}
\vspace{-0.6cm}
\begin{equation}
\left.-\frac{1}{2m}\text{Tr}\left[\nabla{\bf{\Psi}}^{\dagger}\cdot \nabla \bf{\Psi}\right] - m \Phi \text{Tr}\left[{\bf{\Psi}}^{\dagger}\bf{\Psi}\right]   \right]\,, 
\end{equation}
\vspace{-0.4cm}
\begin{equation}
i\frac{\partial}{\partial t}{\bf{\Psi}} = -\frac{1}{2m}\nabla^2{\bf{\Psi}} + m\bm\Psi{\Phi}_{\text{N}}\,,\,\,\nabla^2\Phi_{\text{N}}=\frac{m}{2m^2_{\text{pl}}}\text{Tr}\left[{\bf{\Psi}^{\dagger}}{\bf{\Psi}}\right]\,,    
\end{equation}
where $\text{Tr}\left[{\bf{\Psi}}^{\dagger}{\bf{\Psi}}\right] = \psi^{\dagger}_i \psi_i$ and
$\Phi_{\text{N}}$ is the Newtonian gravitational potential. This effective action is valid in the weak-field Newtonian regime. It applies to the non-relativistic modes of the dark photon field, provided their wavenumbers are significantly smaller than the particle mass, e.g., $k \ll m$.

%%%%%%%%%%%%%%%%%%%%%%%%%%%%%%%%%%%%%%%%%%%%%%%%%%%%%%%%%%%%%%%%%%%%%%%
\subsection{Symmetries of the theory}

The different symmetries present in the effective non-relativistic action, Eq.\,(\ref{Eq:Snr}), 
are linked to several conservation laws. In particular,
the particle number ($N$), rest mass ($M=mN$), energy ($E$), spin angular momentum (${\bm{S}}$), and orbital angular momentum (${\bm{L}}$) are separately conserved according to\,\cite{Jain:2021pnk, Amin:2022pzv} 
\begin{align}
&N = \int d^3x \text{Tr}\left[ {\bf{\Psi}^{\dagger}}{\bf{\Psi}} \right]\,,\label{eq:N}\\
&E = \int d^3x \left[ \frac{1}{2m} \text{Tr}\left[ \nabla {\bf{\Psi}^{\dagger}}\cdot\nabla{\bf{\Psi}} \right]+\right.\nonumber\label{eq:E}\\
&\left.\frac{m^2}{4m^2_{\text{pl}}}\text{Tr}\left[ {\bf{\Psi}^{\dagger}}{\bf{\Psi}} \right] \int \frac{d^3 y}{4\pi|{\bm{x}}-{\bm{y}}|}\text{Tr}\left[ {{\bf{\Psi}}^{\dagger}({\bm{y}})}{{\bf{\Psi}}({\bm{y}})}\right] \right]\,,\\
&{\bm{S}} = \int d^3x\,\mathbb{R}( i{\bf{\Psi}}\times{\bf{\Psi}^{\dagger}})\,,\label{eq:S}\\
&{\bm{L}} = \int d^3x\,  \mathbb{R}(i{\bf{\Psi}}^{\dagger}\nabla{\bf{\Psi}}\times{\bm{x}}).
\end{align}

%%%%%%%%%%%%%%%%%%%%%%%%%%%%%%%%%%%%%%%%%%%%%%%%%%%%%%%%%%%%%%%%%%%%%%%
\subsection{Vector solitons}

A massive spin-1 field holds states with spin multiplicity for some particular direction $\hat{n}$. Such states are labelled by $\lambda \in \{-1,0,1\}$
and defined by the set $\{ {\bm{\epsilon}}^{\lambda}_{\hat{n}}\}$, so that the field can be decomposed as
\begin{equation}
{\bf{\Psi}}(t,{\bm{x}}) = \sum_{\lambda} \psi^{(\lambda)}(t,{\bm{x}}){\bm{\epsilon}}_{\hat{n}}^{\lambda}.\label{eq:Psivec}     
\end{equation}
Here $\psi^{(\lambda)}({\bm{x}},t)$ is understood as the field holding a polarization $\lambda$ in the $\hat{n}$ direction. 

The lowest energy solitons, which are of our interest, are spherically symmetric and defined by a constant $\mu$, which can be interpreted as the chemical potential\,\cite{Jain:2021pnk, Schiappacasse:2017ham}  as 
\begin{equation}
\psi^{(\lambda)}(t,{\bf{x}}) = c_{\lambda}\psi(r)e^{i(\mu t-\varphi_{\lambda})}\,,\label{eq:gs}  
\end{equation}
where $\varphi_{\lambda}$ is an angular phase. In the framework of the non-relativistic approximation, the soliton exhibits oscillation at an angular frequency $\omega$ that is approximately equivalent to the mass of the spin-1 field, specifically $\omega\approx m$. The chemical potential encloses a slight adjustment to this frequency, represented as $\omega = m - \mu$, where $\mu/m \ll 1$. 

\begin{figure}
\centering
  \includegraphics[scale=0.28]{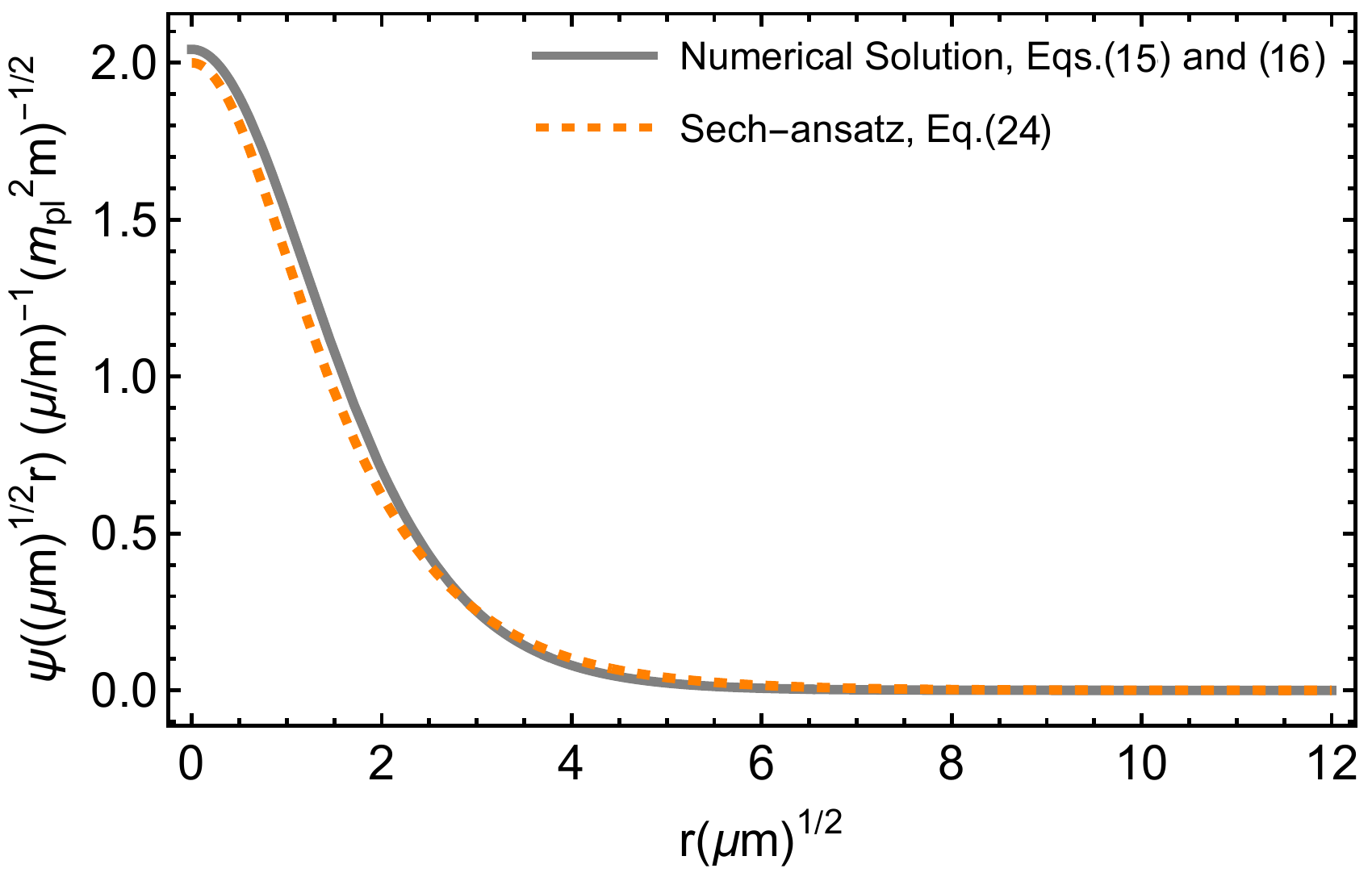}
\caption{The ``universal'' profile for non-relativistic ground state vector solitons obtained by numerically solving Eqs.~(\ref{eq:S1}) and ~(\ref{eq:S2}) indicated by a solid gray line. Ansatz shown in Eq.~(\ref{eq:sech-ans}) indicated by a dashed orange line.}
 \label{Fig:Univprofile}
\end{figure}

Without loss of generality, we take
$\hat{n}=\hat{z}$ and consider the orthonormal basis
\begin{equation}
    {\bm{\epsilon}}_{\hat{z}}^{(\pm 1)} = \frac{1}{\sqrt{2}} 
    \begin{bmatrix}
    1 & \pm i & 0
    \end{bmatrix}^T
    \hspace{0.1cm}
    \text{and}
    \hspace{0.1cm}
     {\bm{\epsilon}}_{\hat{z}}^{(0)} = 
    \begin{bmatrix}
    0 & 0 & 1
    \end{bmatrix}^T\,,\label{eq:vec}
\end{equation}
where $T$ represents the transpose operator.
Using Eqs.~(\ref{eq:Psivec})-(\ref{eq:vec}),
the real-value vector field ${\bf{W}}$ in Eq.~(\ref{eq:W}) is written as
\begin{equation}
\begin{adjustbox}{max width=218pt}
$
{\bf{W}}({\bm{x}},t)= \frac{\psi(r)}{\sqrt{m}} \left( c_1 \begin{bmatrix}
           \text{cos}(\omega t_1) \\
            \text{sin}(\omega t_1)\\
           0
         \end{bmatrix}  + 
         c_{-1} \begin{bmatrix}
           \text{cos}(\omega t_{-1}) \\
            -\text{sin}(\omega t_{-1})\\
           0
         \end{bmatrix}+ 
         c_0\,\sqrt{2} \begin{bmatrix}
           0 \\
           0\\
            \text{cos}(\omega t_0)
         \end{bmatrix} \right),\label{eq:W2}
         $
\end{adjustbox}
\end{equation}
where $c_1^2 +c_{-1}^2 +c_0^2 = 1$, $t_{\lambda}\equiv t+\varphi_{\lambda}/\omega$,  and  the field $\psi({\bm{x}})=\psi(r)$ satisfies the (time-independent) SP system according to
\begin{align}
&-\mu \psi = -\frac{1}{2m} \frac{1}{r^2}\frac{\partial}{\partial r}\left( r^2 \frac{\partial\psi}{\partial r} \right)+ m\Phi \psi\,,\label{eq:S1}\\
&\frac{1}{r^2}\frac{\partial}{\partial r}\left( r^2 \frac{\partial \Phi}{\partial r} \right) = \frac{m}{2 m_{pl}^2} \psi^2\,.\label{eq:S2}
\end{align}
Under the fields and coordinate transformations
$\psi = (\mu/m)m_{\text{pl}}m^{1/2}\psi'$, 
$\Phi_{\text{N}} = (\mu/m)\Phi_{\text{N}}'$, and $r = (\mu m)^{-1/2}r'$, the above system of differential equations becomes only dependent on $\psi', \Phi_{\text{N}}' $ and $r'$. By numerically solving such a system, we find the ``universal" non-relativistic field profile for (lowest energy) soliton solutions as shown in Fig. \ref{Fig:Univprofile}, solid gray line. From this profile, we may obtain the ``universal" total mass, energy and radius configuration, which, for the case of extremally polarized solitons, read as~\cite{Jain:2021pnk, Amin:2023imi, Schiappacasse:2025mao}
\begin{align}
M_{\text{sol}} &\approx 62.3\,\frac{m_{\text{pl}}^2}{m} \left( \frac{\mu}{m} \right)^{1/2}\,, \nonumber\\
&\sim 10^{-9}\,M_{\odot} \left( \frac{10^{-6}\,\text{eV}}{m} \right)  \left( \frac{\mu/m}{10^{-11}} \right)^{1/2}\,,\label{eq:Msol} \\
E_{\text{sol}} &\approx -20.8 \frac{m_{\text{pl}}^2}{m} \left( \frac{\mu}{m} \right)^{3/2}\,,\nonumber\\
&\sim -4 \times 10^{45} \,\text{eV} \left( \frac{10^{-6}\,\text{eV}}{m} \right) \left( \frac{\mu/m}{10^{-11}} \right)^{3/2}\,,\label{eq:Esol} \\
R^{0.95}_{\text{sol}} &\approx 3.16 \frac{1}{m} \left( \frac{\mu}{m} \right)^{-1/2}\,,\nonumber\\
&\sim 200\,\text{km}\, \left( \frac{10^{-6}\,\text{eV}}{m} \right) \left( \frac{\mu/m}{10^{-11}} \right)^{-1/2}\,,\label{eq:Rsol095}\\
R^{0.24}_{\text{sol}} &\approx 1.1 \frac{1}{m} \left( \frac{\mu}{m} \right)^{-1/2}\,,\nonumber\\
&\sim 70\,\text{km}\, \left( \frac{10^{-6}\,\text{eV}}{m} \right) \left( \frac{\mu/m}{10^{-11}} \right)^{-1/2}\,.\label{eq:Rsol024}
\end{align}
Here \( R^{0.95}_{\text{sol}} \) represents the full width at half maximum, which is the radius that encloses about $95\%$ of the total number of particles, while \( R^{0.24}_{\text{sol}} \) encloses around $24\%$ of the total number of particles. The unique mass-radius relation is obtained by combining Eq.\,(\ref{eq:Msol}) with  Eq.\,(\ref{eq:Rsol095}) or Eq.\,(\ref{eq:Rsol024}) to obtain
\begin{align}
M_{\text{sol}} &\approx 197 \frac{m_{\text{pl}}^2}{m^2} (R^{0.95}_{\text{sol}})^{-1}\,,\nonumber\\
& \sim 10^{-9}\,M_{\odot} \left( \frac{10^{-6}\,\text{eV}}{m} \right)^2 \left( \frac{200\,\text{km}}{R^{0.95}_{\text{sol}}} \right)\,,
\label{eq:massradius095}
\\
M_{\text{sol}} &\approx 68 \frac{m_{\text{pl}}^2}{m^2} (R^{0.24}_{\text{sol}})^{-1}\,,\nonumber\\
& \sim 10^{-9}\,M_{\odot} \left( \frac{10^{-6}\,\text{eV}}{m} \right)^2 \left( \frac{70\,\text{km}}{R^{0.24}_{\text{sol}}} \right)\,.
\label{eq:massradius}
\end{align}

The numerical radial profile $\psi(r)$ shown in Fig.\,\ref{Fig:Univprofile} is characterized by 
its exponential decay ($\text{lim}_{r\rightarrow\infty}\,\psi(r) = 0$) and null derivative at the center ($\text{lim}_{r\rightarrow 0}\,\text{d}\psi(r)/dr = 0$). Both properties are well captured by a sech-ansatz of the kind\,\cite{Schiappacasse:2017ham}
\begin{equation}
\psi(r) = \sqrt{\frac{3 N_{\text{sol}}}{\pi^3 (R^{0.24}_{\text{sol}})^3}} \text{sech}(r/R^{0.24}_{\text{sol}})\,,\label{eq:ans}    
\end{equation}
where the prefactor ensures that the total number of particles of the system is $N_{\text{sol}}$ through Eq.\,(\ref{eq:N}). Using the universal expressions for the soliton mass and radius, Eqs.\,(\ref{eq:Msol}) and (\ref{eq:Rsol024}), where $N_{\text{sol}} = M_{\text{sol}}/m$, we may obtain an universal ansatz for the soliton radial profile,
\begin{equation}
\psi(r) \approx 2 (m_{\text{pl}} m^{1/2})\left( \frac{\mu}{m} \right) \text{sech}(r (\mu m)^{1/2}/1.1)\,,\label{eq:sech-ans} 
\end{equation}
which closely agrees with the numerical result as shown by the dashed orange line in Fig.\ref{Fig:Univprofile}. For simplicity, we will use this ansatz in the following sections when addressing the phenomena of dipole radiation and kinetic mixing in vector solitons.

The value for the \((\mu/m)\) ratio, Eq.\,(\ref{eq:Msol}), is linked to
the vector soliton formation mechanism. For example, for the soliton 
nucleation within virialized halos\,\cite{Jain:2023ojg}, as the size of the halo increases, so does the soliton mass. Therefore, the mass function of halos determines the soliton mass function for a specific light vector boson mass\,\cite{Du:2023jxh}. Here, we will adopt a phenomenological approach to maintain generality, treating the $\mu/m \ll 1$ ratio as a free parameter.

As the chemical potential drops in relation to the mass \(m\) of the spin-1 boson, the angular frequency of the soliton trends towards \(m\), guiding us further into the non-relativistic domain. As a result, the soliton becomes increasingly diluted, indicating that it has less mass and occupies a larger spatial volume, e.g. $M_{\text{sol}}m \ll 1$ from Eq.\,(\ref{eq:Msol}) and $R^{0.24}_{\text{sol}}m \gg 1$ from Eq.\,(\ref{eq:Rsol024}).

%%%%%%%%%%%%%%%%%%%%%%%%%%%%%%%%%%%%%%%%%%%%%%%%%%%%%%%%%%%%%%%%%%%%%%%
%%%%%%%%%%%%%%%%%%%%%%%%%%%%%%%%%%%%%%%%%%%%%%%%%%%%%%%%%%%%%%%%%%%%%%%
\section{Dipole radiation from vector solitons embedded in electromagnetic fields}
\label{Sec:DR}

We work in the EFT framework for the interacting Lagrangian density in Eq.~(\ref{Eq:Odipoleradiation}), recalling that the validity of the EFT is given by $g^2_{W\gamma}\bar{W}^2 \ll 1$.

%%%%%%%%%%%%%%%%%%%%%%%%%%%%%%%%%%%%%%%%%%%%%%%%%%%%%%%%%%%%%%%%%%%%%%%
\subsection{Power emission from dipole radiation}
\label{sec:PEDR}

The modified Maxwell equations read as follows,
\begin{align}
&\nabla \cdot {\bf{B}} = 0\,,\label{eq:MMW1}\\
&\dot{\bf{B}} + \nabla \times {\bf{E}} = 0\,,\label{eq:MMW2}\\
&\nabla \cdot {\bf{E}} + g_{W\gamma}^2 \nabla \text{Tr}[{\bf{W W}}]\cdot {\bf{B}} = 0\,,\label{eq:MMW3}\\ 
&{\dot{\bf{E}}} - \nabla \times {\bf{B}} + g_{W\gamma}^2 \left( {\bf{B}}\,\partial_t \text{Tr} [{\bf{W W}}]\right.\nonumber\\
&\textcolor{white}{xxxxxxxxxxxxxx}\left.- {\bf{E}}\times \nabla \text{Tr}[{\bf{W W}}] \right)=0\,,\label{eq:MMW4}
\end{align}
where the vector soliton field ${\bf{W}}$ is giving by Eq.\,(\ref{eq:W2}) for the general case. 

We observe that the soliton field appears in the equations only through its trace, which indicates that its indices are not ``entangled" with the electromagnetic field stress tensor. Consequently, circularly polarized solitons will lose their time dependence in Eqs.\,(\ref{eq:MMW3}) and (\ref{eq:MMW4}), as \(\text{Tr}[{\bf{W W}}] = \psi^2(r)/m\). This prevents them from experiencing dipole radiation phenomena. 

In contrast, linearly polarized vector solitons behave differently, as indicated by \(\text{Tr}[{\bf{W W}}] = (2/m)\psi^2(r) \cos^2(\omega t_0)\). For this reason, we will focus on an extremally linearized soliton represented by \({\bf{W}}(\bm{x},t) = \phi(\bm{x}) \cos(\omega t)\hat{\bm z}\), where \(\phi(\bm{x}) \equiv \sqrt{2/m}\psi(r)\) and \(\varphi_0 = 0\). The field amplitude at the soliton center takes the parametric form $\phi_0  \sim m_{\text{pl}}/(mR^{0.24}_{\text{sol}})^2$, which makes the EFT validity become\,\cite{Amin:2023imi}
\begin{equation}
g^2\bar W^2  \ll 1 \rightarrow
gm_{\text{pl}} \ll (mR^{0.24}_{\text{sol}})^2\,\,\,\text{or}\,\,\,
gm_{\text{pl}} \ll \left(\frac{\mu}{m}\right)^{-1}\,,
\label{eq:EFTvalidity}
\end{equation}
where we have used $\bar W = \phi_0$ and  Eq.\,(\ref{eq:sech-ans}).

 Note that we can define effective charge and current densities in Eqs.~(\ref{eq:MMW3}) and (\ref{eq:MMW4}), respectively. Thus, the modified equations now take the standard form as
\begin{align}
&\nabla \cdot {\bf{E}} - \rho = 0\,,\label{eq:fo1}\\
&{\dot{\bf{E}}} - \nabla \times {\bf{B}} + {\bf{J}} = 0\,,\label{eq:fo2}
\end{align}
where
\begin{align}
&\rho({\bm{x}},t) = - g_{W\gamma}^2 \text{cos}^2(\omega t) \nabla (\phi({\bm{x}})^2)\cdot {\bf{B}}\,,\\
&{\bf{J}}({\bm{x}},t)  = -g_{W\gamma}^2\, \omega\, \text{sin}(2\omega t) \phi^2({\bm{x}})\,{\bf{B}}\,\nonumber\\
&\textcolor{white}{xxxxxxxxxxxxx}+ g_{W\gamma}^2\text{cos}^2(\omega t)\nabla (\phi({\bm{x}})^2)\times {\bf{E}}\,.
\end{align}
Using Eqs.~(\ref{eq:MMW1}) and (\ref{eq:MMW2}), we can rewrite Eqs.~(\ref{eq:fo1}) and (\ref{eq:fo2}) to transform them into second-order differential equations for the electromagnetic field as
\begin{align}
&\ddot{{\bf{E}}} - \nabla^2 {\bf{E}} = - \nabla \rho - \dot{\bf{J}}\,,\label{eq:weq1}\\
&\ddot{{\bf{B}}} - \nabla^2 {\bf{B}} = \nabla \times {\bf{J}}\,,\label{eq:weq2}
\end{align}
where it is understood that the 4-current $J^{\mu} = (\rho, {\bf{J}})$ is spatially confined due to the vector soliton finite extension. In the small amplitude/coupling regime, the fields, currents, and densities can be expanded in terms of $g_{W\gamma}^2\phi^2$ as 
\begin{align}
&{\bf{E}} =  {\bf{E}}_{(0)} +    {\bf{E}}_{(2)} +  {\bf{E}}_{(4)}  ...\,,\\
&{\bf{B}} =  {\bf{B}}_{(0)} +    {\bf{B}}_{(2)} +  {\bf{B}}_{(4)}...\,,\\
&{\rho} =  {\rho}_{(0)} +    {\rho}_{(2)} +  {\rho}_{(4)}...\,,\\
&{{\bf{J}}} =  {\bf{J}}_{(0)} +    {\bf{J}}_{(2)} +  {\bf{J}}_{(4)}...\,,
\end{align}
where ${X}_{(n)} \propto (g_{W\gamma}\phi)^{n}$ with $X = ({\bf{E}},{\bf{B}},{\bf{J}}, \rho )$. 
For easy notation, we write ${\bf{E}}_{0} = \bar{\bf{E}}$ and ${\bf{B}}_{0} = \bar{\bf{B}}$ with the bar indicating the spatio-temporally constant external background fields, which are sourced by ${\rho}_{(0)}$ and ${\bf{J}}_{(0)}$. The background fields are independent of the vector field configuration and can be those associated with the interstellar medium or the magnetospheres of neutron stars or white dwarfs, for example. We essentially assume that the size of the vector soliton is significantly smaller than the coherence length of the background fields. Additionally, we consider that changes in the background fields occur on timescales much longer than the duration over which the configuration remains in that specific region of the fields. To the second order in 
 $(g_{W\gamma}\phi)$, we have 
\begin{align}
&\ddot{{\bf{E}}}_{(2)} - \nabla^2 {\bf{E}}_{(2)} = - \nabla \rho_{(2)} - \dot{\bf{J}}_{(2)}\,,\label{eq:weq1c}\\
&\ddot{{\bf{B}}}_{(2)} - \nabla^2 {\bf{B}}_{(2)} = \nabla \times {\bf{J}}_{(2)}\,,\label{eq:weq2c}
\end{align}
and the soliton and the background fields produce an effective charge and current densities as
 \begin{align}
&\rho_{(2)}({\bm{x}},t) = \mathbb{R} [\varrho_{(2)}({\bm{x}})e^{-2i\omega t}+ \varrho_{(2)}({\bm{x}})]\,,\label{eq:rho2}\\
&{\bf{J}}_{(2)}({\bm{x}},t) = \mathbb{R} [{\bf{j}}_{(2)}({\bm{x}})e^{-2i\omega t}+ {\bf{j}}_{(2)}({\bm{x}})]\,,\label{eq:j2}
 \end{align}
 where
 \begin{align}
  & \varrho_{(2)} ({\bm{x}}) = -\frac{g_{W\gamma}^2}{2}\nabla(\phi^2({\bm{x}})){\cdot{\bar{\bf{B}}}}\,,
  \label{eq:varrho_x}\\  
 & {\bf{j}}_{(2)}({\bm{x}})= -i\, \frac{g^2_{W\gamma}}{2}\phi^2({\bm{x}})(2\omega) \bar{\bf{B}} +\frac{g_{W\gamma}^2}{2}\nabla(\phi^2({\bm{x}}))\times{\bar{\bf{E}}}  \,. 
 \end{align}
 Note that Eqs.~(\ref{eq:weq1c}) and (\ref{eq:weq2c}) have the form of inhomogeneous wave equations, both of which are solvable using a standard Green function method. We express ${\bf{E}}_{(2)}({\bm{x}},t)$ and ${\bf{B}}_{(2)}({\bm{x}},t)$ in terms of  the retarded Green's function as
 \begin{align}
{\bf{E}}_{(2)}(x)&=\int dt'd^3{\bm{x}}'G(x;x') \left[ \nabla \rho_{(2)}(x') + \dot{\bf{J}}_{(2)}(x')\right]\,,\label{eq:E2}\\
{\bf{B}}_{(2)}(x)&=\int dt'd^3{\bm{x}}'G(x;x') \left[ -\nabla \times {\bf{J}}_{(2)}(x')\right]\,,\label{eq:B2}
 \end{align}
 with
 \begin{align}
  G(t,{\bm{x}};t',{\bm{x}'})&=\int \frac{dk_0}{2\pi}\frac{d^3{\bm{k}}}{(2\pi)^3}\frac{e^{-ik_0(t-t')}e^{i{\bm{k}}\cdot(\bm{x}-{\bm{x'}})}}{(k_0+i\epsilon)^2-{\bm{k}}^2}\,,\label{eq:G}\\
  & =-\frac{\delta((t-t')-|{\bm{x}-{\bm{x}'}}|)}{4\pi|{\bm{x}}-{\bm{x}}'|}\Theta(t-t')\,,\label{eq:Gtheta}
 \end{align}
 where it is understood, taking the limit $\epsilon
\rightarrow 0$ at the end of the whole calculation. We will skip explicit calculations, but we kindly invite the reader to see the App. A in Ref.~\cite{Amin:2021tnq}.
 
 For an observation point far away from the soliton configuration, we may write $|{\bm{x}}-{\bm{x}}'| \approx |{\bm{x}}|-\hat{\bm{x}}\cdot{\bm{x}'}$ with $\hat{\bm{x}}\equiv {\bm{x}}/|{\bm{x}}|$. Under this approximation and expressing $\varrho_2({\bm{x}})$ and ${\bm{j}}_{(2)}({\bm{x}})$  in terms of their Fourier transforms, we obtain
 \begin{align}
{\bf{E}}_{(2)}& \approx -\mathbb{R} \left[ \frac{e^{-2i\omega( t - |{\bm{x}}|)}}{4\pi |{\bm{x}}|} \left( i{\bm{k}}\tilde{\varrho}_{(2)}({\bm{k}})-2i\omega \tilde{{\bf{j}}}_{(2)}({\bm{k}}) \right) \right] \,,\\  
{\bf{B}}_{(2)}& \approx -\mathbb{R} \left[ \frac{e^{-2i\omega( t - |{\bm{x}}|)}}{4\pi |{\bm{x}}|} \left( -i{\bm{k}}\times \tilde{{\bf{j}}}_{(2)}({\bm{k}}) \right) \right]\,,
 \end{align}
 where ${\bm{k}}\equiv 2\omega \hat{\bm{x}}$. We note that the time-independent parts of the effective charge and current densities, Eqs.~(\ref{eq:rho2}) and (\ref{eq:j2}), disappear during the calculation. At a location ${\bm{x}}$ far away from the soliton and $t \gg 0$, the emitted power per unit solid angle, 
 $dP_{(4)}/d\Omega = |{\bm{x}}|^2\hat{\bm{x}}\cdot {\bf{S}}_{(4)}$ with ${\bf{S}}_{(4)} \equiv {\bf{E}}_{(2)}\times{\bf{B}}_{(2)}$ as the Poynting flux, reads as
 \begin{equation}
    \frac{dP_{(4)}}{d\Omega} = \frac{4\omega^2}{32\pi^2}\left[ -|\tilde{\varrho}_{(2)}({\bm{k}})|^2 + |\tilde{{\bf{j}}}_{(2)}({\bm{k}})|^2 + \text{oscillating terms} \right]\,,\label{eq:dPdOmegaa} 
 \end{equation}
with
\begin{align}
\phi^2({\bm{x}}) &= \int \frac{d^3{\bm{k}}}{(2\pi)^3}\tilde{\varphi}({\bm{k}})e^{i{\bm{k}}\cdot{\bm{x}}}\,,\\
\tilde{\varrho}_{(2)}({\bm{k}}) &= -i \omega g^2_{W\gamma}\tilde{\varphi}(2\omega)\hat{\bm{x}}\cdot \bar{{\bf{B}}}\,,\label{eq:rhoka}\\
\tilde{\bf{j}}_{(2)}({\bm{k}})&= -i\omega g^{2}_{W\gamma}\tilde{\varphi}(2\omega)\bar{{\bf{B}}} + i \omega g^2_{W\gamma} \tilde{\varphi}(2\omega)\hat{\bm{x}}\cdot \bar{{\bf{E}}}\,,\label{eq:jka}
\end{align}
where we have used the relation $2\omega \tilde{\varrho}_{(2)}({\bm{k}})={\bm{k}}\cdot \tilde{j}_{(2)}({\bm{k}})$ from the continuity equation.

We plug Eqs.~(\ref{eq:rhoka}) and (\ref{eq:jka}) into Eq.~(\ref{eq:dPdOmegaa}), perform the angular integration, and average in time. For the sake of simplicity, we take $\bar{\bf{E}}=0$ and $\bar{\bf{B}}=\bar{B}\hat{\bf{z}}$. The total time-averaged power emitted  is calculated to be  
\begin{align}
&\frac{dP_{(4)}}{d\Omega}=\frac{4\omega^4 g^4_{W\gamma}\tilde{\varphi}^2(2\omega)}{32\pi^2}
\left[(\hat{\bm{x}}\times \bar{\bf{E}})^2-2\hat{\bm{x}}\cdot(\bar{\bf{E}}\times{\bar{\bf{B}}})\right.\nonumber\label{eq:oSC}\\
&\textcolor{white}{xxxxx}\left.+\bar{\bf{B}}^2-(\hat{x}\cdot\bar{\bf{B}})^2 \right](1+\text{oscillating terms})\,,\\
&\langle P_{(4)} \rangle_t = \frac{ g^4_{W\gamma} \omega^4 \tilde{\varphi}^2(2\omega)}{3\pi} \bar{B}^2 \,, \label{eq:P4a}   
\end{align}
where
\begin{align}
&\tilde{\varphi}({\bm{k}}) = \int d^3{\bm{x}}\, \phi^2_0\, \text{sech}^2(r/R^{0.24}_{\text{sol}})\,e^{-i {\bm{k}}\cdot {\bm{x}}}\,,\\ 
&=\frac{4\pi \phi^2_0}{|{\bm{k}}|}\int_0^{\infty} dr\, r\, \text{sech}^2(r/R^{0.24}_{\text{sol}}) \,\text{sin}(|{\bm{k}}| r)\,,\\
&= \frac{(\pi R \phi_0)^2}{|{\bm{k}}|} \,\text{csch}(  |{\bm{k}}|\pi R^{0.24}_{\text{sol}}/2)\times\nonumber\\
&\hspace{2 cm}\left[-2+  |{\bm{k}}|\pi R\, \text{coth}( |{\bm{k}}|\pi R^{0.24}_{\text{sol}}/2)\right]\,.\label{eq:ft}
\end{align}
In Eq.~(\ref{eq:oSC}) the oscillating terms average to zero and
the angular integration is performed in spherical coordinates, taking advantage of the 
trigonometric identities such that $\bar{\bf{B}}^2-(\hat{x}\cdot\bar{\bf{B}})^2 = \bar{\bf{B}}^2\text{sin}^2(\theta)$. 

The emitted power is proportional to the square of the Fourier transform of the square of the soliton radial profile, and the radiation is emitted at a frequency $2\omega$. That is $\langle P_{(4)}\rangle_t\propto \left\{\mathcal{F}[\phi^2(r)](2\omega)\right\}^2=\tilde{\varphi}^2(2\omega)$. Apart from the $(1/4)$-factor, the scalar case is recovered under the replacement $2\omega \rightarrow \omega$ and $\tilde{\varphi}^2(2\omega) \rightarrow \tilde{a}^2(\omega)$, where $\tilde{a}^2(\omega)$ is now the Fourier transform of the axion star radial profile. That is $\langle P_{(2)}\rangle_t\propto \left\{\mathcal{F}[a(r)](\omega)\right\}^2=\tilde{a}^2(\omega)$\,\cite{Amin:2021tnq}.

The Fourier transform $\tilde{\varphi}({\bm{k}})$ can be approximated as
\begin{equation}
\tilde{\varphi}(2\omega) \approx \frac{2}{\omega^3}\phi^2_0 (\pi \omega R^{0.24}_{\text{sol}})^3 e^{-\pi \omega R^{0.24}_{\text{sol}}}\,, \label{eq:FTphi}   
\end{equation}
for $\omega R^{0.24}_{\text{sol}} \gtrsim 4$, where we have used $\text{coth}(\alpha)\approx 1$ and $\text{csch}(\alpha)\approx e^{-\alpha}$. Using Eq.\,(\ref{eq:FTphi}), $\phi_0 \sim m_{\text{pl}}/(mR^{0.24}_{\text{sol}})^2$, and $\omega \approx m$ due to the non-relativistic regime, the time-averaged radiated power associated with the soliton dipole radiation in the presence of a background field $\bar B$ is approximated to be
\begin{align}
    &\langle P_{(4)} \rangle_t \sim (g_{W\gamma} m_{\rm pl})^4\frac{\bar B^2}{m^2}\frac{e^{-2\pi m R^{0.24}_{\text{sol}}}}{(mR^{0.24}_{\text{sol}})^2}\,,\label{eq:PDRa} \\
   &\sim 10^{37}{\rm ergs }\,\rm s^{-1}\left(\frac{g_{W\gamma}}{m_{\rm pl}^{-1}}\right)^{\!\!4}\left(\frac{B}{4.5\times 10^{12}\rm{G}}\right)^{\!\!2}  \nonumber \label{eq:PDRreviseda} \\
   &\hspace{1 cm}\times\left(\frac{10^{-6}\rm eV}{m}\right)^{\!\!2}\left(\frac{3.5}{mR^{0.24}_{\text{sol}}}\right)^{\!\!2}
  e^{-2\pi m R^{0.24}_{\text{sol}}}\,,
\end{align}
where the chosen value for $m R^{0.24}_{\text{sol}}$ implies $(\mu/m) = 10^{-1}$ to barely satisfy the non-relativistic regime. We have taken $g_{W\gamma}m_{\rm pl} = 10^{-1} \times (\mu/m)^{-1}$ as the dark photon-photon strength coupling in consistency with the EFT validity, Eq.\,(\ref{eq:EFTvalidity}). This strength for the coupling also ensures that the phenomenon of parametric resonance of photons in vacuum is absent\,\cite{Amin:2023imi}. We will return to this point in Sec.\,\ref{Sec:Unbounded}.  For the parameter values, the radiated power is 
about $10^{19}\,\rm erg\, \rm s^{-1}$, after taking into account the exponential suppression term, which is already $10^{15}$ times weaker than the Sun's radiation power. For more dilute configurations, e.g. $m R^{0.24}_{\text{sol}} \gtrsim O(10^2)$, the radiation power is extremely suppressed by the exponential term, being essentially zero for $m R^{0.24}_{\text{sol}} \gtrsim O(10^3)$.
 However, the presence of a (homogeneous) background plasma triggers a resonant dark photon to photon conversion when the plasma frequency $\omega_p \approx 2 m$, \textit{killing} the exponential suppression, as was already established for the case of axion stars\,\cite{Amin:2021tnq}. We will analyze this point in the following subsection. 

%%%%%%%%%%%%%%%%%%%%%%%%%%%%%%%%%%%%%%%%%%%%%%%%%%%%%%%%%%%%%%%%%%%%%%%
\subsection{Effective photon mass}
\label{sec:photonmassdipole} 

The previous analysis has assumed photons to be massless, neglecting the presence of any background medium. This picture is only exact in a vacuum. Otherwise, photons acquire an effective mass equal to the plasma frequency, which may come from the interstellar medium. In such a case, we have\,\cite{Carlson:1994}
\begin{equation}
\omega_p^2 = \frac{4 \pi \alpha n_e}{m_e} = \frac{n_e}{0.03\,\text{cm}^{-3}}(6.4\times10^{-12}\,\text{eV})^2\,,  
\end{equation}
where $\alpha$ is the fine structure constant, $m_e$ is the electron mass, and  $n_e$ indicates the number density of free electrons.  The typical value for such a number density today in the galactic halo is $n_e \sim 0.03\, \text{cm}^{-3}$. We see, for example, that for a typical vector boson mass $m \sim \mathcal{O}(10^{-5})\,\text{eV}$, one finds $m \gg \omega_p$. Even though electrons are spatially inhomogeneous and solitons are also moving in the galactic halo, we take such a quantity constant so that we end up with a constant effective photon mass $\omega_p$. Indeed, this assumption is consistent with the assumption that the external background electric and magnetic fields are constant. 

Following Ref.\,\cite{Amin:2021tnq}, in the regime $\omega_p < 2\omega$, we estimate the effect of the plasma frequency on the soliton dipole radiation by making the replacement $\nabla^2 \rightarrow \nabla^2 - \omega_p^2$ in Eqs.\,(\ref{eq:weq1}) and (\ref{eq:weq2}), so that the system of equations to be solved read as
\begin{align}
&\ddot{{\bf{E}}} - \left( \nabla^2 - \omega_p^2 \right) {\bf{E}} = - \nabla \rho - \dot{\bf{J}}\,,\label{eq:weq1wp}\\
&\ddot{{\bf{B}}} - \left( \nabla^2 - \omega_p^2\right) {\bf{B}} = \nabla \times {\bf{J}}\,.\label{eq:weq2wp}
\end{align}
The retarded Green function now includes a massive propagator due to the presence of the plasma frequency in the dispersion relation.  Making the replacement ${\bm{k}}^2 \rightarrow {\bm{k}}^2 + \omega_p^2$ in the denominator of Eq.\,(\ref{eq:G}) and performing the $k_0$-integral, one finds
\begin{align}
G(x,x')&=-\frac{\Theta(t-t')}{2\pi^2|{\bm{x}}-{\bm{x}}'|}\int_0^{\infty}\frac{dk\,k}{\sqrt{k^2+\omega_p^2}}\nonumber\\
&\times\text{sin}(k|{\bm{x-x'}}|)\text{sin}(\sqrt{k^2 + \omega_p^2}(t-t'))\,,    
\end{align}
which needs to be replaced in Eq.\,(\ref{eq:E2}) and (\ref{eq:B2}). Note that
in the vacuum limit, i.e. $\omega_p \rightarrow 0$, we recover Eq.\,(\ref{eq:Gtheta}) as expected. Following the same procedure as before, one finds the following expression for the emitted power per unit of solid angle:
\begin{equation}
    \frac{dP_{(4)}}{d\Omega} = \frac{2\omega\kappa}{32\pi^2}\left[ -|\tilde{\varrho}_{(2)}({\bm{k}})|^2 + |\tilde{{\bf{j}}}_{(2)}({\bm{k}})|^2 + \text{oscillating terms} \right]\,,\label{eq:dPdOmega} 
 \end{equation}
with
\begin{align}
\phi^2({\bm{x}}) &= \int \frac{d^3{\bm{k}}}{(2\pi)^3}\tilde{\varphi}({\bm{k}})e^{i{\bm{k}}\cdot{\bf{x}}}\,,\\
\tilde{\varrho}_{(2)}({\bm{k}}) &= -i  \frac{g^2_{W\gamma}}{2}\kappa\tilde{\varphi}(\kappa)\hat{\bm{x}}\cdot \bar{{\bf{B}}}\,,\label{eq:rhok}\\
\tilde{\bf{j}}_{(2)}({\bm{k}})&= -i \frac{g^{2}_{W\gamma}}{2}(2\omega)\tilde{\varphi}(\kappa)\bar{{\bf{B}}} + i \frac{g^2_{W\gamma}}{2} \kappa \tilde{\varphi}(\kappa)\hat{\bm{x}}\cdot \bar{{\bf{E}}}\,,\label{eq:jk}
\end{align}
where ${\bm{k}}=\kappa \hat{\bm{x}}$, $\kappa \equiv \sqrt{(2\omega)^2 - \omega_p^2}$, and the Fourier transform of the square of the soliton radial profile is given by Eq.\,(\ref{eq:ft}). Plugging  Eqs.\,(\ref{eq:rhok})-(\ref{eq:jk}) into Eq.\,(\ref{eq:dPdOmega}), one finds
\begin{align}
& \frac{dP_{(4)}}{d\Omega} = \frac{2g^4_{W\gamma}\tilde{\varphi}^2(\kappa)}{32\pi^2} \left[ \frac{\omega_p^2\kappa \omega}{4} \bar{\bf{B}}^2 + \frac{\kappa^3\omega}{4} (\hat{\bm{x}}\times\bar{\bf{B}})^2\,\nonumber \right.\\
& \left.+\frac{\kappa^3\omega}{4} (\hat{\bm{x}}\times{\bf{E}})^2-\kappa^2\omega^2\hat{\bm{x}}\cdot(\bar{\bf{E}}\times \bar{\bf{B}}) + \text{oscillating terms}\right]\,.
\end{align}
We take again $\bar{\bf{E}}=0$ and $\bar{\bf{B}}=\bar{B}\hat{\bf{z}}$. Then, the total time-averaged
power emitted by the vector boson soliton is calculated to be
\begin{equation}
\langle P_{(4)} \rangle_t = \frac{g^4_{W\gamma}\omega\kappa\omega_p^2\tilde{\varphi}^2(\kappa)}{16\pi}\left[{\bar{B}^2 + \frac{2}{3}\left(\frac{\kappa}{\omega_p}\right)^2}\bar{B}^2\right] \,.\label{eq:P4}   
\end{equation}
In the limit $\omega_p \rightarrow 0$, we have $\kappa = 2\omega$, and we recover the previous result shown in  Eq.\,(\ref{eq:P4a}), where the presence of a medium is not considered. On the other hand,  we are within the resonant conversion regime when $\omega_p \rightarrow 2\omega$. In such a regime, the emitted power is enhanced. The total time-averaged power emitted in Eq.\,(\ref{eq:P4}) in the resonant domain takes the form
\begin{align}
\langle P_{(4)} \rangle_t =  & \frac{\pi^{-1}(\phi_0 g_{W\gamma})^4\bar{B}^2}{9\sqrt{2}\omega^2}\left( \pi \omega R^{0.24}_{\text{sol}}\right)^6\left(1-\frac{\omega_p^2}{4\omega^2} \right)^{1/2}\nonumber\\
&\textcolor{white}{XXXXXXXX}+\mathcal{O}\left(1-\frac{\omega_p^2}{4\omega^2} \right)^{3/2}\,.
\label{eq:PDRplasma}
\end{align}
Using Eq.\,(\ref{eq:PDRplasma}), $\phi_0 \sim m_{\text{pl}}/(mR^{0.24}_{\text{sol}})^2$, and $\omega \approx m$ due to the non-relativistic regime, the time-averaged radiated power associated with the soliton dipole radiation in the presence of a background field $\bar B$ and a background medium is approximated to be
\begin{align}
&\langle P_{(4)} \rangle_t  \sim  (g_{W\gamma} m_{\rm pl})^4\frac{\bar B^2}{m^2}\frac{(\pi^5/(9 \sqrt{2}))}{(mR^{0.24}_{\text{sol}})^2}
\left(1-\frac{\omega_p^2}{4m^2} \right)^{1/2}\,,\label{eq:PDRplasma2c}\\
 &\sim 10^{39}{\rm ergs }\,\rm s^{-1}\left(\frac{\textit{g}_{\textit{W}\gamma}}{m_{\rm pl}^{-1}}\right)^{\!\!4}\left(\frac{B}{4.5\times 10^{12}\rm{G}}\right)^{\!\!2}  \nonumber
 \\
&\hspace{1 cm} \times\left(\frac{10^{-6}\rm eV}{m}\right)^{\!\!2}\left(\frac{3.5}{mR^{0.24}_{\text{sol}}}\right)^{\!\!2}
  \left(1-\frac{\omega_p^2}{4m^2} \right)^{1/2}\,,
  \label{eq:PDRplasma2b}
\end{align}
where we have set $(\mu/m) = 10^{-1}$ and $g_{W\gamma}m_{\rm pl} = 10^{-1} \times (\mu/m)^{-1}$ to barely satisfy the non-relativistic regime and the EFT validity condition, as before. 
\begin{figure}
\centering
  \includegraphics[scale=0.26]{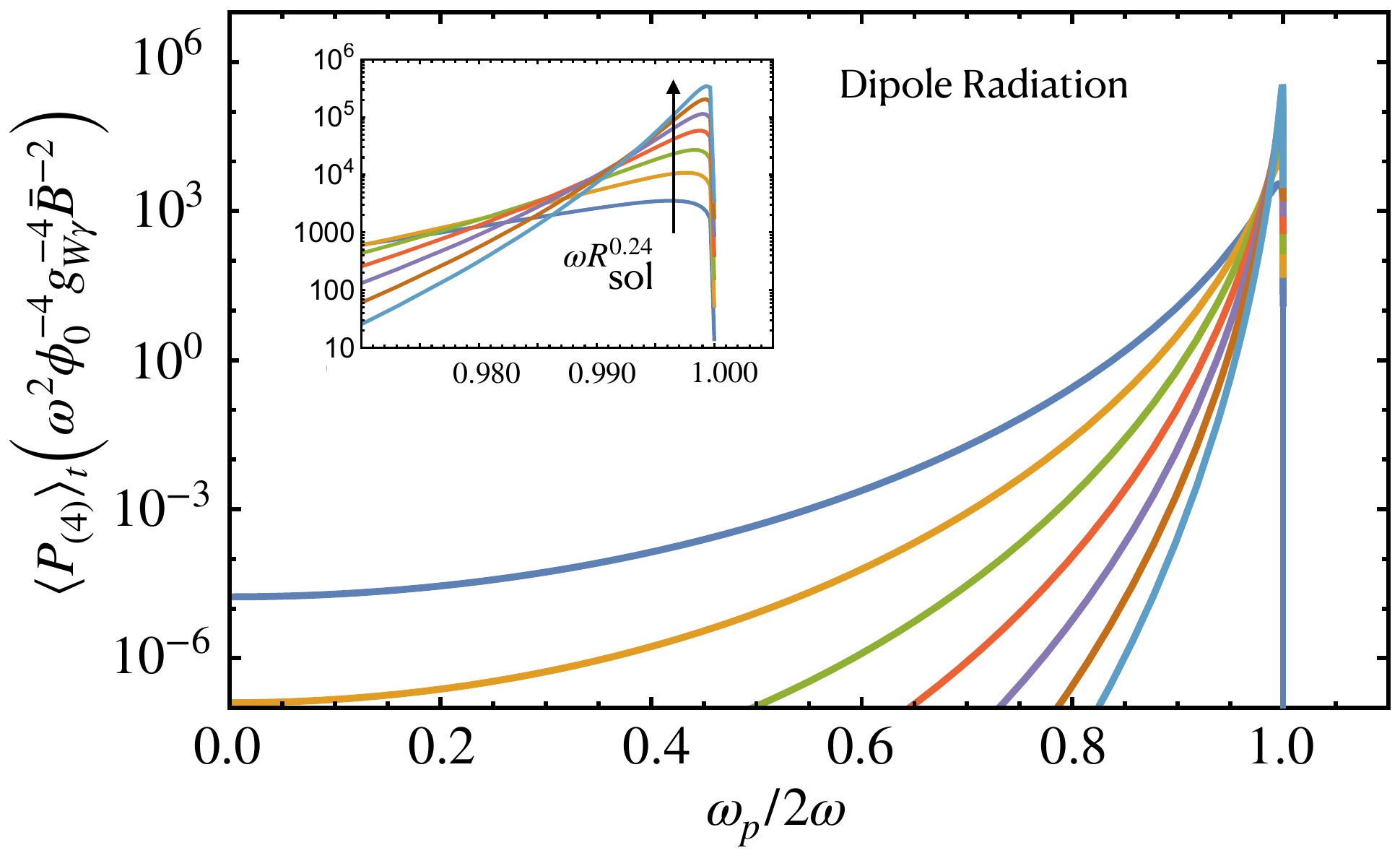}
  \includegraphics[scale=0.274]{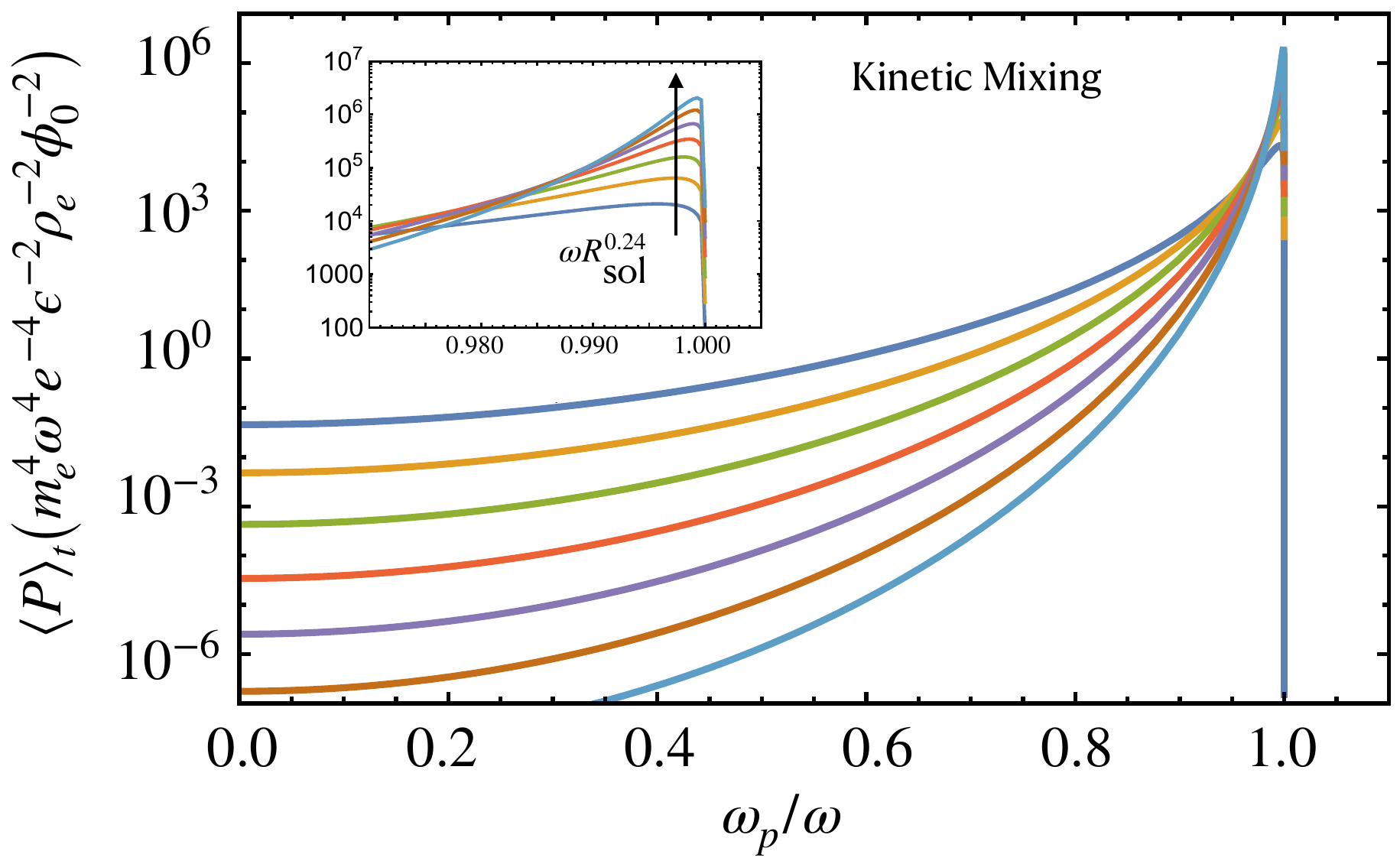}
\caption{Time averaged emitted power by the soliton configuration in terms of the plasma frequency $\omega_p$ and soliton scale length  $R^{0.24}_{\text{sol}}$. The plasma frequency is assumed to be constant in scales comparable to the soliton radius. We vary $\omega R^{0.24}_{\text{sol}}$  between 4 (light blue line) and 10 
(dark blue line). Top panel: soliton dipole radiation due to an external magnetic field, Eq.\,(\ref{eq:P4}). Bottom panel: soliton radiation produced by a kinetic mixing between the massive vector boson and electromagnetic current of ordinary matter, Eq.\,(\ref{eq:PtKM}) with $c_1 = 1$.}
 \label{Peak}
\end{figure}

Figure\,\ref{Peak} (top panel) shows the behavior of the time-averaged radiated power by the soliton in terms of the ratio $\omega_p/2\omega$ and  $\omega R^{0.24}_{\text{sol}}$. We have varied $\omega R^{0.24}_{\text{sol}}$ for 4 (light blue line) to 10 (dark blue line). There exists a significant enhancement of the power as the plasma frequency approaches twice the soliton oscillation frequency from below, e.g.  $\omega_p \rightarrow 2\omega$, which corresponds to the resonant phenomenon. Note that the production of photons via resonance is significantly more efficient as solitons become more diluted, namely as the factor $\omega R^{0.24}_{\text{sol}}$ increases. 

%%%%%%%%%%%%%%%%%%%%%%%%%%%%%%%%%%%%%%%%%%%%%%%%%%%%%%%%%%%%%%%%%%%%%%%
\subsection{Massive vector dark matter background}
\label{DMbckg1}

In the previous subsection, we discussed soliton dipole radiation in the presence of an electromagnetic background field. To assess the relevance of solitons as potential electromagnetic sources in galactic halos, we compare these results with the expected emissivity from a massive vector boson dark matter background in the presence of a magnetic background field. 

Consider again the four-vector potential $A^{\mu} = (A^0,{\bf{A}})$. The already obtained Eqs.\,(\ref{eq:MMW3}) and (\ref{eq:MMW4}) can be written in terms of only one equation for the four-vector current  $J^{\mu} = (J^0, {\bf{J}})$ as 
\begin{equation}
 \ddot{A}^{\mu} - \nabla^2 A^{\mu} + \omega_p^2 A^{\mu} = J^{\mu}\,, 
 \label{eq:Amu}
\end{equation}
where $\omega_p$ is the plasma frequency and
\begin{align}
J^{0} & = g^2_{W\gamma}\nabla(W^{\alpha}W_{\alpha})\cdot {\bf{B}}\,,\\
{\bf{J}} &= g^2_{W\gamma}\partial_0(W^{\alpha}W_{\alpha}){\bf{B}} + g^2_{W\gamma}\nabla(W^{\alpha}W_{\alpha})\times {\bf{E}}\,.
\end{align}
At zero order, we replace the electric and magnetic fields by background fields as before, i.e. ${\bar{\bf{E}}}$ and ${\bar{\bf{B}}}$. By simplicity, we take ${\bar{\bf{E}}} = 0$. 

In the absence of any background field, the massive vector boson satisfies the Klein-Gordon equation, and we can express it in terms of $W^{\alpha}_{\bm{k}}$ modes in the $k$-space as 
\begin{equation}
W^{\alpha}({\bm{x}},t) = \int \frac{d^3 k}{(2\pi)^3}\left(W^{\alpha}_{\bm{k}} e^{-i mt + i {\bm{k}}\cdot{\bm{x}}} + \text{c.c.}\right)\,,    
\label{eq:Wk}
\end{equation}
where c.c. indicates complex conjugate, and we have approximated the vector boson energy as
$(m^2 + k^2)^{1/2} \simeq m$ for $k \sim \mathcal{O}(10^{-3})m$ as the typical velocity within the galactic halo. The goal is to replace this expansion into Eq.\,(\ref{eq:Amu}), so that the massive vector boson acts as a source for photon emission. Using Eq.\,(\ref{eq:Wk}), the four-vector potential $J^{\mu}$ takes the form
\begin{align}
&J^0 =\,  g^{2}_{W\gamma} \int \frac{d^3k}{(2\pi)^3} \frac{d^3k'}{(2\pi)^3}\times  \,\nonumber\\
& \left[i({\bm{k}}+{\bm{k}}')\cdot {\bf{\bar{B}}}\,W^{\alpha}_{{\bm{k}}}W_{ {{\bm{k}}'}\alpha}\,e^{i({\bm{k}}+{\bm{k}}')\cdot{\bm{x}}-2im t} \right. \,\nonumber\\  
& + i({\bm{k}}-{\bm{k}}')\cdot {\bf{\bar{B}}}\,W^{\alpha}_{\bm{k}}W^*_{{\bm{k}'}\,\alpha} \,e^{i({\bm{k}}-{\bm{k}}')\cdot{\bm{x}}}\nonumber\\
& -i({\bm{k}}-{\bm{k}}')\cdot {\bf{\bar{B}}}\,W^{\alpha *}_{\bm{k}}W_{ {\bm{k}'}\,\alpha} \,e^{-i({\bm{k}}-{\bm{k}}')\cdot{\bm{x}}}\nonumber\\
& -\left. i({\bm{k}}+{\bm{k}}')\cdot {\bf{\bar{B}}}\,W^{\alpha *}_{\bm{k}}W^*_{{\bm{k}'}\,\alpha}\,e^{-i({\bm{k}}+{\bm{k}}')\cdot{\bm{x}}+2imt} \right]\,,\\
&=J^{0}_2 + J^0_{0}+ J^{0}_{-2}\,,
\end{align}
\begin{align}
&{\bf{J}} = g^2_{W\gamma} \int \frac{d^3k}{(2\pi)^3}\frac{d^3k'}{(2\pi)^3} \times\,\nonumber\\
&\left[ -i2m\,W^{\alpha}_{{\bm{k}}}W_{{\bm{k}'}\alpha}\,e^{i({\bm{k}}+{\bm{k'}})\cdot{\bm{x}}-2imt}\right.\,\nonumber\\
&+\left. i2m\,W^{\alpha\,*}_{{\bm{k}}}W^*_{{\bm{k}'}\alpha}\,e^{-i({\bm{k}}+{\bm{k'}})\cdot{\bm{x}}+2imt}\right]{\bar{\bf{B}}}\,\\
&={\bf{J}}_{2}+{\bf{J}}_{-2}\,,
\end{align}
where the $n$ subscript in $J^{\mu}_n$ indicates the $e^{-i n mt}$-term(s) in $J^{\mu}$. For simple kinematics, the condition $\Theta(2m - \omega_p)$ needs to be satisfied to have radiation. Thus, only $J^{\mu}_2$ will contribute to the process. After performing a time average over several oscillation cycles, the total emissivity is calculated from the radiated power as
\begin{align}
\langle\epsilon \rangle_t &= \frac{\langle P \rangle_t}{\text{Volume}}\,,\label{eq:Pq0}\\
&= \frac{1}{\text{Volume}}\int d\Omega \frac{2m \kappa}{8\pi^2} J^{\mu}_2({\bm{k}})J^{*}_{\mu\,2}({\bm{k}})\Theta(2m - \omega_p)\,, \label{eq:Pq}
\end{align}
where ${\bm{k}}=\kappa {\bm{\hat{x}}}$ and  $\kappa = \sqrt{(2m)^2-\omega_p^2}$ as before,
and the Fourier transform of $J^{\mu}_2$ reads as
\begin{align}
J^{0}_2({\bm{q}}) &= g^{2}_{W\gamma} \int \frac{d^3k}{(2\pi)^3} i{\bm{q}}\cdot {\bar{\bf{B}}}\,W^{\alpha}_{\bm{k}} W_{\bm{q-k}\,\alpha} \,e^{-2im t}\,,\label{eq:Jq1}\\ 
{\bf{J}}_{2}({\bm{q}})& = -g^2_{W\gamma} \int \frac{d^3k}{(2\pi)^3} i2m\,{\bar{\bf{B}}}\,W^{\alpha}_{\bm{k}} W_{\bm{q-k}\,\alpha} \,e^{-2im t}\,.\label{eq:Jq2}
\end{align}
Since we have assumed that the field fluctuations are invariant under spatial translations, the source is considered to fill all of space. Therefore, the emissivity is calculated by treating the volume factor in Eq.\,(\ref{eq:Pq0}) as three-dimensional Dirac delta functions.

Replacing Eqs.\,(\ref{eq:Jq1}) and (\ref{eq:Jq2}) into Eq.\,(\ref{eq:Pq}), one finds
\begin{align}
&\langle \epsilon \rangle_t  = \frac{g^4_{W\gamma}2m\kappa}{8\pi^2}\int d\Omega \left[ 4m^2 {\bf{B}}^2 - \kappa^2 ({\bm{\hat{x}}}\cdot{\bar{\bf{B}}})^2 \right]\,\nonumber\\ 
 &\times\int  \int \frac{d^3k'}{(2\pi)^3} \frac{d^3k''}{(2\pi)^3} \langle W^{\alpha}_{\bm{k}'}W_{ {\bm{k}}-{\bm{k}'}\alpha} W^{\beta *}_{\bm{k}''}W^*_{ {\bm{k}}-{\bm{k}''}\beta}\rangle\,.
\label{eq:P}
\end{align}

We treat the mode amplitudes $W^{\alpha}_{\bm{k}}$ as stochastic variables expressing their expectation values in terms of the power spectrum $P_{{\bm{k}}}$. We take the spatial components of the four-vector boson to be uncorrelated among themselves. The time component $W^0_{\bf{k}}$ is related to the spatial components by $\partial_{\mu}W^{\mu}= 0$, so that
$W^0_{\bm{k}}=m^{-1}{\bm{k}}\cdot {\bm{W}}_{\bm{k}}$. Two-point correlation functions between different vector field components read as
\begin{align}
\langle W^{i}_{{\bm{k}}}W^*_{ {\bm{k}' }j}\rangle &= (2\pi)^3\delta^{(3)}({\bm{k}}-{\bm{k}'})P_{\bm{k}}\delta_{ij}\,,\label{WW1}\\
\langle W^{i}_{{\bm{k}}}W_{ {\bm{k}'}j}\rangle &= (2\pi)^3\delta^{(3)}({\bm{k}}+{\bm{k}'})P_{\bm{k}}\delta_{ij}\,,\label{WW2}\\
\langle W^0_{{\bm{k}}}W^*_{ {\bm{k}'}0}\rangle &= -\frac{{\bm{k}}\cdot{\bm{k}'}}{m^2}(2\pi)^3\delta^{(3)}({\bm{k}}-{\bm{k}'})P_{\bm{k}}\,,\label{WW3}\\
\langle W^0_{{\bm{k}}}W_{ {\bm{k}'}0}\rangle &= -\frac{{\bm{k}}\cdot{\bm{k}'}}{m^2}(2\pi)^3\delta^{(3)}({\bm{k}}+{\bm{k}'})P_{\bm{k}}\,,\label{WW4}\\
\langle W^0_{{\bm{k}}}W^*_{ {\bm{k}'}i}\rangle &= \frac{k_i}{m}(2\pi)^3\delta^{(3)}({\bm{k}}-{\bm{k}'})P_{\bm{k}}\,,\label{WW5}\\
\langle W^0_{{\bm{k}}}W_{ {\bm{k}'}i}\rangle &= \frac{k_i}{m}(2\pi)^3\delta^{(3)}({\bm{k}}+{\bm{k}'})P_{\bm{k}}\,.\label{WW6}
\end{align}
Assuming that the vector boson field has Gaussian fluctuations, the use of Wick's Theorem\,\cite{PhysRev.80.268}
leads to
\begin{align}
     \langle W^{\alpha}_{\bm{k}'}W_{ {\bm{k}}-{\bm{k}'}\alpha} &W^{ \beta *}_{\bm{k}''}  W^*_{{\bm{k}}-{\bm{k}''}\beta }\rangle = \nonumber\\
    & \langle W^{\alpha}_{ \bm{k}'} W_{{\bm{k}}-{\bm{k}'}\alpha }\rangle \langle W^{\beta * }_{\bm{k}''}W^{*}_{ {\bm{k}}-{\bm{k}''}\beta}\rangle\nonumber\\
   & + \langle W_{\alpha\bm{k}'} W^{\beta *}_{ {{\bm{k}''}}}\rangle \langle W^{\alpha}_{ {\bm{k}}-{\bm{k}'}}W^{*}_{ {\bm{k}}-{\bm{k}''}\beta}\rangle \nonumber\\
  &+  \langle W^{\alpha}_{\bm{k}'} W^*_{ {\bm{k}}-{\bm{k}''}\beta}\rangle \langle W_{{\bm{k}}-{\bm{k}'}\alpha}W^{\beta *}_{ {\bm{k}''}}\rangle\,.\label{eq:wick} 
\end{align}
Replacing Eqs.\,(\ref{WW1})-(\ref{eq:wick}) in Eq.\,(\ref{eq:P}), we obtain the four-point correlation function for the vector field. Each resulting term is proportional to a pair of Dirac delta functions. One of these cancels out due to the volume factor, while the other is used to integrate the ${\bm{k}}''$-integral. Thus, Eq.\,(\ref{eq:P}) becomes 
\begin{align}
&\langle \epsilon \rangle_t= \frac{g^4_{W\gamma}m \kappa}{2\pi^2}\int d\Omega\left[4m^2 {{\bf{\bar{B}}^2}} - \kappa^2({\bm{\hat{x}}}\cdot \bar{\bf{B}})^2\right] \times \,\\
&\int d^3k'  P_{\bm{k}'}P_{\bm{k}-\bm{k}'}\left[ 3-2\frac{\bm{k}'\cdot({\bm{k}}-{\bm{k}'})}{m^2}+\frac{{\bm{k}}'^2({\bm{k}}-{\bm{k}'})^2}{m^4} \right]\,.
\end{align}
Here $d\Omega = d^2\hat{\bm{x}}$ integrates over the ${\bm{k}}$ orientation, i.e. ${\bm{k}}=\kappa {\bm{\hat{x}}}$, and $d^3k'=2\pi d\text{cos}(\theta')dk' k'^2$ with $\theta'$ as the angle between ${\bm{k}}$ and ${\bm{k}'}$ vectors. As before, we take the background magnetic field as $\bar{\bf{B}}=\bar{B}\hat{\bm{z}}$, so that $({\bm{\hat{x}}}\cdot \bar{\bf{B}}) = \bar{B}\text{cos}(\theta)$.
Regarding the field power spectrum, we assume that this
depends only on the magnitude of ${\bm{k}}$ as
$P_{\bm{k}} = P_{\kappa} = \bar{P} e^{-\kappa^2/(2\sigma^2)}$, where $\sigma$ is the standard deviation of the fluctuations in $k$-space and $\bar{P}$ can be expressed in terms of the time-averaged expectation value of the vector field energy, $\bar{\rho}_W \equiv \langle \rho_W \rangle_t \approx m^2 \langle W^i W_i \rangle_t/2 $, as $\bar{P} = (2\pi)^{3/2}\bar{\rho}_W/(3m^2 \sigma^3)$. 

The total emissivity is obtained to be
\begin{align}
\langle \epsilon \rangle_t =& \frac{\pi^{7/2}g_{W\gamma}^4\bar{\rho}_W^2\bar{B}^2\kappa}{27m^7 \sigma^3}e^{-\kappa^2/(4\sigma^2)}\,\nonumber\\
&\times (8m^2+\omega_p^2)(32 m^4+\omega_p^4 +\mathcal{O}(\sigma^2))\,,
\label{eq:Em}
\end{align}
when the photon acquires an effective mass coming from electric charges in the interstellar medium. Note that when $\omega_p \approx 2 m$, the emissivity undergoes an enhancement. Expanding Eq.\,(\ref{eq:Em}) around this value, the emissivity takes the form
\begin{align}
&\langle \epsilon \rangle_t = \frac{512 \pi^{7/2}\bar{B}^2g_{W\gamma}^4\bar{\rho}_W^2m}{3\sqrt{2}\sigma^3\omega_p}\left(1 - \frac{\omega_p^2}{4m^2}\right)^{1/2}\nonumber\\
&\textcolor{white}{XXXXXXXXXXXXX}+\mathcal{O}\left(1 - \frac{\omega_p^2}{4m^2}\right)^{3/2}
\end{align}

The case at which $\omega_p = 0$ is readily obtained from Eq.\,(\ref{eq:Em}) as
\begin{equation}
\langle \epsilon \rangle_t = \frac{512 \pi^{7/2}g_{W\gamma}^4\bar{\rho}_W^2\bar{B}^2}{27\sigma^3}e^{-m^2/\sigma^2}(1+\mathcal{O}(\sigma^2))\,.
\end{equation}

The emitted radiation associated with the vector boson background is negligible in comparison to the soliton dipole radiation. Take, for example, the case at which $\omega_p=0,$ Eq.\,(\ref{eq:PDRa}). If we consider the background emissivity in a volume comparable to the soliton volume, i.e., $\sim (R^{0.24}_{\text{sol}})^3$, the ratio of the powers radiated by both sources can be expressed as follows:
\begin{align}
\frac{\langle P_{(4)} \rangle_t}{\langle \epsilon \rangle_t (R^{0.24}_{\text{sol}})^3} &\sim \frac{m_{\text{pl}}^4 m \sigma^3}{(m R^{0.24}_{\text{sol}})^5 \bar{\rho}_{\text{DM}}^2}e^{\frac{m^2}{\sigma^2}-2\pi m R} \,,\\
&\sim 10^{59}\,\left(\frac{m}{10^{-5}\,\text{eV}} \right)\left(\frac{10^6}{m R^{0.24}_{\text{sol}}} \right)^5\left(\frac{\sigma}{10^{-3}m} \right)^3\nonumber\\
&\hspace{2cm}\times  \left(\frac{0.3\,\text{GeV/cm}^3}{\bar{\rho}_{\text{DM}}} \right)^2\,,
\end{align}
where the time-averaged background vector boson density is taken to be the DM background, i.e. $\bar{\rho}_W = \bar{\rho}_{\text{DM}}$. 

%%%%%%%%%%%%%%%%%%%%%%%%%%%%%%%%%%%%%%%%%%%%%%%%%%%%%%%%%%%%%%%%%%%%%%%
\subsection{Parametric resonance of photons and unbounded radiation}
\label{Sec:Unbounded}

We have examined dipole radiation induced by a dimension-6 operator based on the Lagrangian density in Eq.\,(\ref{Eq:Odipoleradiation}). As mentioned in the introduction, such an operator could also facilitate the parametric resonance of photons under certain conditions\,\cite{Amin:2023imi}. Without requiring external electromagnetic fields or charged particles, dark photon solitons can radiate electromagnetic waves very efficiently due to the exponential growth of the photon occupancy number. This phenomenon can be analyzed using Floquet theory\,\cite{Amin:2014eta}. The parametric resonance condition stipulates that the inverse of the characteristic length scale of the vector soliton must be on the order of or smaller than the photon growth rate. This results in a minimal (critical) value of the dark photon-photon coupling necessary to trigger parametric resonance, as detailed in \,\cite{Amaral:2025fcd, Amin:2023imi}:
\begin{equation}
g_{W\gamma}m_{\text{pl}} > g_{W\gamma, \text{crit}}m_{\text{pl}} \equiv \left( \frac{\mu}{m} \right)^{-3/4}  \approx (mR^{0.24}_\text{sol})^{3/2}\,,
\label{eq:PRcond}
\end{equation}
or, equivalently, to a minimum (critical) soliton mass,
\begin{equation}
    M_{\text{sol}} > M_{\text{sol, crit}} 
    \approx 62.3 \left( \frac{m_{\text{pl}}^2}{m} \right)(mR^{0.24}_\text{sol})^{-1}\,.
\end{equation}
When external electromagnetic fields are present, two distinct behaviors can be observed based on the value of \( g_{W\gamma} \): (1) If \( g_{W\gamma} < g_{W\gamma, \text{crit}} \ll (\mu/m)^{-1} m_{\text{pl}}^{-1} \), periodic solutions are expected to arise due to the dipole radiation phenomenon.
(2) Conversely, if \( g_{W\gamma, \text{crit}} < g_{W\gamma} \ll (\mu/m)^{-1} m_{\text{pl}}^{-1} \), we anticipate exponentially growing solutions driven by the parametric resonance phenomenon, which will dominate the radiation output.\,\footnote{This behavior has also been predicted for axion stars and oscillons, as noted in Ref.\,\cite{Amin:2021tnq}.}. Regarding the second case, the radiation emitted during the parametric resonance phenomenon would extract energy from the dark photon soliton, resulting in the release of excess particles until the soliton mass approaches its critical value \( M_{\text{sol}} \rightarrow M_{\text{sol,crit}} \). Once the parametric resonance process ceases, the dipole radiation phenomenon would become the sole remaining radiative process.

%%%%%%%%%%%%%%%%%%%%%%%%%%%%%%%%%%%%%%%%%%%%%%%%%%%%%%%%%%%%%%%%%%%%%%%
\section{Kinetic mixing}
\label{Sec:KM}

\subsection{Power emission from kinetic mixing}
\label{Sec:KMPW}

Here, we consider the possibility that photons couple only to the Standard Model particles, whereas the massive vector boson couples to the electromagnetic current of ordinary matter (apart from the dark sector). The interacting Lagrangian density of interest is given by Eq.\,(\ref{eq:LKM}).

We picture the soliton configuration in the presence of a mean external charge density. Charge particles, such as electrons, feel a Lorentz force inside the soliton, which comes from the dark electromagnetic field generated by the oscillating vector boson field. The generated oscillating current density is associated with a space-time-dependent charge density, and both source the radiation. Thus, the electromagnetic fields satisfy the following differential equations (see App.\,\ref{App:KM} for further details):  
\begin{align}
&\ddot{{\bf{E}}} - \nabla^2 {\bf{E}} = - \nabla \rho - \dot{\bf{J}}\,,\label{eq:weq1b}\\
&\ddot{{\bf{B}}} - \nabla^2 {\bf{B}} = \nabla \times {\bf{J}}\,,\label{eq:weq2b}
\end{align}
where the charge and current densities read as
\begin{align}
&\rho({\bm{x}},t) \sim \left(\frac{\epsilon e^2\rho_e}{m^2_e}\right) \int^t \nabla \cdot {\bf{W}}({\bm{x}},t')\,dt' + \text{constant} \,,\label{eq:rhomixing}\\
&{\bf{J}}({\bm{x}},t)\sim -\left(\frac{\epsilon e^2\rho_e}{m^2_e}\right){\bf{W}}({\bm{x}},t) 
+ \text{constant}\,.\label{eq:jmixing}
\end{align}
Here we have neglected the gradients of the field within the non-relativistic regime in Eq.\,(\ref{eq:jmixing}), use the continuity equation in Eq.\,(\ref{eq:rhomixing}), and  $e$, $m_e$,  $\rho_e$ indicate the electron charge,  mass, and mean density,  respectively. 

For simplicity, let us consider a linearly polarized soliton, $c_{\pm 1}=0$, $c_0=1$, and $\varphi_0=0$ in Eq.\,(\ref{eq:W2}). However, note that the charge and current densities are proportional to the real-value vector field ${\bf{W}}$ rather than the trace of their contraction $\text{Tr}[{\bf{W}}{\bf{W}}]$, as in the previous case. This difference makes such quantities time-dependent, regardless of the soliton polarization, i.e., linear, circular, or a linear combination of both. For a linearly polarized soliton, Eqs.\,(\ref{eq:jmixing}) and (\ref{eq:rhomixing})  become 
\begin{align}
\rho({\bm{x}},t)&= \mathbb{R}\left[\varrho({\bm{x}})e^{-i\omega t} \right]\,,\label{eq:source1}\\
{\bf{J}}({\bm{x}},t)& = \mathbb{R}\left[j({\bm{x}})e^{-i\omega t} \right]\,,\label{eq:source2}
\end{align}
where
\begin{align}
   & \varrho({\bm{x}}) = i\left(\frac{\epsilon e^2 \rho_e}{\omega m_e^2} \right) \nabla \phi({\bm{x}})\cdot {\bm{\epsilon}}_{\hat z}^{(0)}\,,\label{eq:resource1}\\
    &{\bf{j}}({\bm{x}}) = -\left(\frac{\epsilon e^2 \rho_e}{m_e^2} \right) \phi({\bm{x}}){\bm{\epsilon}}_{\hat z}^{(0)}\,,\label{eq:resource2}
\end{align}
 where ${\bm{\epsilon}}_{\hat z}^{(0)}$ is defined in Eq.\,(\ref{eq:vec}) and we have dropped the constant term in the charge density from Eq.\,(\ref{eq:rhomixing}) due to its gradient vanishes in Eq.\,(\ref{eq:weq1}). 
 
 Following the same procedure as before, we use a standard Green function method to solve the inhomogeneous wave equations in Eqs.\,(\ref{eq:weq1}) and (\ref{eq:weq2}). The electric and magnetic field components read as
 \begin{align}
{\bf{E}}& = -\mathbb{R} \left[ \frac{e^{-i\omega( t - |{\bf{x}}|)}}{4\pi |{\bf{x}}|} \left( i{\bf{k}}\tilde{\varrho}({\bf{k}})-i\omega \tilde{{\bf{j}}}({\bf{k}}) \right) \right] \,,\\  
{\bf{B}}& = -\mathbb{R} \left[ \frac{e^{-i\omega( t - |{\bf{x}}|)}}{4\pi |{\bf{x}}|} \left( -i{\bf{k}}\times \tilde{{\bf{j}}}({\bf{k}}) \right) \right]\,.
 \end{align}
 where ${\bm{k}}\equiv \omega \hat{\bm{x}}$.

Far away from the soliton and $t \gg 0$, the emitted power per unit solid angle reads as
 \begin{equation}
    \frac{dP}{d\Omega} = \frac{\omega^2}{32\pi^2}\left[ -|\tilde{\varrho}({\bm{k}})|^2 + |\tilde{{\bf{j}}}({\bm{k}})|^2 + \text{oscillating terms} \right]\,,\label{eq:dPdOmegab} 
 \end{equation}
with
\begin{align}
\phi({\bm{x}}) &= \int \frac{d^3{\bm{k}}}{(2\pi)^3}\tilde{\phi}({\bm{k}})e^{i{\bm{k}}\cdot{\bf{x}}}\,,\\
\tilde{\varrho}({\bm{k}}) &= - \left( \frac{\epsilon e^2 \rho_e}{m_e^2} \right)  \tilde{\phi}(\omega)    \hat{\bm{x}}\cdot {\bf{\epsilon}^{(0)}}\,,\label{eq:rhokb}\\
\tilde{\bf{j}}({\bm{k}})&= -\left( \frac{\epsilon e^2 \rho_e}{m_e^2} \right) \tilde{\phi}(\omega){\bf{\epsilon}}^{(0)}\,,\label{eq:jkb}
\end{align}
where we have used the relation $\omega \tilde{\varrho}({\bm{k}})={\bm{k}}\cdot \tilde{\bf{j}}({\bm{k}})$ from the continuity equation. Replacing Eqs.\,(\ref{eq:rhokb}) and (\ref{eq:jkb}) into Eq.\,(\ref{eq:dPdOmegab}), the total time-averaged power emitted by the vector boson soliton due to kinetic mixing is calculated to be
\begin{align}
&\frac{dP}{d\Omega} = \frac{\omega^2 \epsilon^2 e^4 \rho_e^2 \tilde{\phi}^2(\omega)}{32\pi^2 m_e^4}    \left[ |{\bm{\epsilon}}_{\hat z}^{(0)}|^2 - |\hat{\bm{x}}\cdot {\bm{\epsilon}}_{\hat z}^{(0)}|^2\right]\nonumber\\
&\textcolor{white}{xxxxxxxxxxxxx}\times(1+\text{oscillating terms})\,,\\
&\langle P \rangle_{t} = \frac{ \omega^2 \epsilon^2 e^4 \rho_e^2 \tilde{\phi}^2(\omega)}{12\pi m_e^4}\,,
\end{align}
 where we have used $|{\bf{\epsilon}}^{(0)}|^2-|\hat{\bf{x}}\cdot{\bf{\epsilon}}^{(0)}|^2=\text{sin}^2(\theta)$.
The emitted power is proportional to the square of the Fourier transform of the soliton profile, and the radiation is emitted at a frequency $\omega$. That is is $\langle P \rangle_t\propto \left\{\mathcal{F}[\phi(r)](\omega)\right\}^2=\tilde{\phi}^2(\omega)$. 

We previously showed that a total or partial linear polarization of the soliton is required to have dipole radiation for the dimension-6 operator considered in Eq.\,(\ref{Eq:Odipoleradiation}). By contrast, the presence of kinetic mixing allows the system to undergo dipole radiation regardless of the soliton polarization. The reason is the following. Now the quantities of interest are proportional to the real-value vector field ${\bf{W}}({\bf{x}},t)$ rather than the trace of its contraction 
$\text{Tr}[{\bf{W}}{\bf{W}}]$, so that the \textit{soliton time-dependence is not lost throughout the calculation}. For the case of a circularly polarized soliton with $c_{1}=1$ or $c_{-1}=1$, the previous equations hold under the replacement ${\bm{\epsilon}}_{\hat z}^{(0)} \rightarrow {\bm{\epsilon}}_{\hat z}^{(\pm 1)}$, Eq.\,(\ref{eq:vec}). Thus, the emitted power per unit of solid angle is now given by  
\begin{align}
&\frac{dP}{d\Omega} = \frac{\omega^2 \epsilon^2 e^4 \rho_e^2 \tilde{\phi}^2(\omega)}{32\pi^2 m_e^4}    \left[ |{\bf{\epsilon}}^{(\pm 1)}|^2 - |\hat{\bm{x}}\cdot {\bf{\epsilon}}^{(\pm 1)}|^2\right.\nonumber\\
&\textcolor{white}{xxxxxxxxxxxxxxxxxxx}\left.+\text{oscillating terms}\right]\,,\\
&\langle P \rangle_{t} = \frac{ \omega^2 \epsilon^2 e^4 \rho_e^2 \tilde{\phi}^2(\omega)}{24\pi m_e^4}\,,
\end{align}
where
\begin{align}
&\tilde{\phi}(|{\bm{k}}|) = \int d^3{\bm{x}}\, \phi_0\, \text{sech}(r/R^{0.24}_{\text{sol}})\,e^{-i {\bm{k}}\cdot {\bm{x}}}\,,\\ 
&=\frac{4\pi \phi_0}{|{\bm{k}}|}\int_0^{\infty} dr\, r\, \text{sech}(r/R^{0.24}_{\text{sol}}) \,\text{sin}(|{\bm{k}}| r)\,,\\
&= \frac{4\pi^3 (R^{0.24}_{\text{sol}})^2 \phi_0}{|{\bm{k}}|}\text{csch}^2(|{\bm{k}}|\pi R^{0.24}_{\text{sol}})\text{sinh}^3(|{\bm{k}}|\pi R^{0.24}_{\text{sol}})\,.\label{eq:ft2}
\end{align}
The total time-averaged emitted power by a generic fractionally polarized soliton, Eq.\,(\ref{eq:W2}), is readily calculated as
\begin{equation}
    \langle P \rangle_t = \frac{\omega^2\epsilon^2e^4\rho^2_{e}\tilde{\phi}^2(\omega)}{12\pi m_e^4}\left(
    \frac{c_{1}^2}{2}+ \frac{c_{-1}^2}{2}+ c_{0}^2\right)\,.\label{eq:Pt}
\end{equation}
%\MV{--- Checked up to here ---} 
In the regime $\omega R \gg 1$, the emitted power is exponentially suppressed due to the behavior of the Fourier transform $\tilde{\phi}(|{\bf{k}}|)$, which is approximated to be
\begin{equation}
\tilde{\phi}(\omega) \approx \frac{2\pi}{\omega^3}\phi_0(\pi \omega R^{0.24}_{\text{sol}})^2 e^{-\pi \omega R^{0.24}_{\text{sol}}/2}\,.\label{eq:ftphi}
\end{equation}
for $\omega R^{0.24}_{\text{sol}} \gtrsim 2$. Replacing Eq.\,(\ref{eq:ftphi}) into Eq.\,(\ref{eq:Pt}), and using 
$\phi_0 \approx m_{\text{pl}}/(mR^{0.24}_{\text{sol}})^2$ as before,
the radiated power is estimated to be
\begin{align}
    \langle P \rangle_t &\sim  \epsilon^2 \frac{m_{\rm pl}^2n_e^2}{m^4 m_e^2} e^{-\pi m R^{0.24}_{\text{sol}}}\,,
    \label{eq:KMnowpl} \\
    &\sim 10^{17}{\rm ergs}\, s^{-1}\left(
    \frac{n_e}{0.03\,{\rm cm}^{-3}}\right)^2\left(\frac{\epsilon}{10^{-11}}\right)^2\,
    \nonumber\\
    &\textcolor{white}{XXXXXXXX}\times\left(\frac{10^{-6}{\rm eV}}{m}\right)^4 e^{-\pi m R^{0.24}_{\text{sol}}}\,,
\end{align}
 where we have dropped the prefactor $(\pi^5e^4/3)\approx 1$, with $e \approx 0.303$ in natural units. We will see in the next subsection that including the effects of the plasma mass removes the exponential suppression in the radiative power.

%%%%%%%%%%%%%%%%%%%%%%%%%%%%%%%%%%%%%%%%%%%%%%%%%%%%%%%%%%%%%%%%%%%%%%%
\subsection{Effective photon mass}
\label{sec:photonmassmixing}

Here, we consider an effective mass for the photon in the same way as was done in Sec.\,\ref{sec:photonmassdipole}, where the soliton dipole radiation in the presence of external electric and magnetic fields is calculated. In the regime $\omega_p < \omega$, we make the replacement $\nabla^2 \rightarrow \nabla^2 - \omega^2_p$ in Eqs.\,(\ref{eq:weq1b}) and (\ref{eq:weq2b}), with sources given by Eqs.\,(\ref{eq:source1}) and (\ref{eq:source2}). To work out the general case of polarization, we make the replacement $\bm{\epsilon}^{(0)}_{\hat z} \rightarrow \bm{\epsilon}_{\hat z}^{\lambda}$ in Eqs.\,(\ref{eq:resource1}) and (\ref{eq:resource2}), where $\lambda = \{0,1,-1\}$ as shown in Eq.\,(\ref{eq:W2}).

The emitted power per unit of solid angle is readily obtained as
 \begin{equation}
    \frac{dP}{d\Omega} = \frac{\omega\kappa}{32\pi^2}\left[ -|\tilde{\varrho}({\bm{k}})|^2 + |\tilde{{\bf{j}}}({\bm{k}})|^2 + \text{oscillating terms} \right]\,,\label{eq:dPdOmegab2} 
 \end{equation}
with
\begin{align}
\phi(|{\bm{x}}|) &= \int \frac{d^3{\bm{k}}}{(2\pi)^3}\tilde{\phi}(|{\bm{k}}|)e^{i{\bf{k}}\cdot{\bm{x}}}\,,\\
\tilde{\varrho}({\bm{k}}) &= - \left( \frac{\epsilon e^2 \rho_e}{m_e^2} \right)\left(\frac{\kappa}{\omega} \right)  \tilde{\phi}(\kappa)    \hat{\bm{x}}\cdot {\bm{\epsilon}_{\hat z}^{(0)}}\,,\label{eq:rhokb2}\\
\tilde{\bf{j}}({\bm{k}})&= -\left( \frac{\epsilon e^2 \rho_e}{m_e^2} \right) \tilde{\phi}(\kappa){\bm{\epsilon}}_{\hat z}^{(0)}\,,\label{eq:jkb2}
\end{align}
 where ${\bm{k}}\equiv \kappa \hat{\bm{x}}$,   $\kappa^2 = \omega^2 - \omega_p^2$,
 and the Fourier transform of the soliton profile, $\tilde{\phi}(\kappa)$, is given by Eq.\,(\ref{eq:ft2}).

 Replacing Eqs.\,(\ref{eq:rhokb2}) and (\ref{eq:jkb2}) into Eq.\,(\ref{eq:dPdOmegab2}), we obtain
 \begin{align}
\frac{dP}{d\Omega} = \frac{\omega\kappa \epsilon^2 e^4 \rho_e^2 \tilde{\phi}^2(\kappa)}{32\pi^2 m_e^4}  &  \left[ |{\bm{\epsilon}}_{\hat z}^{\lambda}|^2 -\left( \frac{\kappa}{\omega} \right)^2 |\hat{\bm{x}}\cdot {\bm{\epsilon}}_{\hat z}^{\lambda}|^2\right.\nonumber\\
&\left.+\text{oscillating terms}\right]\,.
\end{align}
For a generic fractionally polarized soliton, the total time-averaged emitted power reads as
\begin{equation}
\langle P \rangle_t =\frac{\omega \kappa \epsilon^2 e^4 \rho_e^2 \tilde{\phi}^2(\kappa)}{8\pi m_e^4} \left[ 1 - \frac{2}{3}\left( \frac{\kappa}{\omega}\right)^2\left(c_1^2  + c_{-1}^2 + \frac{c_0^2}{2}\right) \right]\,,
\label{eq:PtKM}
\end{equation}
where we recover Eq.\,(\ref{eq:Pt}) in the limit $\omega_p \rightarrow 0$, i.e.  
$\kappa = \omega$. By contrast, when $\omega_p \rightarrow \omega$, we are in the resonant conversion domain where the emitted power is enhanced. In such a limit, the total time-averaged power emitted in Eq.\,(\ref{eq:P4}) takes the form.
\begin{align}
&\langle P \rangle_t = \frac{\pi \left( \epsilon e^2 \rho_e \phi_0/m_e^2    \right)^2}{16\sqrt{2}\omega^4} \left(\pi \omega R \right)^6\left(1-\frac{\omega_p^2}{\omega^2} \right)^{1/2}\nonumber\\
&\textcolor{white}{XXXXXXXXXXXX} + \mathcal{O}\left[ (1-\omega_p^2/\omega^2)^{3/2} \right]\,.  
\end{align}
Taking $\phi_0 \sim m_{\text{pl}}/(mR^{0.24}_{\text{sol}})^2$, and $\omega \approx m$ as before, we have
\begin{align}
&\langle P \rangle_t \sim \epsilon^2 \frac{m_{\rm pl}^2n_e^2}{m^4 m_e^2} \left( m R_{\text{sol}}^{\text{0.24}} \right)^6\left(1-\frac{\omega_p^2}{\omega^2} \right)^{1/2}\nonumber\\
&\textcolor{white}{XXXXXXXXXXXX} + \mathcal{O}\left[ (1-\omega_p^2/\omega^2)^{3/2} \right]\,, 
\label{eq:KMyeswpl}\\ 
    & \sim 10^{18}{\rm ergs}\, s^{-1}\left(
    \frac{n_e}{0.03\,{\rm cm}^{-3}}\right)^2\left(\frac{\epsilon}{10^{-11}}\right)^2\,
    \nonumber\\
    & \textcolor{white}{XX}\times\left(\frac{10^{-6}{\rm eV}}{m}\right)^4  \left(\frac{mR_{\rm sol}^{0.24}}{3.5}\right)^2
    \left(1-\frac{\omega_p^2}{\omega^2} \right)^{1/2}\,,
    \label{eq:kmapp}
\end{align}
where we have dropped the prefactor $e^4\pi^7/(16\sqrt{2})\approx 1$ and set $(\mu/m) = 10^{-1}$ in $mR_{\rm sol}^{0.24}$ to barely satisfy the non-relativistic regime and the EFT validity condition, as before.

Figure\,\ref{Peak} (bottom panel) shows the behavior of the time-averaged radiated power by the soliton in terms of $\omega_p$ and $\omega R_{\rm sol}^{0.24}$. We have varied $\omega R_{\rm sol}^{0.24}$ for 4 (light blue line) to 10 (dark blue line). The resonance phenomenon, i.e., the significant enhancement of the power, is favored as solitons become more diluted when $\omega_p$ approaches $\omega$.

%%%%%%%%%%%%%%%%%%%%%%%%%%%%%%%%%%%%%%%%%%%%%%%%%%%%%%%%%%%%%%%%%%%%%%%
\subsection{Massive vector dark matter background}
\label{DMbckg2}

In the previous subsection, we have calculated the radiated power from a soliton due to kinetic mixing. To assess the relevance of this process, we compare this result with the expected emissivity from a massive vector boson dark matter background.
We follow a similar procedure to that in Sec.\,\ref{DMbckg1}. Using the modes expansion of the four-vector massive boson in Eq.\,(\ref{eq:Wk}), we express the four-vector current $J^{\mu}$, Eqs.\,(\ref{eq:rhomixing}) and (\ref{eq:jmixing}), as
\begin{align}
J^0 &= \frac{i\epsilon e^2 \rho_e}{m_e^2}\int dt \int \frac{d^3k}{(2\pi)^3}\left( k_i W^i_{\bm{k}} e^{-im t + i{\bm{k}}\cdot{\bm{x}}}\right.\,\nonumber\\
&\textcolor{white}{XXXXXXXXXXX}\left.- k_i {W}^{i*}_{\bm{k}} e^{im t - i{\bm{k}}\cdot{\bm{x}}}\right)\,,\\
&=J^0_1 + J^0_{-1}\,\nonumber\\
{\bf{J}} &= -\frac{ \epsilon e^2 \rho_e}{m_e^2}\int \frac{d^3k}{(2\pi)^3}\left( {\bf{W}}_{\bm{k}}e^{-im t + i{\bm{k}}\cdot {\bm{x}}}\,\right.\nonumber\\
&\textcolor{white}{XXXXXXXXXXX}\left.+ {\bf{W}}^{*}_{\bm{k}}e^{im t - i{\bm{k}}\cdot {\bm{x}}} \right)\,,\\
&={\bf{J}}^i_{1} + {\bf{J}}^i_{-1}\,.
\end{align}
For kinematics, the condition $\Theta(m-\omega_p)$ needs to be satisfied to have radiation. Thus, only $J^{\mu}_1$ contributes to the process. After applying a time average over many oscillation cycles, the total emissivity reads as
\begin{equation}
\langle \epsilon\rangle_t  = \frac{1}{\text{Volume}} \int d\Omega \frac{m \kappa}{8\pi^2}J^{\mu}_1({\bm{k}})J^*_{\mu 1}({\bm{k}})\Theta(m-\omega_p)\,,\label{eq:emiss} 
\end{equation}
where ${\bm{k}}=\kappa {\hat{\bm{x}}}$ with $\kappa = \sqrt{m^2 - \omega^2_p}$. The corresponding Fourier transforms of the time and spatial components of the four-vector current are given by
\begin{align}
    J^0_1({\bm{k}})&=-\frac{\epsilon e^2 \rho_e}{m_e^2m}{\bm{k}}\cdot {\bf{W}}_{\bm{k}}  e^{-im t}\,,\label{eq:J01}\\
    {\bf{J}}_1({\bm{k}})&=-\frac{\epsilon e^2 \rho_e}{m_e^2}{\bf{W}}_{\bm{k}} e^{-im t}\,,\label{eq:Ji1}
\end{align}
where we have dropped the constant term in Eq.\,(\ref{eq:J01}). Replacing these two last equations into Eq.\,(\ref{eq:emiss}), one finds
\begin{align}
\langle \epsilon \rangle_t& =\frac{m \kappa \epsilon^2e^4\rho_e^2}{8\pi^2m^4_e}\int d\Omega \left( \langle |W_{\bm{k}}|^2\rangle-\frac{\langle |{\bm{k}}\cdot {\bf{W}}_{\bm{k}}|^2\rangle }{m^2} \right)\,, \\
& =\frac{8\sqrt{2}\pi^{7/2}\epsilon^2e^4n_e^2\omega_p^2\bar{\rho}_w}{3m_e^2m^3\sigma^3}(m^2-\omega_p^2)^{1/2}e^{\frac{-(m^2-\omega_p^2)}{2\sigma^2}}\,,\label{eq:Emisbkg}
\end{align}
where we have used the two-point function relation of Eqs.\,(\ref{WW1}) and the same Gaussian power spectrum as before, i.e. $P_{\bm{k}} = \bar{P}e^{-\kappa^2/(2\sigma^2)}$ with $\bar{P}=(2\pi)^{3/2}\bar{\rho}_W/(3m^2\sigma^3)$. The emissivity vanishes in the absence of an effective photon mass because angular momentum cannot be conserved in such a case. By contrast, when $m \approx \omega_p$, the exponential suppression shuts off, and the emissivity is enhanced. In such a regime, Eq.\,(\ref{eq:Emisbkg}) takes the form
\begin{align}
\langle \epsilon \rangle_t \sim \frac{16 \pi^{7/2}e^4 \epsilon^ 2n_e^2 \bar{\rho}_W}{3m_e^2 \sigma^3}\frac{m}{\omega_p}(1-\omega_p^2/m^2)^{1/2} +\,\nonumber\\
\textcolor{white}{XXXXXXXXXXXX}\mathcal{O}[(1-\omega_p^2/m^2)^{3/2}]\,.    
\end{align}

If we compare the radiated power from the soliton in the regime at which $m \approx \omega_p$, Eq.\,(\ref{eq:kmapp}), with that from
a vector boson background in a volume $(R_\text{sol}^{0.24})^3$, one finds
\begin{align}
&\frac{\langle P\rangle_t}{\langle \epsilon_t\rangle (R_\text{sol}^{0.24})^3} \sim  \frac{\sigma^3 m_{\text{pl}}^2}{m^2 \bar{\rho}_W R_\text{sol}^{0.24}}\,,\\
&\sim 10^{35}\,\left(\frac{\sigma}{10^{-3}m} \right)^3\left(\frac{10^{-5}\,\text{eV}}{m} \right)\left( \frac{10^6}{m R_\text{sol}^{0.24}}\right)\left( \frac{0.3\,\frac{\text{GeV}}{\text{cm}^3}}{\bar{\rho}_{\text{DM}}}\right)\,.
\end{align}
As before, the radiated power from the vector dark matter background is negligible in comparison to that from the soliton.

%%%%%%%%%%%%%%%%%%%%%%%%%%%%%%%%%%%%%%%%%%%%%%%%%%%%%%%%%%%%%%%%%%%%%%%
%%%%%%%%%%%%%%%%%%%%%%%%%%%%%%%%%%%%%%%%%%%%%%%%%%%%%%%%%%%%%%%%%%%%%%%
\section{Astrophysical Signatures}
\label{Sec:AS}

The primary purpose of this work is to explore the emission of electromagnetic radiation from vector solitons in relation to the physical phenomena of dipole radiation and kinetic mixing. In this section, we examine the primary characteristics of such electromagnetic radiation, with a focus on some detection considerations. Specifically, we estimate the expected spectral flux density and assess the feasibility of vector solitons producing radiation in favorable astrophysical scenarios, particularly when they are embedded in the magnetospheres of highly magnetized compact objects, such as neutron stars and white dwarfs.

To detect using radio telescopes located on Earth, the mass of the spin-1 particle should rest within the range of $\sim(10^{-7}-10^{-3})\,\text{eV}$. The lower limit corresponds to a signal frequency of approximately \( 30 \text{ MHz} \). Detecting frequencies below this limit is particularly difficult due to the ionosphere's absorption and scattering of low-frequency photons. When we account for projected space-based facilities, the minimum mass limit could decrease further. For instance, the Orbiting Low Frequency Antennas for Radio Astronomy Mission (OLFAR) \cite{2020AdSpR..65..856B, 2016ExA....41..271R}, which aims to deploy thousands of nanosatellites on the Moon's far side, would allow for the detection of signals with frequencies as low as \( 0.30 \text{ MHz} \)\,\cite{2025rajan}. OLFAR would expand the range of interest for the mass of spin-1 particles to $\sim(10^{-9}-10^{-3})\,\text{eV}$\,\footnote{Detection from space has been discussed in various contexts relating to indirect searches for dark matter. For example, refer to Ref.\,\cite{Choi:2022btl} concerning the detection of ultra-long radio waves for axion-photon conversion during interactions between axion self-similar minihalos and neutron stars.}.

Following Ref.\,\cite{Amin:2021tnq}, for the compact objects of our interest, neutron stars and white dwarfs, we adopt the Goldreich–Julian model (GJ)\,\cite{1969ApJ...157..869G} to describe their magnetospheres\,\footnote{While the GJ model provides a foundational framework, the magnetospheres of highly magnetic white dwarfs, which is the group of our interest regarding white dwarfs, may have non-dipolar field structures with higher-order multipoles\,\cite{2015SSRv..191..111F}. Our goal is to give a rough estimate of the signal as a starting point, highlighting that more advanced models are needed to capture these complexities. We leave this compelling task for future work.}. This model provides the minimum plasma density at which particles on the dipole magnetic field lines are constrained to co-rotate with the star. The GJ model has been adopted in previous studies related to axion dark matter\,\cite{Millar:2021gzs, Foster:2020pgt, Leroy:2019ghm, Safdi:2018oeu,  Huang:2018lxq, Safdi:2018oeu, Pshirkov:2007st}, axion stars, miniclusters, and minihalos around primordial black holes\,\cite{Witte:2022cjj, Choi:2022btl, Nurmi:2021xds, Edwards:2020afl, Buckley:2020fmh}  and axion dark radiation\,\cite{Long:2024qvd}. 

The compact object in question is represented as a sphere of radius $R_{\mathrm{co}}$ that spins with angular velocity $\bf \Omega$ about an axis passing through its center. The magnitude of the angular velocity is inversely proportional to the rotation period $P$, such that $|{\bf\Omega}| =\Omega = 2\pi/P$. In the frame where  ${\bf\Omega}$ is fixed, the magnetic field in the magnetosphere of the compact object, ${\bf B} ({\bf r}, t)$, is well described by a dipole configuration, with a magnetic dipole ${\bf m}(t)$ with varying orientation but constant magnitude $|{\bf m}(t)| =m$. The magnetic field takes the form
\begin{equation}
{\bf  B} ({\bf r}, t) = B_0 {\bm{\psi}}_B(\hat{\bf r}, t) 
\left( \frac{r}{R_{\text{co}}} \right)^{-3}\,,
\label{eq:B}
\end{equation}
where $r = |{\bf r}|$ denotes the radial distance from the compact object's center, $B_0 = 2m/R_{\text{co}}^3$ is the magnetic field amplitude at the surface poles, and 
${\bm{\psi}}_B(\bm{\hat r}, t)$ encodes the magnetic field angular structure as
\begin{align}
 \bm{\psi_B}(\bm{\hat r}, t)=\frac{3}{2}\left(\bm{\hat m}(t)\cdot \bm{\hat r}\right)\bm{\hat r} -\frac{1}{2}\bm{\hat m}(t)\,.
\end{align}
Within the GJ description, the plasma frequency is determined by the density of  charged particles $n_c(\bm{r},t)$, such that
\begin{equation}
n_c(\bm{r},t) = \frac{2\bm{\Omega}\cdot{\bm B}(\bm r, t)}{e}\,,
\end{equation}
where $e \approx 0.303$ is the unit of the electromagnetic charge.
Away from polar regions, where the assumption of a non-relativistic electron-proton plasma is well justified, the plasma frequency is approximately given by the electron number density $n_{\text{e}}({\bf{r}}, t)$ as
\begin{equation}
\omega^2_{\text{pl}}(\bm{r},t) = \frac{4\pi\alpha n_{\text{e}}(\bm{r},t)}{m_e}\,,
\label{eq:omegapl}
\end{equation}
where $m_{\text{e}} \approx 0.511\,\text{MeV}$ is the electron rest mass and 
$\alpha =e^2/(4\pi)\approx 1/137$ is the electromagnetic fine structure constant. Taking the number density of  charged particles as an estimate of the electron number density,  $n_{\text{e}}({\bf{r}}, t)
\approx |n_{\text{c}}({\bf{r}}, t)|$ and using Eqs.\,(\ref{eq:B}-\ref{eq:omegapl}), we have
\begin{align}
\label{eq:barwpl}
    & \omega_{\text{pl}}(\bm r,t) 
    = \omega_{\text{pl},0} \ 
    \psi_\omega(\bm r,t) \ 
    \biggl( \frac{r}{R_{\text{co}}} \biggr)^{\! \! -3/2}\,,
    \\ & \quad \text{where} \quad 
    \psi_\omega(\bm r,t) = \bigl| 2 \bm\Omega \cdot \bm\psi_B(\bm r,t) \bigr|^{1/2}\,,\nonumber \\
 & \quad \text{and} \quad 
    \omega_{\mathrm{pl},0} \approx 
    \bigl( 70 \mu\text{eV} \bigr) 
    \biggl( \frac{B_0}{10^{14} \text{G}} \biggr)^{\! \! 1/2} 
    \biggl( \frac{P}{1 \sec} \biggr)^{\! \! -1/2}\,. 
    \nonumber
\end{align}
This description is valid only within the light cylinder radius, $R_{\mathrm{LC}}$, which takes the form 
\begin{equation}
        R_{\mathrm{LC}}\approx 4.8\times 10^4\,\mathrm{km}\, \,\left(\frac{P}{1\,\mathrm{sec}}\right)\,.
  \label{eq:RLC}      
\end{equation}
The GJ model remains reliable at a radius $r$ from the center of the compact object that satisfies the condition $R_{\mathrm{co}} < r \ll R_{\mathrm{LC}}$ and is far enough from the polar regions, where the plasma experiences a boost\,\cite{Safdi:2018oeu, Long:2024qvd}.  
\begin{figure}[t]
\centering
 \includegraphics[width= 70mm]{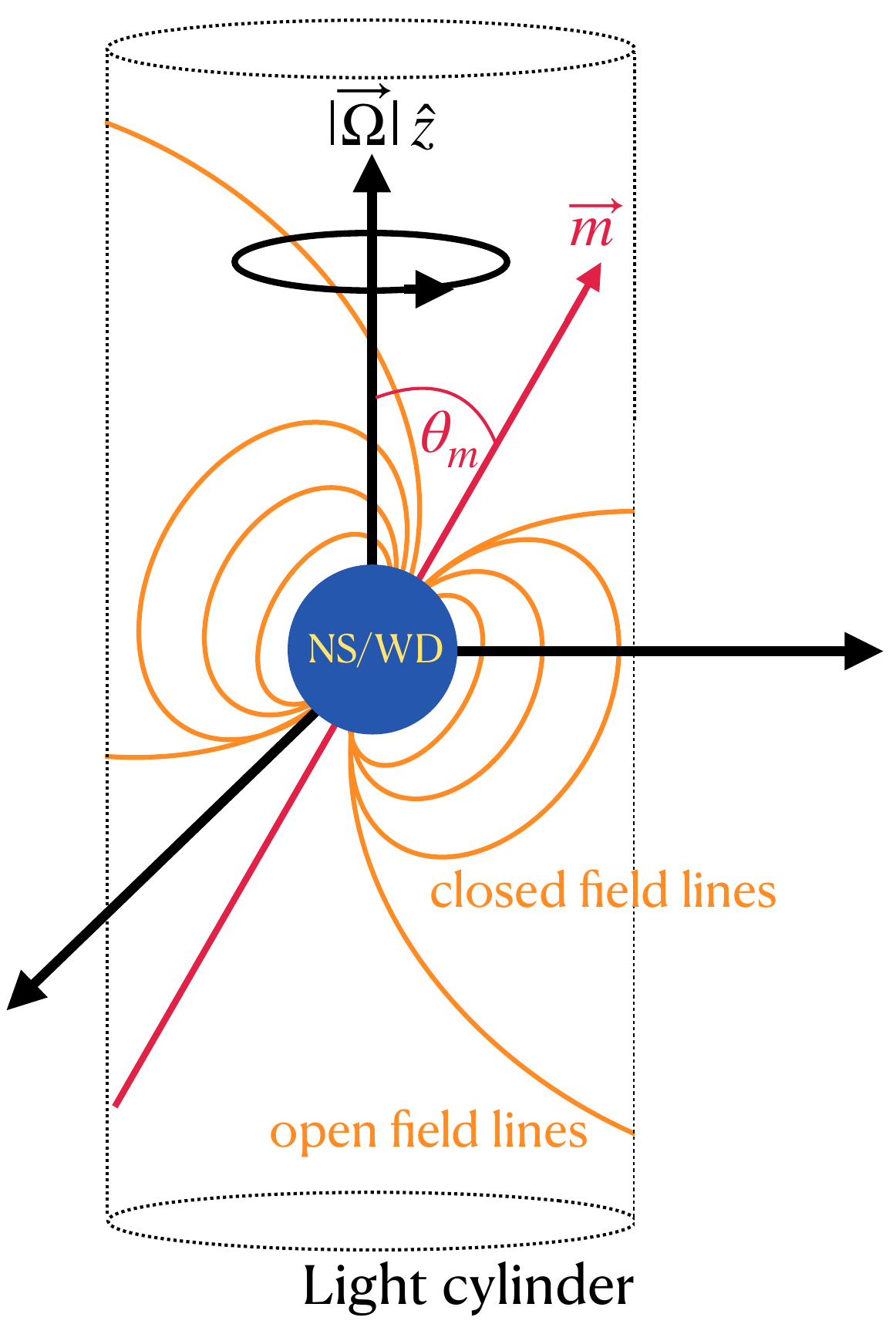}
\caption{Diagram for the magnetic field lines around a rotating neutron star or white dwarf, holding an angular velocity along the $z$-axis and $\hat\Omega\cdot\hat m = \text{cos}(\theta_m)$.}
 \label{fig:NSdiagram}
\end{figure}
Figure\,\ref{fig:NSdiagram} shows a diagram of a neutron star/white dwarf and its magnetosphere, where the compact object's rotation axis is along with the $z$-axis and the magnetic dipole formed an angle $\theta_m$ with it. 

For aligned compact objects, $\theta_m = 0$ and $\bf \Omega || \bf m$, the equations above simplify considerably. In such a case, the electron number density and the plasma frequency, for example, take the simple form
\begin{align}
\label{eq:nealigned}
&n_e(r) \approx 3 \times 10^{10}\,\text{cm}^{-3}
\left( \frac{B_0}{10^{12}\,\text{G}}\right)
\left( \frac{1\,\text{s}}{P}\right)
\left( \frac{r}{R_\text{co}}\right)^{-3}\nonumber\\
&\hspace{1cm}\times|3\text{cos}^2(\theta)-1|\,,\\
&\omega_{\text{pl}}(r) 
    = \omega_{\text{pl},0} \  
    \biggl( \frac{r}{R_{\text{NS}}} \biggr)^{\! \! -3/2}
    |3\text{cos}^2(\theta)-1|^{1/2}\,,
    \label{eq:wplaligned}
\end{align}
where both quantities are now constant in time.

For the case of neutron stars, galactic pulsar surveys and simulations show that the neutron star spin period and magnetic field at poles are well represented by a normal and log-normal distribution, respectively, with a mean and dispersion given by $\langle\text{(P/s)}\rangle = 0.3,\, \sigma_{\text{P}} = 0.15 $,  and $\langle\text{log}_{10}\text{(B}_0\text{G)}\rangle = 12.95,\, \sigma_{\text{B}_0} = 0.55$, respectively\,\cite{Safdi:2018oeu, Faucher-Giguere:2005dxp}.  

Regarding white dwarfs, they represent the most prevalent conclusion of stellar evolution\,\cite{2010A&ARv..18..471A}, and it is thought that more than 97$\%$ of the stars in the Galaxy will ultimately become white dwarfs\,\cite{2001PASP..113..409F}. Volume\,\cite{2007ApJ...654..499K, 2012ApJS..199...29G, 2014AJ....147..129S}  and limited-surveys\,\cite{Liebert:2002qu, 2013MNRAS.429.2934K, 2016MNRAS.455.3413K} suggest that around (10-20)$\%$ and (2-5)$\%$ of them, respectively, have measurable
magnetic fields with strengths ranging from $10^3$ to $10^9$ G\,\cite{2015SSRv..191..111F}. High-field magnetic white dwarfs, which have magnetic fields ranging from approximately \(10^6\) to \(10^9\) G, make up the majority of known magnetic white dwarfs. This group has an average mass of about \(0.93\, \text{M}_\odot\) \cite{2005MNRAS.356..615F} and a characteristic radius on the order of \(10^3 \, \text{km}\)\,\cite{Acevedo:2023xnu}. The rotation periods of magnetic white dwarfs vary widely, ranging from approximately $10^3$ seconds to several decades. An analysis of the distribution of magnetic fields in relation to spin periods suggests the presence of two distinct groups. One group is defined by ``short" periods that range from hours to weeks, while the other is characterized by rotation periods extending from decades to centuries, based on observations of objects monitored over many years. It appears that magnetic white dwarfs with very long rotation periods tend to have high magnetic fields. In contrast, those with short rotation periods do not exhibit a specific preferred field strength \cite{2015SSRv..191..111F}. 

%%%%%%%%%%%%%%%%%%%%%%%%%%%%%%%%%%%%%%%%%%%%%%%%%%%%%%%%%%%%%%%%%%%%%%%
\subsection{Spectral flux density}

We begin by estimating the expected spectral flux density associated with the two phenomena under investigation. As previously noted, the most favorable scenario occurs when vector solitons are embedded within the magnetosphere of a highly magnetized compact object, such as a neutron star or a white dwarf. \textcolor{black}{In the following subsections, we proceed to estimate the spectral flux density for some fiducial parameters.}

%%%%%%%%%%%%%%%%%%%%%%%%%%%%%%%%%%%%%%%%%%%%%%%%%%%%%%%%%%%%%%%%%%%%%%%
\subsubsection{Dipole Radiation}

In the case of dipole radiation, a direct coupling between the dark photon field and photons occurs through a dimension-6 operator, Eq.\,(\ref{Eq:Odipoleradiation}). This interaction enables vector solitons to emit electromagnetic radiation when they are situated within external electromagnetic fields. Specifically, when considering an external spatio-temporally constant magnetic field \(\bar{B}\), the time-averaged power assumes a different form depending on whether we account for the presence of a medium at the background level or not. The presence of a medium would provide photons with an effective mass. Recasting Eqs.\,(\ref{eq:PDRa}) and (\ref{eq:PDRplasma2c}), we have 
\begin{equation}
\label{eq:P4dipoleres}
\langle P_{(4)}\rangle_t \sim (gm_{\text{pl}})^4\frac{\bar B^2}{m^2}\frac{1}{(mR^{0.24}_{\text{sol}})^2} \times \mathcal{F}\left(m R^{0.24}_{\text{sol}}, \frac{\omega_p}{m}\right)\,,
\end{equation}
where
\begin{equation}
\mathcal{F}\left(m R^{0.24}_{\text{sol}}, \frac{\omega_p}{m}\right) =
\begin{cases}
    & e^{-2\pi m R^{0.24}_{\text{sol}}}\,,\,\,\,\,\,\,\,\,\,\,\omega_p \approx 0\\
    &\frac{\pi^5}{9\sqrt{2}} (1-\frac{\omega_p^2}{4m^2})^{1/2}\,,\,\omega_p \approx 2m
\end{cases}
\label{eq:F}
\end{equation}
encodes the soliton shape and plasma effects. The radiation is emitted at a frequency 
\begin{equation}
\nu \approx \frac{2m}{2\pi} = 0.5\,\text{GHz} \left( \frac{m}{10^{-6}\,\text{eV}} \right)\,.
\end{equation}
The strength of the electromagnetic signal can be characterized by its spectral flux density $S_B$. Assuming the source is located at a cosmic distance $d$ and emits a time-averaged total power $\langle P_{(4)} \rangle_t$, the spectral flux density observed at Earth is expressed as follows:
\begin{align}
S_B&=\frac{   \langle P_{(4)} \rangle_t }{4\pi\; \Delta\nu\;d^2}\,, \nonumber \\
 &\sim (gm_{\text{pl}})^4\frac{\bar B^2}{m^3}\frac{\pi}{(2mR^{0.24}_{\text{sol}}d)^2} \times \mathcal{F}\left(m R^{0.24}_{\text{sol}}, \frac{\omega_p}{m}\right)\,, 
\end{align}
where we have taken $\Delta \nu \sim \nu$ following the estimate performed in Ref.\,\cite{Amin:2021tnq}.
Plugging numbers into the above equation, we have 
\begin{align}
S_B &\sim 3 \times 10^2\,\text{Jy}\,  \left( \frac{g_{W\gamma}}{m_{\text{pl}}} \right)^4 \left( \frac{\bar B}{4.5 \times 10^{12}\,\text{G}} \right)^2\left( \frac{10^{-6}\,\text{eV}}{m} \right)^3\nonumber \\
&\left( \frac{3.5}{mR^{0.24}_{\text{sol}}} \right)^2
\left( \frac{0.1\,\text{Mpc}}{d} \right)^2\times 
\mathcal{F}\left(m R^{0.24}_{\text{sol}}, \frac{\omega_p}{m}\right)\,,
\label{eq:SB}
\end{align}
where Jy stands for a Jansky, a standard radio unit for spectral flux density defined as Jy = 10$^{-23}$  erg s$^{-1}$ cm$^{-2}$ Hz$^{-1}$. We have taken $(\mu/m) = 10^{-1}$ and $g_{W\gamma}m_{\rm pl} = 10^{-1} \times (\mu/m)^{-1}$ to satisfy the non-relativistic regime and the EFT validity condition barely (but still outside of the parametric resonance regime, Eq.\,(\ref{eq:PRcond})).

The expression for \( S_B \) given in Eq.\,(\ref{eq:SB}), which shows the expected spectral flux density based on factors such as the dark photon-photon coupling strength, the dark photon mass, and the soliton shape, needs to be compared with those from Earth or space-based facilities, depending on the dark photon mass. Radio telescopes typically have sensitivities around \( \sim 100\, \mu\text{Jy} \) at approximately \( \sim 100\, \text{kHz} \), with a resolution bandwidth of about \( \sim \text{kHz} \) (see, for example, Table 1 in Ref.\,\cite{SKA}). Regarding space-based facilities reserved for low frequencies below $30$ MHz, the projected sensitivity of the OLFAR mission would be limited by cosmic confusion (unresolved background sources), not by receiver noise. For the sake of analysis, we may consider a typical sensitivity similar to that of Earth facilities.

 In a vacuum environment, the expected spectral flux density is exponentially suppressed as shown in Eq.\,(\ref{eq:F}), unless the soliton configuration is particularly dense, e.g., when \( mR^{0.24}_{\text{sol}} \lesssim \mathcal{O}(10^2) \). Even in such a case, the average magnetic field strength across the Milky Way is approximately \( \sim (1-10)\, \mu\text{G} \) \cite{2001SSRv...99..243B}, resulting in a typical spectral density that is tens of orders of magnitude lower than a Jansky.

However, when considering the presence of a non-negligible medium, the spectral flux density not only avoids exponential suppression but can also be significantly enhanced as the plasma frequency approaches twice the dark photon mass, as illustrated in Fig.\,\ref{Peak} (top panel). This resonance phenomenon is likely to occur in highly magnetized environments, such as those found around neutron stars or white dwarfs, which hold strong magnetic fields in the range $(10^{10}-10^{14})$\,G\,\cite{Faucher_Giguere_2006}  and $(10^{4}-10^{9})$\,G\,\cite{2000PASP..112..873W}, respectively. In these scenarios, the spectral flux density may exceed a Jansky by tens of orders of magnitude. 

\subsubsection{Kinetic Mixing}

In the kinetic mixing case, the vector field directly couples to the electromagnetic current of ordinary matter. The expression for the time-averaged power changes when the effective photon mass is considered, compared to when it is ignored. Recasting Eqs.\,(\ref{eq:KMnowpl}) and (\ref{eq:KMyeswpl}), we have 
\begin{equation}
\label{eq:Pkineticres}
\langle P\rangle_t \sim \frac{\epsilon^2 m_{\text{pl}}^2n_e^2}{(m_e m^{2})^2} \times \mathcal{F}\left(m R^{0.24}_{\text{sol}}, \frac{\omega_p}{m}\right)\,,
\end{equation}
where
\begin{equation}
\mathcal{F}\left(m R^{0.24}_{\text{sol}}, \frac{\omega_p}{m}\right) =
\begin{cases}
    & e^{-\pi m R^{0.24}_{\text{sol}}}\,,\,\,\,\,\,\,\,\,\,\,\omega_p \approx 0\\
    &(m R^{0.24}_{\text{sol}})^2 (1-\frac{\omega_p^2}{m^2})^{1/2}\,,\,\omega_p \approx m
\end{cases}
\label{eq:Fmix}
\end{equation}
encodes the soliton shape and plasma effects. The radiation is emitted at a frequency 
\begin{equation}
\nu \approx \frac{m}{2\pi} = 0.25\,\text{GHz} \left( \frac{m}{10^{-6}\,\text{eV}} \right)\,.
\end{equation}
We estimate the typical spectral flux density as
\begin{align}
S_B&=\frac{   \langle P \rangle_t }{4\pi\; \Delta\nu\;d^2}\,, \nonumber \\
 &\sim \frac{\epsilon^2 m^2_{\text{pl}}n_e^2}{2 m_e^2 m^5 d^2}
  \times \mathcal{F}\left(m R^{0.24}_{\text{sol}}, \frac{\omega_p}{m}\right)\,, 
\end{align}
where we have taken $\Delta\nu \sim \nu$ as before. Plugging numbers into this equation yields the following:
\begin{align}
S_B  &\sim 10^{-16}\,\text{Jy}\,
\left( \frac{n_e}{0.03\,\text{cm}^{-3}} \right)^2
\left( \frac{\epsilon}{10^{-11}} \right)^2
\left( \frac{10^{-6}}{m} \right)^5
\nonumber \\
&\hspace{1.5 cm}\times \left( \frac{0.1\,\text{Mpc}}{d} \right)^2 \times \mathcal{F}\left(m R^{0.24}_{\text{sol}}, \frac{\omega_p}{m}\right)\,.
\label{eq:SBKM}
\end{align}
If we consider the typical electron number density of the interstellar medium, $n_e \sim 0.03 \text{cm}^{-3}$, and take massless photons, the spectral flux density is typically several orders of magnitude weaker than the typical facilities sensitivity, even in the case when $mR_{\text{sol}}^{0.24} \sim 1$, where the exponential suppression in Eq. (\ref{eq:Fmix}) is still mild. 

When considering the effective mass of a photon close to the dark photon mass, the exponential suppression is eliminated, and the signal is significantly enhanced due to a resonance phenomenon as illustrated in Fig.\,\ref{Peak} (bottom panel). This scenario is particularly relevant when vector solitons are close to neutron stars or magnetars, where the motion of charged constituents generates free charge and current densities in their magnetospheres. The typical electron charge density around neutron stars and white dwarfs may reach values as high as of  $\sim(10^{9}-10^{13})\,\text{cm}^{-3}(10^{-1}\,\text{s}/P)$  and $\sim(10^{-2}-10^{3})\,\text{cm}^{-3}(10^4\,\text{s}/P)$, respectively, as shown Eq.\,(\ref{eq:nealigned}) for the case of aligned compact objects. Considering this electron charge density 
and the resonant enhancement driven by the factor 
$\mathcal{F}\left(m R^{0.24}_{\text{sol}}, \frac{\omega_p}{m}\right)$, especially for $m R^{0.24} \gg 1$, the predicted flux density may easily reach a few tens of Jansky.

\textcolor{black}{Finally, we emphasize that our results for both phenomena, dipole radiation and kinetic mixing, rely on the assumption that the magnetosphere is described by the GJ model, which is expected to accurately characterize the plasma distribution along closed magnetic field lines. Deviations from this picture arise primarily along open field lines and in return-current regions, where pair production can enhance the plasma density. This effect is particularly efficient in young pulsars.
However, these regions occupy only a small fraction of the stellar surface \cite{Belyaev_2016}, and are therefore geometrically suppressed. As a result, while local deviations from the GJ density can be significant, their impact on our results is limited, with the dominant effect arising from localized shifts in the conversion region rather than large global changes.}

\subsection{Solitons-Neutron Stars/White dwarfs: tidal disruption}

 Our analysis of the emitted radiation power via resonance conversion of dark photons within solitons (embedded in the magnetosphere of a neutron star or white dwarf) assumes that the soliton remains intact and is not disrupted by the compact object's gravitational field, as described in Eqs. (\ref{eq:PDRplasma2b}) and (\ref{eq:kmapp}). To determine the threshold for tidal disruption, we require that the Jacobi radius, $R_\mathrm{J}$,\,\footnote{We have used the Jacobi radius as a good first-order approximation to the scale of tidal forces. We have numerically checked that results remain closely the same if we use the Roche limit to analyze tidal disruption\,(see page 431 in Ref.\,\cite{shu1982}).} equals the soliton’s physical radius, $R_\text{sol}^{0.95}$, in Eq.\,(\ref{eq:Rsol095}). Then, the minimum distance $R_{\mathrm{0}}$ at which the soliton can approach the compact object without being subjected to tidal disruption is given by numerically solving (see Eq. 8.89 in Ref.\,\cite{binney2011galactic})
 \begin{align}\label{eq:roche}
    &\frac{M_{\mathrm{co}}}{(R_{\mathrm{0}}-R_\text{sol}^{0.95})^2}-\frac{M_\text{sol}}{(R_\text{sol}^{0.95})^2}-\nonumber\\
   &\hspace{0.6cm} \frac{M_{\mathrm{co}}+M_\text{sol}}{R_0^3}\left(\frac{M_{\mathrm{co}} R_{\mathrm{0}}}{M_{\mathrm{co}}+ M_\text{sol}}-R_\text{sol}^{0.95}\right)=0\,,
\end{align} %eq.(8.90) in the textbook
with $M_\text{sol}$ and $M_{\mathrm{co}}$ as the soliton mass, Eq.\,(\ref{eq:Msol}), and the mass of the compact object, respectively.

For a given dark photon mass and the properties of a compact object, the conversion radius can be calculated using Eq.\,(\ref{eq:barwpl}). This requires setting $\omega_{\text{pl}}$ to $2m$ or $m$, depending on whether we are analyzing dipole radiation or the kinetic mixing phenomena, respectively. As long as the conversion radius exceeds $R_0$, the effects of tidal disruption can be reasonably ignored.

%%%%%%%% FIG.4 %%%%%%%%
\begin{figure*}[!htbp]
\centering
 \includegraphics[width= 0.66\textwidth]{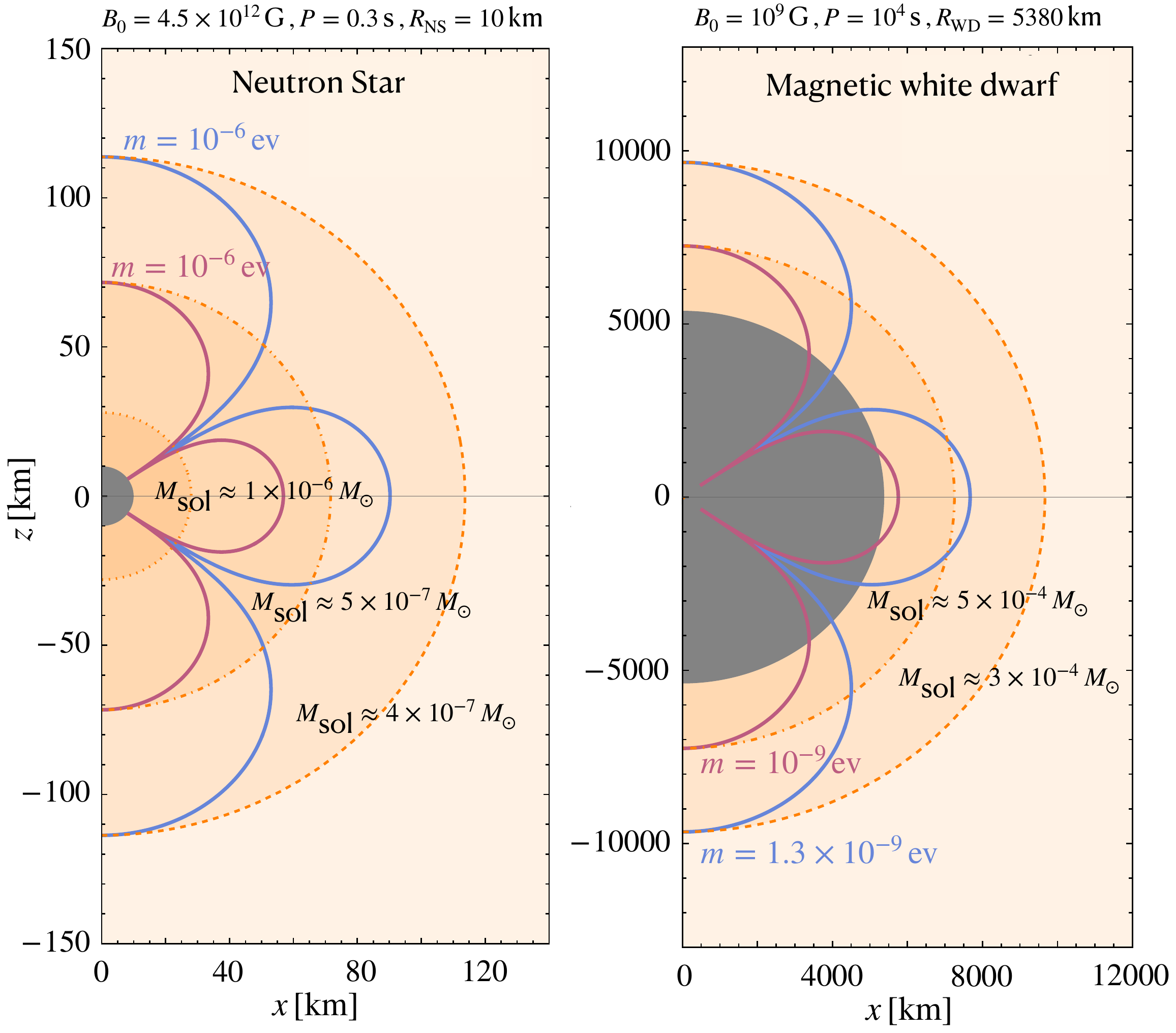}
\caption{Parameter space in cylindrical coordinates for the conversion radius within a magnetosphere via dipole radiation (purple) or kinetic mixing (blue) phenomena. The origin is located at the center of the compact object (gray circle).\textit{ Left:} aligned neutron star with properties $R_{\rm{NS}}=10\;\rm{km}$,  $B_0=4.5\times10^{12}$ G, and $P=0.3$ s. Dark photon mass $m = 10^{-6}\,\text{eV}$. \textit{Right:} aligned white dwarf with properties $B_0=10^{9}$ G, $P=10^{4}$ s and $R_{\rm{WD}}=5380\;\rm{km}$. Dark photon mass $m = 10^{-9}\,\text{eV}$ (dipole radiation) and $m = 1.3\times 10^{-9}\,\text{eV}$ (kinetic mixing). Orange regions indicate the minimum distance that solitons can approach the compact object before beginning to undergo tidal disruptions for certain values of $M_{\rm{sol}}$.}
 \label{fig:polartwo}
\end{figure*}
%%%%%%%%%%%%

For the case of neutron star-soliton encounters, the plasma frequency within neutron star magnetospheres may reach values about the dark photon masses in the range of masses about $(10^{-7}-10^{-3})$ eV, leading to an output of radio waves via resonance in the $(0.050-50+)$\,GHz regime within the range of radio telescopes\,(see Table 1 in \cite{SKA}).  In the left panel in Fig.~\ref{fig:polartwo}, we show the polar dependence of the conversion radius within a soliton which is composed of dark photons with a mass $m=10^{-6}\,\rm{eV}$, embedded in the magnetosphere of an aligned neutron star with radius $r=10\;\rm{km}$ (gray central circle), and magnetic field at poles and spin period equal to $B_0=4.5\times10^{12}$ G and $P=0.3$ s, respectively. The two interaction mechanisms of our interest are considered: dipole radiation (purple lines) and kinetic mixing (blue line). The difference between the two cases arises from the distinct resonance conditions. For kinetic mixing, conversion occurs when the plasma frequency satisfies $\omega_{\rm{pl}} = m$, while for dipole radiation, the condition is $\omega_{\rm{pl}} = 2 m$. Higher plasma frequencies are found closer to the neutron star, which means the $\omega_{\rm{pl}}= 2 m$ condition is satisfied at smaller radii. This results in a smaller conversion radius for the dipole radiation scenario than for the kinetic-mixing case. \textcolor{black}{Using Eq.\,(\ref{eq:wplaligned}) for aligned compact stars, the radius at which the resonance condition is satisfied takes the form\,\cite{Long:2024qvd}
\begin{align}
 r_\text{res} \approx 17 &R_0\, \left( \frac{B_0}{10^{14}\,\text{G}} \right)^{1/3} \left( \frac{P}{1\,\text{s}} \right)^{-1/3} \nonumber  \\
&\times \left( \frac{mf  }{10^{-6}\,\text{eV}} \right)^{-2/3} |3\text{cos}^2(\theta)-1|^{1/3}
\label{eq:rres}
\end{align}
where $f =(1,2)$ for kinetic mixing and dipole radiation, respectively.}
%%%%%%%%%%%%

The asymmetry in the purple and blue lines with respect to the equatorial axis arises from the angular dependence in the plasma frequency via the factor $|3\cos^2(\theta)-1|^{1/2}$, Eq.\,(\ref{eq:wplaligned}). Interestingly, the plasma frequency approaches zero at $\theta\approx \pm 0.96\;\rm{rad}$, when $(3\cos^2(\theta)-1) = 0$, which prevents resonant conversion at any radius along these directions.

The orange-shaded regions indicate tidal disturbance zones based on the soliton mass or, equivalently, the soliton radius from Eq.\,(\ref{eq:massradius095}). For instance, all solitons with masses $M_{\text{sol}}\gtrsim 4\times 10^{-7}\,M_{\odot}$ and $M_{\text{sol}}\gtrsim 5\times 10^{-7}\,
M_{\odot}$ will experience resonance due to kinetic mixing and dipole radiation mechanisms, respectively, before disruption. 

For the case of a soliton near a white dwarf, we consider an aligned highly magnetic white dwarf, with mass, radius, magnetic field at poles, and rotation period equal to $M_\text{WD}=M_\odot$,  $R_\text{WD}=5380$ km, $B_0=10^9$ G, and $P = 10^4$ s, respectively. Due to the relatively weak magnetic field and slow rotation compared to typical neutron stars, the plasma frequency outside the star remains low, making it insufficient to match larger dark photon masses in the interesting range for radio waves output. In our particular example, it is needed that the dark photon mass satisfies
the condition $m < 2.2\times 10^{-9}\,\text{eV}$ (kinetic mixing) and $m < 1.1\times 10^{-9}\,\text{eV}$ (dipole radiation) so that the resonance occurs at a radius $r > R_{\text{WD}}$. On the other hand, such light masses produce signals at frequencies that are
screened by the Earth’s ionosphere and thus cannot be
detected by terrestrial radio telescopes. As mentioned previously, one option is to consider space-based facilities such as OLFAR\,\cite{2020AdSpR..65..856B, 2016ExA....41..271R}, which will operate in the range $\sim(0.3-30)\,\rm{MHz}$\,\cite{2025rajan}. In this frequency range, the mass range for dark photons of our interest would be  $1.25\times 10^{-9}\;\rm{eV} \leq \textit{m} \leq 1.25\times 10^{-7}\;\rm{eV}$ (kinetic mixing) and $0.625\times 10^{-9}\;\rm{eV} \leq \textit{m}  \leq 0.625\times 10^{-7}\;\rm{eV}$ (dipole radiation). Consequently, there will be a small frequency mass window at which the resonance process can occur in the white dwarf magnetosphere, producing radiation in the frequency range projected by OLFAR. Although the small size of this window may imply that white dwarfs are less relevant to the potential detectability of the predicted signal, it is still worthwhile to consider these compact objects. This is due to their greater abundance in the Galaxy compared to neutron stars\,\footnote{Dipole radiation from axion stars and oscillons situated within white dwarf magnetospheres was investigated in Ref.\,\cite{Amin:2021tnq}. However, the authors did not address the expected detectability issues we are highlighting here for the case of vector solitons.}.

The right panel in Fig.\,\ref{fig:polartwo} shows the polar dependence for the conversion radius via kinetic mixing and dipole radiation processes, where we have considered a dark photon mass equal to $m = 1.3\times10^{-9}\,\text{eV}$ and $m = 10^{-9}\,\text{eV}$, respectively. We observe that solitons with masses $M_{\text{sol}}\gtrsim 3\times10^{-4}\,M_{\odot}$ and $M_{\text{sol}}\gtrsim 5\times10^{-4}\,
M_{\odot}$ will experience resonance under kinetic mixing and dipole radiation mechanisms, respectively, before they undergo disruption. 

In both panels, lighter solitons than those indicated in Fig.\,\ref{fig:polartwo} would start to undergo tidal disruption due to the gravitational well potential of the compact object before they experience resonant dark photon-photon conversion through dipole radiation or kinetic mixing. In this case, Eqs.\,(\ref{eq:P4dipoleres}) and (\ref{eq:Pkineticres}) should be regarded as an upper limit for the emitted radiation power.  

\subsection{Validity regime of our analytic estimates}

Our estimates assume that the plasma frequency remains approximately spatially constant on the typical soliton length scale and varies over timescales that are much longer than the duration for which the soliton is present in the region of interest.\,\footnote{This assumption aligns with spatio-temporally constant external background fields (dipole radiation analysis) and charged density (kinetic mixing analysis).} 

According to Eq.\,(\ref{eq:Rsol095}), which determines the soliton radius, and Eqs. (\ref{eq:omegapl}), (\ref{eq:nealigned}), and (\ref{eq:wplaligned}), which characterize the magnetosphere in the GJ model, our estimates for the emitted radiation power and spectral flux density may still provide a reasonable first approximation for aligned neutron stars and white dwarfs, and denser soliton configurations. In the first case, we have a temporally constant plasma frequency in a co-rotating frame with the star. In the second scenario, we may approximate the plasma frequency at the soliton's scale length as being spatially uniform.
For instance, taking $\mu/m \lesssim 0.1$, the soliton radius reads as 
$R^{0.95}_{\text{sol}} \gtrsim 2 \times 10^{-3}\,\text{km}$. We see that spatially smaller solitons hold typical length scales much smaller than the typical light cylinder radius, Eq.\,(\ref{eq:RLC}), especially for long periods of star rotations. 

\textcolor{black}{The plasma frequency, electron number density, and magnetic field in the magnetosphere, given in Eqs.\,(\ref{eq:nealigned}), (\ref{eq:wplaligned}), and (\ref{eq:B}), respectively,
can be described by power-law radial profiles of the form $Q(r) = Q_0\, r^{-p}$, where $Q(r)$ denotes the quantity of interest and $Q_0$ is its radius-independent part. For the three cases above one has $p = (3/2,3,3)$, respectively. The corresponding fractional radial variation  $|dQ/Q|= p\, |dr|/r$. Since the emitted power, and hence the spectral flux density, scales as $\bar B^2$ for dipole radiation and as $n_e^2$ for kinetic mixing, one correspondingly has  $|dS_B/S_B| = 2\,|dQ/Q|=2p|dr|/r$. Therefore, a useful indicator of the regime in which our approximation remains reasonably reliable is  $\Delta r/r \ll 1$, where $\Delta r$ denotes the spatial scale over which the background quantities are evaluated.}

\textcolor{black}{To make this criterion more quantitative, we estimate the fractional variation at the conversion radius and take $\Delta r \sim R^{0.95}_\text{sol}$, so that the relevant expression is $R^{0.95}_\text{sol}/r_\text{res}$.  Using Eqs.\,(\ref{eq:Rsol095}) and (\ref{eq:rres}), we have the condition
 \begin{align}
&\frac{R^{0.95}_\text{sol}}{r_\text{res}} \approx 12 (\text{km}/R_0)\,f^{2/3}
\left(\frac{10^{-6}\,\text{eV}}{m}\right)^{1/3}\left(\frac{\mu/m}{10^{-11}}\right)^{-1/2}\nonumber\\
&\times
\left(\frac{B_0}{10^{14}\,\text{G}}\right)^{-1/3}\left(\frac{P}{1\,\text{s}}\right)^{1/3}
|3\text{cos}^2(\theta)-1|^{-1/3} \ll 1\,,
\label{eq:CondReliable}
 \end{align}
where the soliton compactness is controlled by the ratio $\mu/m$. By inspection, we see that the denser soliton, the more reliable the approximation becomes.}
 
\textcolor{black}{ We calculate Eq.\,(\ref{eq:CondReliable}) for all allowed polar angles, namely those for which  resonance occurs at a radius $r_\text{res} > R_0$. The allowed set of angles depends on the compact-star parameters and on the dark photon mass according to $\theta \in [0,\theta_1)\, \cup (\theta_2,\theta_3)\, \cup\, (\theta_4,\pi)$, where \,\cite{Long:2024qvd, Nurmi:2021xds} 
 \begin{equation}
 \resizebox{1\columnwidth}{!}{
$\theta_{1, 2} = \text{arccos}\left[ \sqrt{\frac{1}{3}\left( 1 \pm \frac{1}{22.4^3} \left( \frac{m}{0.66\,\mu \text{eV}}\right)^2\left( \frac{P}{1\, \text{s}}\right) 
\left( \frac{10^{14}\,\text{G}}{B_0}\right)\right)} \right]\,,$}
 \end{equation}
 and $\theta_3 = \pi - \theta_2$, $\theta_4 = \pi - \theta_1$, and 
 $0 \leq \theta_1 < \theta_2 < \theta_3 < \theta_4 \leq \pi$.}
 %
 %%%%%%%% FIG.5 %%%%%%%%
\begin{figure*}[!htbp]
\centering
 \includegraphics[width= 1.0\textwidth]{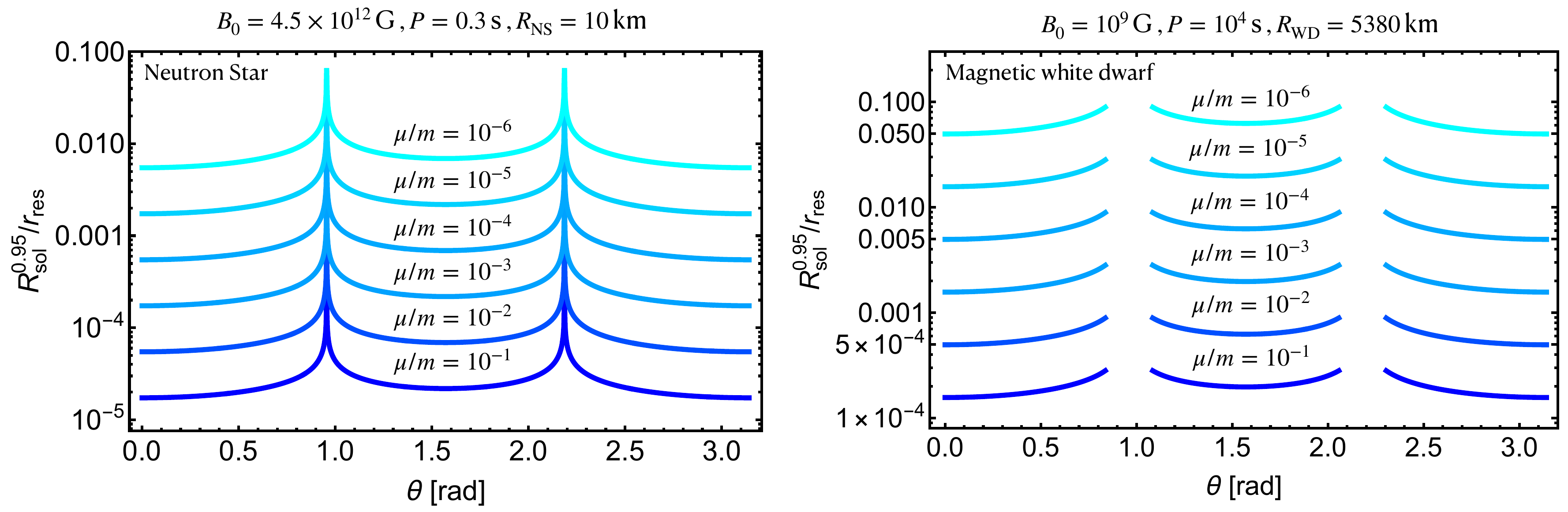}
\caption{\textcolor{black}{Ratio $R_\text{sol}^{0.95}/r_\text{res}$ in terms of the polar angle for the kinetic mixing phenomenon. \textit{ Left:} aligned neutron star with properties $R_{\rm{NS}}=10\;\rm{km}$,  $B_0=4.5\times10^{12}$ G, and $P=0.3$ s. Dark photon mass $m = 10^{-6}\,\text{eV}$. \textit{Right:} aligned white dwarf with properties $B_0=10^{9}$ G, $P=10^{4}$ s and $R_{\rm{WD}}=5380\;\rm{km}$. Dark photon mass $m = 1.3 \times 10^{-9}\,\text{eV}$. In both panels, the ratio $\mu/m$ goes from $10^{-1}$ (darkest blue) to $10^{-6}$ (lightest blue).}}
 \label{fig:RegimeValidity}
\end{figure*}
%%%%%%%%%%%%

\textcolor{black}{Figure\,\ref{fig:RegimeValidity} shows the ratio $R_\text{sol}^{0.95}/r_\text{res}$ for $r_\text{res} > R_0$ as a function of the polar angle for a soliton near a neutron star (left panel) and a magnetic white dwarf (right panel). We have used the same compact-star properties and dark photon mass adopted in Fig.\,\ref{fig:polartwo} for the kinetic mixing case (the dipole radiation case follows a similar pattern).
We see that for $\mu/m \gtrsim 10^{-3}$, we have  $R_\text{sol}^{0.95}/r_\text{res} \lesssim 10^{-3}$. This indicates that our preliminary estimates are more reliable for denser solitons, although a more detailed analysis, potentially involving numerical simulations, would be beneficial for further exploration.}

\textcolor{black}{We note that in the regime where our estimates are less reliable, the soliton is also more dilute and therefore less resistant to tidal disruption. For the two particular cases studied here, when $\mu/m \gtrsim 10^{-3}$ the soliton undergoes resonance without disruption for all polar angles. As the ratio $\mu/m$ decreases further, however, the tidal radius also decreases. For example, for $\mu/m \lesssim 10^{-6}$ and a soliton/magnetic white dwarf encounter, the tidal radius is always larger than the resonant conversion radius for the fiducial parameters used in the right panel of  Fig.\,\ref{fig:RegimeValidity} ($r_\text{tidal} \gtrsim 10^3\, \text{km}$ and $M_\text{sol}\lesssim 3\times 10^{-4}\,M_\odot$), in agreement with the right panel of Fig.\,\ref{fig:polartwo}.}

\subsection{Solitons-Neutron Stars/White dwarfs: Galactic encounter rate}

After exploring the feasibility of undergoing resonant conversion and determining the typical expected spectral flux density associated with the resonant dark photon-photon conversion, we now shift our focus to estimating the frequency of such events in the Galaxy.

It is shown in Ref.\,\cite{Amin:2021tnq} that the typical encounter rate between solitons and compact objects such as neutron stars or white dwarfs, 
$\Gamma_{\text{coll}}$, reads as
\begin{align}
&\Gamma_{\text{coll}} \sim 10^{-13}\,\text{year}\,\left(    \frac{M_{\text{co}}}{1\,\text{M}_{\odot} }\right)
\left(    \frac{R_{\text{co}}}{0.01\,\text{R}_{\odot} }\right)\nonumber\\
& \times  \left( \frac{\rho_{\text{DM}}}{0.3\,\text{GeV/cm}^3 }\right)
\left(    \frac{M_{\text{sol}}}{10^{-9}\,\text{M}_{\odot} }\right)^{-1}
\left(    \frac{v_{\text{rel}}}{10^{-3} }\right)^{-1}\,,
\label{eq:ratemustafa}
\end{align}
where $M_{\text{co}}$ and $R_{\text{co}}$ are the mass and radius of the compact object, $\bar\rho_{\text{DM}}$ is the dark matter energy density, and $v_{\text{rel}}$ is the relative velocity between the colliding 
astrophysical bodies. The current rate seems unfavorable; however, we can significantly improve it by considering a few key factors. Higher concentrations of neutron stars and white dwarfs, which are typically found at the centers of galaxies, along with the potential presence of dark matter density spikes in these regions, would increase the encounter rate. Furthermore, the geometrical cross-section should be taken proportional to \( (R_{\text{sol}} + R_{\text{co}})^2 \) instead of \( R^2_{\text{co}} \), as considered in Ref.\,\,\cite{Amin:2021tnq}, especially in the case at which \( R_{\text{sol}} \gg R_{\text{co}} \) when \( \mu/m  \) is small enough. %Additionally, the magnetic field extends well beyond the boundary of the star and scales with the radius as \( r^{-3} \) for a magnetic dipole. 
In this subsection, we estimate the rate of encounters between solitons and neutron stars, as well as between solitons and white dwarfs, while accounting for these factors.

The rates of collisions between soliton-neutron stars and white dwarfs are both proportional to the square of the dark soliton number density, \( n_\mathrm{sol} \), indicating that this rate could significantly increase under favorable astrophysical conditions. One such condition is the presence of supermassive black holes (SMBHs) located at the centers of galaxies. It is well known that nearly all large galaxies contain these massive objects~\cite{1995ARA&A..33..581K, 2013ARA&A..51..511K}. The formation of SMBHs may result in a highly concentrated, or ‘spiky’ dark matter profile in the central regions of galaxies~\cite{Gondolo:1999ef, Bertone:2002je, Zhang:2025mdl}, as evidenced by observations\,\cite{Chan:2024yht}. These spiky dark matter profiles have been extensively used to impose significant limits on the Galactic dark matter annihilation rate \,\cite{2000PhLB..494..181G, Bertone:2002je, Fields:2014pia,  2022SCPMA..6569512L, Balaji:2023hmy}. They also provide a potential explanation for the origin of fast radio bursts through interactions between neutron star and primordial black hole interactions\,\cite{Amaral:2023ekd, Kainulainen:2021rbg}, as well as constraining the fraction of dark photon solitons via their decay in the aftermath of merging events\,\cite{Amaral:2025fcd}.

In Ref.~\cite{Gondolo:1999ef} and subsequent studies\,\cite{Bertone:2002je, Zhang:2025mdl, Chan:2024yht}, it was shown that when supermassive black holes (SMBHs) at the centers of large galaxies grow more slowly than the typical relaxation time of dark matter halos (adiabatic growth), a spiky dark matter profile is anticipated to develop around them.
Beginning with a pre-formation profile defined by the power law $\rho_{\text{DM}} \propto r^{-\gamma}$, the outcome is a radial dark matter spiky profile, $\rho_\mathrm{DM,sp}$, of the form~\cite{Zhang:2025mdl}
\begin{equation}
\begin{split}
&\rho_{\text{DM,sp}}(r)= 0.0263\,M_{\odot}\,\mrm{pc^{-3}} \left[
10^b \left( \frac{M_{\text{SMBH}}}{M_{\odot}} \right)^a \right. \\
&~~~~\times \left.\left( \frac{G_\mrm{N} M_{\text{SMBH}}}{r} \right)^{\gamma_\mathrm{sp}} \left( 1-\frac{4G_\mrm{N} M_{\text{SMBH}}}{r} \right)^{\eta}\right]
\,,
\end{split}
\label{eq:nspiky}
\end{equation}
where $r$ is the radial distance from the galactic center, $M_\mrm{SMBH} = 4.3 \times 10^6\,\text{M}_{\odot}$ is the Galactic SMBH (Sagittarius A$^*$) mass\,\cite{2017ApJ...837...30G}, and $\{a, b, \gamma_\mathrm{sp}, \eta\}$ correspond to the fit parameters show in Table \ref{tab:best-fit-params} for different initial power-law indices $\gamma$. 
% ------------------------------------
\begin{table}[t!]
\centering
\renewcommand{\arraystretch}{1.5}
    \centering
    \begin{tabular*}{\columnwidth}{@{\extracolsep{\fill}}lcccc}
    \toprule\midrule
$\gamma$ & $a$   & $b$  & $\gamma_\mathrm{sp}$ 
                & $\eta$      \\
    \midrule
    $1$     & $-1.612$ & $31.35$       & $2.09$  & 
    $2.00$   \\
    %\hline
    $1.25$ & $-1.642$ & $32.05$ & $2.11$ & $2.01$  \\            
    %\hline
    $1.5$ & $-1.677$ & $32.81$ & $2.13$ & $2.01$   \\
    %\hline
    $1.75$ & $-1.714$ & $33.70$ & $2.16$ & $2.04$  \\
    \midrule\bottomrule
  \end{tabular*}
  \caption{Best-fit parameters for the spiky dark matter profile $\rho_\mrm{DM, sp}$ formed around Galactic supermassive black holes\,\cite{Zhang:2025mdl}.}
    \label{tab:best-fit-params}
  \end{table}
% ------------------------------------
Equation (\ref{eq:nspiky}) holds for $2r_{\text{Sch}} \lesssim r \lesssim r_{\text{sp}}$, where 
 $r_{\text{Sch}} \equiv 2G_\mrm{N} M_{\text{SMBH}}$ is the Schwarschild radius and $r_{\text{sp}}(\gamma)$ is the spiky radius. This radius is defined by the radius at which the dark matter spike profile matches the expected one for a standard Navarro-French-White profile\,\cite{Fakhry:2024kjj}.  

For the Milky-Way, we have $r_{\text{h}} \lesssim r_{\text{sp}} \lesssim\, \mathcal{O}(10)\, r_{\text{h}}$ for $0.5 \leq \gamma \leq 2$\,\cite{Fakhry:2024kjj}. Here $r_{\text{h}}$ represents the radius of gravitational influence of the black hole, defined as $r_\mrm{h}  \equiv G_\mrm{N} M_{\text{SMBH}}/\sigma_{*}^2 \approx 3\, \text{pc}$\,\cite{Safdi:2018oeu}, where $\sigma_{*}$ is the typical stellar velocity dispersion of the host bulge. We adopt a conservative approach and calculate the merger rates within the inner spherical region of the Galaxy, defined by $2r_{\text{Sch}} \lesssim r \lesssim r_{\text{h}}$.

\begin{figure}[t]
\centering
  \includegraphics[width= \linewidth]{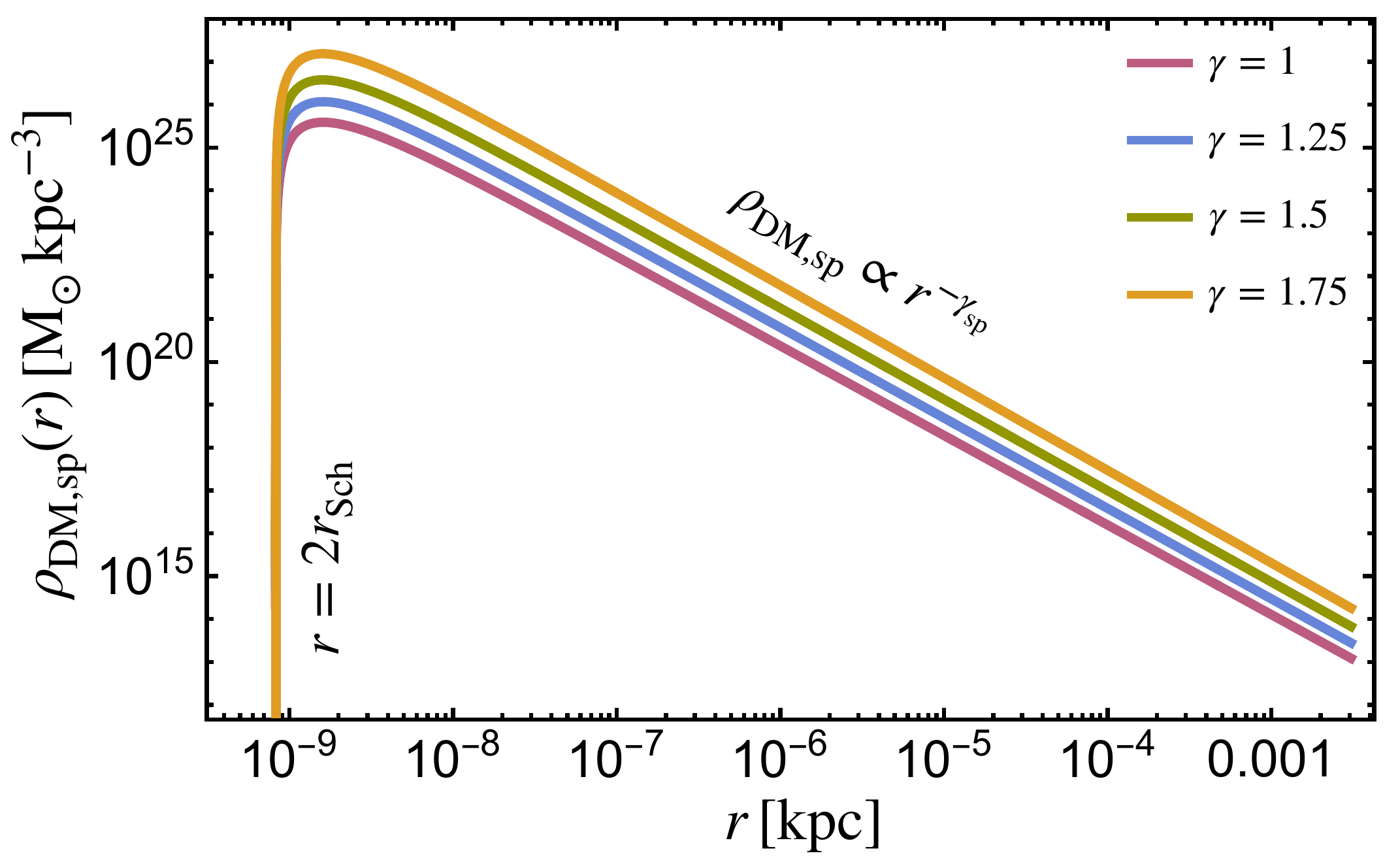}
\caption{ Dark matter spike density profiles $\rho_{\rm{DM,sp}}(r)$ as a function of distance from the galactic center for different power-laws.  The initial power-law index $\gamma$ controls the height of the central density peak. }
 \label{fig:spikeprofile}
\end{figure}

While \(N\)-body dark matter simulations suggest that the value of \(\gamma\) typically falls within the range of \(0.9 < \gamma < 1.5\)~\cite{Diemand:2008in, Navarro:2008kc}, \(N\)-body/hydrodynamical simulations that incorporate dark matter, stars, and gas have found the power-law index to be in the range of \(1.7 < \gamma < 2.1\)~\cite{Gustafsson:2006gr}. Figure \ref{fig:spikeprofile} illustrates the dark matter spikes that form from an initial power-law dark matter halo profile characterized by the index \(\gamma\). A higher initial value of \(\gamma\) results in a greater final dark matter halo energy density at a fixed radius. It is important to note that the density decreases rapidly as the radial distance \(r\) from the center of the supermassive black hole approaches twice the Schwarzschild radius. At smaller radii, particles are captured by the SMBH.

In the presence of a dark matter density spike at the Galactic center, the collision rate in the whole Galaxy is primarily dominated by the central region due to the higher density of solitons, as well as neutron stars and white dwarfs. Then, we focus on the Galactic inner region and express the soliton-neutron star/white dwarf collision rate as follows 
\begin{equation}\label{eq:collrate}
\Gamma_{\mathrm{coll}}=4\pi \!\!\int^{r_{\text{h}}}_{\mathrm{max(2r_{\text{Sch}}},r_\mathrm{tidal})}\!\!r^2\langle \sigma_\mathrm{eff} v_\mathrm{rel}\rangle\;n_{\mathrm{co}}(r)\;n_{\mathrm{sol}}\!\left(r,M_{\text{sol}}\right)\dd r\,,
\end{equation}
where the lower integration limit is set to be the maximum value between twice the Schwarzschild radius and the tidal radius, where the tidal radius is estimated by using the Roche limit defined as
\begin{align}
&r_\mathrm{tidal} = R^{0.95}_\mathrm{sol}\left( 2\frac{M_\mathrm{SMBH}}{M_\mathrm{sol}} \right)^{1/3}\,,\\
&= (2r_\mathrm{Sch})\left( \frac{10^{-6}\,\mathrm{eV}}{m} \right)^2\left( \frac{1.5\times 10^{-9}\,M_{\odot}}{M_\mathrm{sol}} \right)^{4/3}\,.
\label{eq:Rochelimit}
\end{align}
We understand that only lighter solitons formed by lighter dark photon masses will experience disruption in the innermost regions of the Galaxy, so that $r_\mathrm{tidal} > 2r_{\text{Sch}}$. The lower integration limit in Equation (\ref{eq:collrate}) guarantees that we only consider undisrupted solitons, which orbit in the Galactic halo around Sagittarius A$^*$, when calculating the collision rate. Local number densities $n_\mathrm{sol}$ and $n_{\mathrm{co}}$ refer to those from the soliton and compact objects, respectively,  and $\langle \sigma_\mathrm{eff} v_\mathrm{rel}\rangle(r)$ represents the effective cross section averaged by the relative velocity between the colliding objects. The soliton number density is computed under the assumption that solitons are non-relativistic  and constitute a fraction  $f_{\text{dm}}$ of the DM abundance, such that
\begin{align}
    n_\mathrm{sol}=f_{\text{dm}} \frac{\rho_{\text{DM,sp}}}{M_\mathrm{sol}}\,,
 \label{eq:nsol}   
\end{align}
where $\rho_{\text{DM,sp}}$ obeys Eq.\,(\ref{eq:nspiky}) and the soliton mass reads from the mass-radius relation, Eq.\,(\ref{eq:massradius095}).

The dynamics of stellar objects within the inner few pc of the Galactic Center, meaning within the sphere of gravitational influence $r_\text{h}$, are strongly influenced by the presence of Sagittarius A$^*$
at the center of the Galaxy. Stars in the nuclear stellar cluster (NSC)\,\cite{Neumayer2020} could develop cuspy distributions as suggested by the numerical simulations of the Milky Way NSC under the presence of Sagittarius A$^{*}$ performed in Ref.\,\cite{Freitag:2006qf}. These distributions have also received support from recent spectroscopy and photometric surveys\,\cite{2018A&A...609A..26G, 2018A&A...609A..27S, 2019ApJ...872L..15H}, and X-ray observations\,\cite{Hailey2018}. Astrophysically speaking, if the average relaxation time is sufficiently brief, through two-body interactions, stars will interchange orbital energy, facilitating the development of a Bahcall-Wolf density cusp\,\cite{1976ApJ...209..214B} surrounding the supermassive black hole\,\cite{Safdi:2018oeu}.

For the case of neutron stars, numerical simulations\,\cite{Freitag:2006qf} suggest that approximately $10^4$ NSs could be located within 1 pc of $\text{Sagittarius A}^*$ after about $\usim 13\,\text{Gyr}$, maintaining a number density $n_{\text{NS}}(r)$ that can be modeled by an ``eta-model"\,\cite{1993MNRAS.265..250D, 1994AJ....107..634T, Safdi:2018oeu} with $\gamma_{\text{ns}} = 1.3$, as follows
\begin{align}\label{eq:nns}
    n_{\text{NS}}(r)= \frac{\eta N_{\text{Tot} } } {4 \pi R_b^3  } \Big( \
\frac{r}{R_b}\Big)^{\eta-3} \Big( 1 + \frac{r}{R_b}\Big)^{-\eta-1}\,,
\end{align}
which captures the expected central concentration and outer falloff in systems with a massive black hole. Here, $\eta=3-\gamma_{\text{ns}}$, $N_{\text{Tot} } = 749076$, and $R_b$ is defined as the radius of break, estimated in Ref.\,\cite{Freitag_2006} as $R_b\sim 28\,\mathrm{pc} \,(2\eta-1)^{-1}$. 

\textcolor{black}{For the case of white dwarfs, we take a conservative approach, partially following Ref.~\cite{lillian2024milky}, to model their number density profile, $n_{\mathrm{WD}}(r)$,  as follows}
\begin{equation}\label{eq:nwd}
 n_{\mathrm{WD}}=f_\mathrm{MWD}n_{\mathrm{WD,0}}(r_0)\left(\frac{r}{r_0}\right)^{-\gamma_{\text{wd}}}\,,
\end{equation}
where
\begin{align}
\gamma_{\text{wd}}= \begin{cases}
1.4  & \text{for}\quad r\leq r_0\,,  \\
0 & \text{for}\quad r_0<r\leq r_h\,.
\end{cases}
\end{align}
Here $r$ is the radial distance from the Galactic center, $r_0= 1.5$ pc is the inner radius, and  $n_{\mathrm{WD,0}}(r_0) = 3.28\times 10^5 \,\text{pc}^{-3}$
to find about $8.7\times 10^6$ white dwarfs at $r\leq r_0$. The fraction of highly magnetic white dwarfs among all white dwarfs, which is the subset of our interest, is denoted by \( f_\mathrm{MWD} \).

Now that we have established the number densities of the two compact objects under study, neutron stars and white dwarfs, we turn to the analysis of the effective cross section averaged over the relative velocity between solitons and neutron stars/white dwarfs. Such a term reads as
\begin{align}
   \langle \sigma_\mathrm{eff} v_\mathrm{rel}\rangle = \int_0^{2v_{\text{esc}}} P(v_{\text{rel}} )v_\mathrm{rel} \sigma_{\text{{eff}}}(v_{\text{rel}} )\,  \dd v_{\text{rel}}\,.
  \label{eq:Csol,co}
\end{align}
In this context, \( v_{\text{rel}} \) represents the relative velocity, while \( \sigma_{\text{eff}} \) refers to the effective cross section, and  $v_{\text{esc}}(r) \approx \sqrt{2G_{\text{N}}M_{\text{SMBH}}/r}$ is the escape velocity in the Galactic inner region. This effective cross-section accounts for both the geometric surface area and gravitational focusing, with the latter enhancing the probability of a collision under appropriate conditions. Explicitly, it reads as
\begin{align}\label{eq:collision}
    \sigma_{\mathrm{eff}}(v_{\mathrm{eff}})&=\pi \left(R_\mathrm{co}+R_\mathrm{sol}^{0.95} \right)^2 \left[1+\left(\frac{v_\mathrm{esc}^{\mathrm{mutual}}}{v_
    \mathrm{rel}}\right)^2 \right]\,,
\end{align}
where $(v_\mathrm{esc}^{\mathrm{mutual}})^2=2G_N \left(M_\mathrm{co}+ M_\mathrm{sol}\right)/\left(R_\mathrm{co}+R_\mathrm{sol}^{0.95}\right)$ is the mutual escape speed between the soliton and the compact object and $p(v_\mrm{rel})$ refers to the probability density function governing their relative velocities within the Galactic center. We assume this to follow a 3D isotropic Maxwell-Boltzmann distribution, such that
\begin{equation}
\funop{P({v_{\text{rel}}})}\mrm{d}v_{\text{rel}} = 4 \pi v^2_{\text{rel}}\left( \frac{3}{2\pi\sigma^2_{\text{rel}}} \right)^{3/2}  
\exp\left(-\frac{3v^2_{\text{rel}}}{2\sigma^2_{\text{rel}}}\right) \dd v_{\text{rel}}\,,
\label{eq:Pvrel}
\end{equation}
where $\sigma_{\text{rel}}$ is the 3D relative velocity dispersion for soliton-neutron star/white dwarf encounters. The probability density function is peaked at 
$\sigma_{\text{rel}}$, which can be estimated using the spherical Jeans equation at the Galactic center as follows. 

Let us be $n_i(r)$ and $\sigma_{i}$ the local number density and dispersion velocity of the astrophysical object involved in the collision, where $i = \{ \text{sol}, \text{NS}, \text{WD} \}$ for solitons, neutron stars, and white dwarfs, respectively. Assuming that their individual dispersion velocity do not depend on the direction, meaning $\sigma_r = \sigma_\theta=\sigma_\phi \equiv \sigma_{i}$, and noting that $n_i(r) \propto r^{-\omega}$, we have
\begin{equation}
\frac{\partial(r^{-\omega}\sigma_{i}^2)}{\partial r} + r^{-\omega}\frac{\partial \Phi_\mrm{N}(r)}{\partial r} =0 \,,
\end{equation}
where $\Phi_\mrm{N}(r) \approx -G_\mrm{N} M_{\text{SMBH}}/r$ for $r \lesssim r_h$ is the gravitational potential at radius $r$ and the power-law index reads as $\omega = \{ \gamma_{\text{sp}}, \gamma_{\text{ns}}, \gamma_{\text{wd}}\}$ for solitons, neutron stars, and white dwarfs, respectively, as shown Eqs.\,(\ref{eq:nspiky}), (\ref{eq:nns}), and (\ref{eq:nwd}). The solution of this equation is given by
\begin{equation}
\sigma_{i}(r) \approx \left(1+\omega\right)^{-1/2}\left( \frac{G_\mrm{N} M_{\text{SMBH}}}{r}\right)^{1/2}\,.
\label{eq:Jeq2}
\end{equation}
The relative velocity between soliton and neutron stars and white dwarfs for $ r\lesssim r_h$, to be replaced in Eq.\,(\ref{eq:Csol,co}),  reads
\begin{equation}
\sigma_\mathrm{rel} =  
\sqrt{\sigma^2_\mathrm{sol} + \sigma^2_\mathrm{co}} \,,  
\end{equation}
where $\sigma_\mathrm{co} =\{  \sigma_\mathrm{ns},  \sigma_\mathrm{dw}  \}$ according to the collision rate under study.

\begin{figure*}[t]
    \centering
    \includegraphics[width=\linewidth]{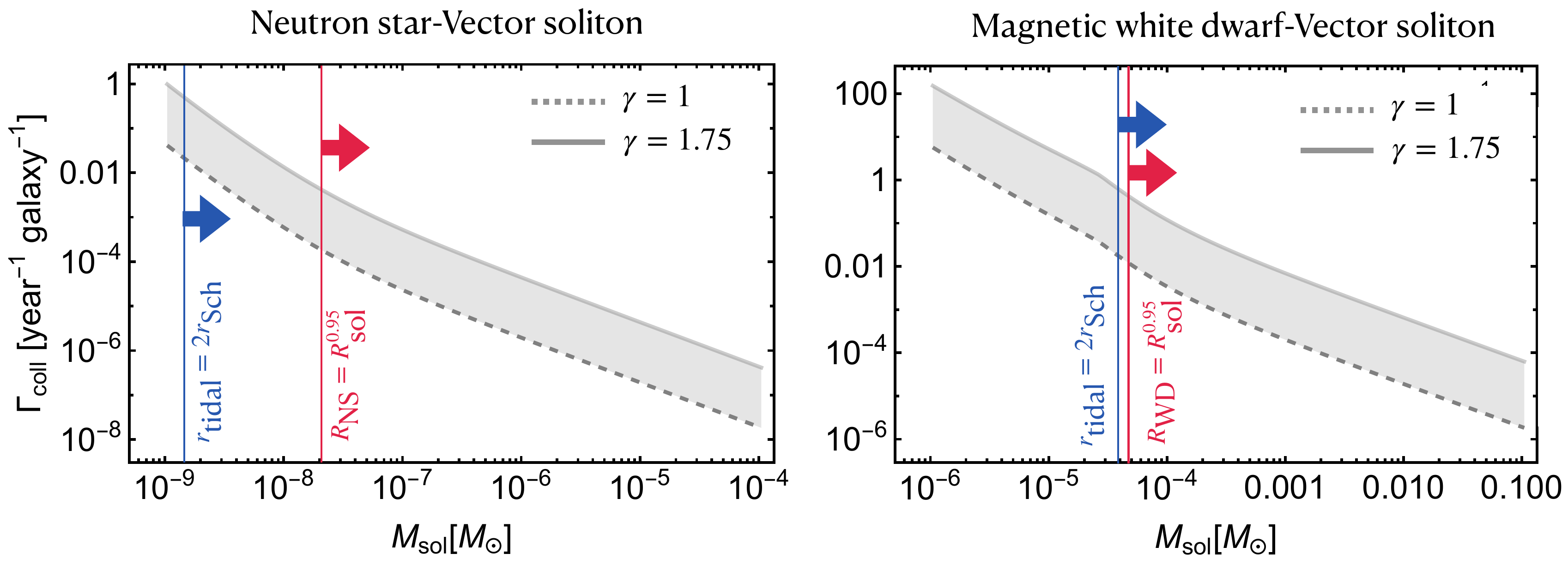}
    \caption{The gray shaded region illustrates the collision rate \(\Gamma_{\rm{coll}}\) between solitons and neutron stars in the left panel, and between solitons and magnetic white dwarfs in the right panel. The masses and radii considered are \((M_\odot, 10 \, \text{km})\) for neutron stars and \((M_\odot, 5380 \, \text{km})\) for magnetic white dwarfs, and the dark photon masses used are \(m = 10^{-6} \, \text{eV}\) and \(m = 10^{-9} \, \text{eV}\), respectively. The width of the shaded band reflects variations in the initial power-law index \(\gamma\) for the density profile of dark matter before the formation of the spike. Blue and red arrows indicate the parameter space where solitons remain stable in the Galactic center, and where the compact object radius exceeds the soliton radius, respectively. We have taken $f_\text{DM} = 1$ and $f_\text{MWD} = 0.05$ in Eqs.\,(\ref{eq:nsol}) and \,(\ref{eq:nwd}), respectively. }
    \label{fig:rate_wd}
\end{figure*}

Figure \ref{fig:rate_wd} shows the collision rate between solitons and white dwarfs as well as neutron stars in the spiky inner region of the Galaxy, expressed in terms of soliton mass. The gray shaded region indicates the number of collisions per year, based on an initial power-law index $\gamma$ for the dark matter halo density profile. This is considered before the adiabatic growth of Sagittarius A$^*$ and within the range of $\gamma = 1$ (represented by the dotted gray line) to $\gamma = 1.75$ (represented by the solid gray line). As the initial value of $\gamma$ increases, the final spike in the dark matter profile within the Galactic inner region becomes denser. Consequently, we observe that $\Gamma_\text{coll}(\gamma = 1.75) \sim 10\,\Gamma_\text{coll}(\gamma = 1)$.

For the case of neutron stars, we have considered a representative star with a mass and radius equal to $M_\text{NS} = M_\odot$ and $R_\text{NS} = 10\,\text{km}$, respectively. We have fixed the dark photon mass to $m= 10^{-6}\,\text{eV}$.
For the case of white dwarfs, we have considered highly magnetic white dwarfs holding a representative mass and radius equal to $M_\text{WD} = M_\odot$ and $R_\text{WD} = 5380\,\text{km}$\,\cite{lillian2024milky}, respectively. We have fixed the dark photon mass to $m= 10^{-9}\,\text{eV}$ and taken $f_\text{MWD} = 0.05$ in Eq.\,(\ref{eq:nwd}). Blue and red arrows indicate the parameter space where solitons remain stable in the Galactic center, and where the compact object radius exceeds the soliton radius, respectively.  For both cases, we have assumed \( f_\text{DM} = 1 \), representing the fraction of dark matter in vector solitons.

The mass-radius soliton relation, Eq.\,(\ref{eq:massradius095}), tells us that the lighter the soliton mass, the larger its radius. In terms of the ratio $\mu/m$, we have that as we move deeper into the non-relativistic regime, e.g., as $\mu/m$ decreases, solitons become wider and lighter. At a particular soliton mass, the star radius matches the soliton radius. Thus, there are clearly two regimes for the collision rate in Figs.\,\ref{fig:rate_wd}: one regime is when $R^{0.95}_\text{sol} \gg R_\text{co}$ and $\sigma_\text{eff} \approx \pi R_\text{sol}^2$, and the other is when $R^{0.95}_\text{sol} \ll R_\text{co}$ and $\sigma_\text{eff} \approx \pi R_\text{co}^2$ (see Eq.\,(\ref{eq:collision})).

It is important to note that tidal disruption begins to occur in the innermost regions of the Galaxy for sufficiently light solitons, as indicated by Equation (\ref{eq:Rochelimit}). For these lighter solitons, the lower limit of the integral in Equation (\ref{eq:collrate}) is determined by the Roche limit. This ensures that solitons are not torn apart by the SMBH's gravitational potential.

As expected, the lighter the soliton, the larger the collision rate due to the increase in the soliton number density. In such a case, $\Gamma_\text{coll} \gg \mathcal{O}(1)\, \text{year}^{-1}$ in both cases, soliton-neutron star and soliton-white dwarf encounters.

\textcolor{black}{In this work we employ dark-matter spike profiles motivated by the idealized limit of adiabatic SMBH growth at the center of the host halo. More realistic SMBH formation histories, however, need not satisfy either adiabaticity or exact centrality, and additional dynamical effects can significantly alter the spike structure, either weakening or enhancing it. For instance, scattering and dynamical heating by dense stellar populations can reduce the inner slope to $\sim 3/2$~\cite{Gnedin:2003rj, 2004PhRvL..92t1304M, 2022PhRvD.106d3018S}, rapid SMBH formation can instead give $\sim 4/3$~\cite{2001PhRvD..64d3504U}, halo mergers can wash out the cusp through dark-matter heating~\cite{PhysRevLett.88.191301}, and sufficiently massive off-center black-hole seeds may also suppress the spike~\cite{2001PhRvD..64d3504U}. On the other hand, processes such as chaotic motion in triaxial halos~\cite{2004ApJ...606..788M} or gravothermal collapse in self-interacting dark matter~\cite{2000PhRvL..84.5258O} may strengthen the central density enhancement.}

\textcolor{black}{While a full treatment of these effects lies beyond the scope of this work, their impact on our final results regarding the Galactic collision rate can be partially inferred from our analysis with different initial DM power-law profiles, $1 \leq \gamma \leq 1.75$, shown in Fig.\,\ref{fig:rate_wd}. A shallower dark matter density profile around the Galactic center would reduce the predicted soliton–neutron star/white dwarf collision rate in Eq.\,(\ref{eq:collrate}), since the dark-spike mass enclosed within a fixed radius decreases as $\gamma$ decreases (see Table II of Ref.\,\cite{Zhang:2025mdl}). }

\section{Summary and outlook}
\label{Sec.Soutlook}

The primary focus of our research is to investigate the emission of electromagnetic radiation in vector solitons—solitons made of spin-1 light particles—when they are placed in the presence of background electromagnetic fields or external charge distributions. We analyze these phenomena analytically, emphasizing how the emitted radiation depends on the properties of the solitons, such as their mass and radius, as well as on the coupling strength between dark photons and electromagnetism. 

In the framework of effective field theory (EFT), when vector solitons pass through external electromagnetic fields, we study 
the dipole radiation phenomenon triggered by a dimension-6 operator which connects dark photons with photons via the interacting Lagrangian density $\mathcal{L}^{(6)}_{\text{int}} = -\frac{1}{4} g^2_{W\gamma} W_{\mu}W^{\mu} F_{\alpha \beta} \tilde{F}^{\alpha \beta}$\,\cite{Amin:2023imi, Jain:2021pnk}. The soliton and the electromagnetic background fields induce effective charge and current densities that both source the radiation.
In the small coupling/field amplitude regime, we perform an expansion in terms of powers of $(g_{W\gamma}\phi)$, where background fields are taken to be spatio-temporally constant.

On the other hand, when vector solitons are embedded in external charge densities, we study the radiative phenomenon associated with the gauge-kinetic mixing via the coupling of the light dark photon field with charged matter through the Lagrangian density  $\mathcal{L}^{(4)}_{\text{int}} = -e\epsilon J_{\mu}W^{\mu}$\,\cite{Fabbrichesi:2020wbt}.
Charged particles, such as electrons, experience a Lorentz force within the soliton. This force arises from the dark electromagnetic field generated by the oscillating vector boson field. Assuming a spatio-temporally constant electron number density background, the oscillating current density is associated with a charge density that varies in space and time, both of which source the output electromagnetic radiation. 

Here are our key findings concerning the radiation output in electromagnetic waves due to the two phenomena under study:

\begin{center}
\textit{External electromagnetic fields (Dipole Radiation)}
\end{center}

\begin{itemize}

\item In the presence of external electromagnetic fields, the coherence oscillation feature of vector solitons, along with the coupling between dark photons and photons, results in emitted radiation peaking at a frequency of approximately \textcolor{black}{\(\nu \approx 2m/(2\pi)\)}. This radiation significantly depends on the soliton's shape and plasma effects, as described in Eq.\,\((\ref{eq:P4dipoleres})\).

\item In relation to soliton polarization, only linearly polarized solitons generate dipole radiation. In contrast, circularly polarized solitons do not exhibit time dependence in the modified Maxwell equations, as represented in Eqs. (\ref{eq:MMW3}) and (\ref{eq:MMW4}). This is because the contraction of indices between the spatial vector field components is a time-independent quantity that is proportional to the square of the soliton's radial profile, given explicitly by \(\text{Tr}[{\bf{W W}}] = \psi^2(r)/m\).

\item In a vacuum environment, the shape of the soliton significantly suppresses the radiated power, unless the vector soliton is particularly dense, meaning \(mR_\text{sol}^{0.24} \lesssim \mathcal{O}(10^2)\). Even in such cases, if we consider the typical average magnetic field strength in the Milky Way, the expected spectral flux density is tens of orders of magnitude smaller than one Jansky in the radio spectrum. This makes detection strategies in the Galaxy quite challenging.

\item Plasma effects can not only alleviate exponential suppression but also significantly enhance the radiated power through resonance when the plasma frequency approaches twice the mass of dark photons. The production of photons via resonance is considerably more efficient as we delve deeper into the non-relativistic regime, namely as the factor $m R^{0.24}_{\text{sol}}$ increases. This resonance is naturally realized in highly magnetized environments, such as those found around neutron stars and white dwarfs. Given the typical astrophysical properties of these compact objects, our preliminary estimates suggest that the spectral flux density could exceed the Jansky scale by tens of orders of magnitude, highlighting the need for a more detailed analysis.

\item Our findings are of interest in the regime in which parametric resonance of photons is blocked in dark photon solitons, e.g. when 
$g_{W\gamma} < (mR^{0.24}_{\text{sol}})^{3/2}$ or, equivalently, $M_{\text{sol}} < 62.3 (m^2_{\text{pl}}/m)(m R^{0.24}_{\text{sol}})^{-1}$. By contrast, if the parametric resonance condition is satisfied, Eq.\,(\ref{eq:PRcond}), the radiation output will be dominated by an unbounded signal due to the exponential growth of the photon occupancy.

\item    \textcolor{black}{It is also instructive to compare our results with the axion star scenario studied in Ref.\,\cite{Amin:2021tnq}. Both cases show similar qualitative behavior: the signal is highly peaked and strongly suppressed in vacuum for dilute configurations but significantly enhanced in resonant plasma environments. The key difference is that axion stars resonantly emit at $\nu \approx m/(2\pi)$, whereas in the vector case the dipole operator leads to resonant emission at $\nu\approx 2m/(2\pi)$. Moreover, for comparable DM mass and soliton radius, the vector-soliton signal is more exponentially suppressed in vacuum than the axion-star signal. This difference originates from the interaction structure. In the scalar case, the radiation power arises at the second order in perturbation theory and is proportional to the square of the Fourier transform of the soliton profile evaluated at frequency $\omega$, e.g. $\langle P_{(2)}\rangle_t\propto \left\{\mathcal{F}[a(r)](\omega)\right\}^2=\tilde{a}^2(\omega)$, where $\tilde a(r)$ is understood as the Fourier transform of the axion star profile. By contrast, the emitted power for the vector case appears only at the fourth order and is proportional to the square of the Fourier transform of the squared soliton radial profile. In this case, the radiation is emitted at frequency $2\omega$, e.g.  $\langle P_{(4)}\rangle_t\propto \left\{\mathcal{F}[\phi^2(r)](2\omega)\right\}^2=\tilde{\varphi}^2(2\omega)$, where $\tilde \varphi(r)$ is understood as the Fourier transform of the squared of the vector soliton profile. Unlike the scalar case, the vector signal depends on the soliton polarization, with circularly polarized configurations not radiating for the operator considered.
 }
\end{itemize}

\begin{center}
\textit{External charge density  (Kinetic Mixing)}
\end{center}

\begin{itemize}

\item Vector solitons influenced by a mean external charge density can emit radiation due to the kinetic mixing between the soliton field and the matter current. This radiation primarily peaks at a frequency of \textcolor{black}{\(\nu \approx m/(2\pi)\)} and is significantly affected by the shape of the soliton and plasma effects, as described in Equation \((\ref{eq:Pkineticres})\). These dependencies are similar to those discussed in the previous case.

\item Unlike the case of dipole radiation, all soliton polarizations can radiate through kinetic mixing. This is because the soliton field-dependent charge and current densities are proportional to the real-valued vector field \({\bf{W}}\), rather than being based on the trace of their contraction, \(\text{Tr}[{\bf{W}}{\bf{W}}]\), as was the case previously.

\item When considering massless photons, the radiated power is significantly suppressed due to the soliton shape. Even if we take into account the typical electron number density in the interstellar medium, and the condition \( mR_{\text{sol}}^{0.24} = \mathcal{O}(1) \)—which means the exponential suppression in Eq. (\ref{eq:Fmix}) is still mild—the estimated spectral flux density in the radio spectrum remains too weak to support an indirect detection strategy.

\item When considering plasma effects, meaning an effective mass for the photon, the exponential suppression is eliminated. Additionally, the radiated power is heavily enhanced via resonance when the effective mass of the photon is close
to the dark photon mass. This situation is particularly pertinent when vector solitons are near neutron stars or magnetars. In these environments, the motion of charged constituents produces free charge and current densities within their magnetospheres. Our preliminary estimate suggests that the predicted flux density could reach several tens of Jansky for typical electron charge densities around neutron stars and white dwarfs. This is particularly true for more dilute configurations where \(m R^{0.24}_{\text{sol}} \ll 1\). This outcome underscores the necessity for a more accurate analysis. 
\end{itemize}

Our estimates of the radiated power associated with dipole radiation and kinetic mixing phenomena are particularly important in the resonance regime. This is especially true in scenarios where dark photon solitons exist within the magnetospheres of highly magnetized compact objects, such as neutron stars and white dwarfs. In these cases, our estimates provide preliminary predictions that require refinement through numerical simulations. 

Our estimates are more reliable for denser soliton configurations, where \( m R^{0.24}_{\text{sol}} = \mathcal{O}(1) \). These configurations typically feature smaller radii and exhibit greater resistance to tidal disruption. In contrast, dilute configurations, where \( m R^{0.24}_{\text{sol}} \gg 1 \), are generally more extended and tend to experience tidal disruption before they can resonantly emit radiation. In such cases, it is essential to address the spatial variation of the plasma frequency, electromagnetic fields, and charge density, as well as soliton disruption, through dedicated numerical simulations.

Typical parameter values for neutron stars suggest that they are ideal astrophysical compact objects for generating radio emission via vector solitons via dipole radiation and kinetic mixing. In contrast, white dwarfs, which have smaller magnetic fields and larger radii than neutron stars, produce radiation output from vector solitons that falls outside the detectable spectrum on Earth. This characteristic makes them particularly interesting for extraterrestrial searches using space-based facilities\,\cite{2020AdSpR..65..856B}.

\textcolor{black}{A potentially interesting observational difference between the two mechanisms is the characteristic emission frequency. As mentioned, for a fixed dark-photon mass, dipole radiation is expected to peak around $\nu \simeq 2m/(2\pi)$, whereas kinetic mixing leads to emission around $\nu \simeq m/(2\pi)$. Therefore, if the characteristic frequency of a detected signal could be identified observationally and the source properties were reasonably well constrained, this factor-of-two difference in the peak frequency, together with the corresponding spectral flux density, could in principle help discriminate between the two channels. In practice, the usefulness of this signature will depend on the signal bandwidth and on astrophysical uncertainties associated with the compact-star environment.}

The typical encounter rate between vector solitons and neutron stars/white dwarfs
was reported in Ref.\,\cite{Amin:2021tnq} to be unfavorable, Eq.\,(\ref{eq:ratemustafa}).
However, such a rate increases by many orders of magnitude when considering the higher neutron-star and white-dwarf concentrations at the centers of galaxies, as well as the potential presence of dark matter density spikes in these regions. The
presence of supermassive black holes located at the centers of galaxies may lead, under favorable conditions, to ‘spiky’ dark matter profiles in the central regions of galaxies~\cite{Gondolo:1999ef, Bertone:2002je, Zhang:2025mdl}. In Milky Way-like galaxies, the rate may exceed unity per year, particularly for diluted configurations (Fig.~\ref{fig:rate_wd}). 

Our findings open multiple paths for further research. We plan to integrate numerical modeling of radiation and gravitation—particularly focusing on tidal disruption—to study the effects of spatially and temporally varying plasma surrounding compact stars. Our work emphasizes the importance of the strong connection among particle physics, cosmology, and astrophysics in exploring new pathways to discover dark matter.

\begin{acknowledgments}
E.D.S. thanks Mustafa A. Amin and Andrew J. Long, both faculty members at Rice University, USA, for enriching discussions during the first stages of this work. E.D.S. acknowledges support from the FONDECYT project N° 1251141 (Agencia Nacional de Investigaci\'on y Desarrollo, Chile). M.V. acknowledges support by the National Science Foundation under Grant N° PHY-2412797.
 \end{acknowledgments}

% ---------- END ----------
%
\appendix
\section{Electromagnetic equation of motions under gauge kinetic mixing}
\label{App:KM}

For clarity, we start with a general Lagrangian density which involves the two Abelian gauge bosons of interest, photons and dark photons, including an interaction between both fields as follows
\begin{align}
\mathcal{L}_{A^\mu, W^\mu} &= -\frac{1}{4}F_{\mu\nu}F^{\mu\nu} - e A_\mu J^\mu 
-\frac{1}{4}W_{\mu\nu}W^{\mu\nu}\nonumber\\
&- e_W W_\mu J_W^\mu - \frac{1}{2}m^2W_{\mu}W^{\mu}
+ \frac{\epsilon}{2}F_{\mu\nu}W^{\mu\nu}\,,
\label{eq:LKMgeneral}
\end{align}
where $F^{\mu\nu} = \partial^\mu A^\nu - \partial^\nu A^\mu$   and $W^{\mu\nu} = \partial^\mu W^\nu - \partial^\nu W^\mu$ are the photon and dark photon field stress tensor, respectively, 
$e_W$ is a dark coupling constant, $J^{\mu}$ and $J^{\mu}_W$ are the fourth current density of the ordinary and dark sectors, respectively, and $\epsilon \ll 1$ is the mixing parameter. In Eq.\,(\ref{eq:LKMgeneral}), the photon and dark photon are ``mixed" through the kinetic term \(\frac{\epsilon }{2}F_{\mu \nu }W^{\mu \nu }\), e.g. they do not propagate as independent particles. A photon would constantly oscillate into a dark photon and vice versa.

If we apply the following field transformations 
\begin{align}
A_\mu &= \tilde{A}_\mu + \frac{\epsilon}{\sqrt{1-\epsilon^2}} \widetilde{W}_\mu\,,\\
W_\mu &=  \frac{1}{\sqrt{1-\epsilon^2}} \widetilde{W}_\mu\,,
\end{align}
in the original Lagrangian density, we can eliminate the kinetic mixing term and allow fields to propagate independently as follows
\begin{align}
\mathcal{L}_{\widetilde{A}^\mu, \widetilde{W}^\mu} &= -\frac{1}{4}\widetilde{F}_{\mu\nu}\widetilde{F}^{\mu\nu} - e \widetilde{A}_\mu J^\mu 
-\frac{1}{4}\widetilde{W}_{\mu\nu}\widetilde{W}^{\mu\nu}\nonumber\\
&- e_W \widetilde{W}_\mu J_W^\mu - \frac{m^2}{2}\widetilde{W}_{\mu}\widetilde{W}^{\mu}
- e \epsilon \widetilde{W}_{\mu}J^{\mu}\,,
\label{eq:LagKM}
\end{align}
where we have taken $m/\sqrt{1-\epsilon^2} \approx m$ and $(e_W J_W^\mu + e \epsilon J^{\mu})/\sqrt{1-\epsilon^2} \approx (e_W J_W^\mu + e \epsilon J^{\mu})$. The original current \(J^{\mu }\) (which only coupled to \(A_{\mu }\)) now couples to both the new photon and the new dark photon. Varying Eq.\,(\ref{eq:LagKM}) with respect to the transformed dark photon field, we have
\begin{align}
{\bf{\dot{\widetilde{E}}}}_W - \nabla \times {\bf{ {\widetilde{B}}}}_W &= m^2 \widetilde{\bf{ W}} + e_W {\bf{ J}}_W + e\epsilon {\bf{ J}}\,\label{eq:WJ}\\
\nabla \cdot {\bf{ {\widetilde{E}}}}_W &= - e_W {\rho}_W - e\epsilon {\rho}\,,
\label{eq:WJ2}
\end{align}
where we have neglected in Eq.\,(\ref{eq:WJ2}) a mass term proportional to the time-component of the soliton field under the non-relativistic approximation. The ordinary current density term present in Equation\,(\ref{eq:WJ}) indicates that an ordinary particle behaves as if it has a tiny ``dark charge". Charged particles, like electrons, will feel a Lorentz force from the dark electromagnetic field, resulting in an ordinary current. In detail, 
\begin{align}
&F_i = -\epsilon e \left[ -\dot{\widetilde{W}}_i - \partial_i {\widetilde{W}}_0 + V_j(\partial_i{\widetilde{W}}_j - \partial_j{\widetilde{W}}_i) \right]\,,\\
&v_i(t)\sim \frac{\epsilon e}{m_e}{\widetilde{W}}_i + \text{constant}\,,
\end{align}
where we have neglected gradients of the soliton field and used $F_i = m_e \dot v_i$. Here $v_i$ is the i$th$-component of the electron velocity. As a result, the ordinary current density reads as
\begin{align}
{\bf J}(\bm x, t)  \sim -\left(\frac{ \epsilon e^2 \rho_e}{m^2_e}\right){\bf {\widetilde{W}}}(\bm x, t) + \text{constant}\,,
\label{eq:J}
\end{align}
where $\rho_e$ is the electron number density. The associated ordinary charged density can be obtained via the continuity equation as
\begin{align}
\frac{\partial \rho}{\partial t }  &= -\nabla \cdot \bf J \,,\\
\rho(\bm x, t)& \sim \frac{\epsilon e^2 \rho_e }{m^2_e} \int^t \nabla \cdot {\bf {\widetilde{W}}}(\bm x, t') dt' + \text{constant}\,.
\label{eq:rho}
\end{align}
These charge and current densities source the ordinary electromagnetic fields via the usual equation of motions obtained by varying the transformed Lagrangian density, Eq.\,(\ref{eq:LagKM}), with respect to the transformed photon field, namely
\begin{align}
&\ddot{{\bf{\widetilde E}}} - \nabla^2 {\bf{\widetilde E}} = - \nabla \rho - \dot{\bf{J}}\,,\\
&\ddot{{\bf{\widetilde B}}} - \nabla^2 {\bf{\widetilde B}} = \nabla \times {\bf{J}}\,.
\label{eq:Btilde}
\end{align}
Equations (\ref{eq:J}), (\ref{eq:rho})--(\ref{eq:Btilde}) match Eqs.\,(\ref{eq:weq1b})--(\ref{eq:jmixing}) under the approximation
$(\bf \widetilde E, \bf \widetilde B)\approx(\bf  E, \bf  B)$ and ${\bf{\widetilde W}} \approx {\bf{ W}}$ for $\epsilon \ll 1$.

\bibliographystyle{JHEP}
\bibliography{arxiv}

\end{document}